\let\clineorig\cline
\documentclass[pdflatex,sn-mathphys-ay]{sn-jnl}
\let\cline\clineorig

\usepackage{graphicx}%
\usepackage{multirow}%
\usepackage{amsmath,amssymb,amsfonts}%
\usepackage{amsthm}%
\usepackage{mathrsfs}%
\usepackage[title]{appendix}%
\usepackage{xcolor}%
\usepackage{textcomp}%
\usepackage{manyfoot}%
\usepackage{booktabs}%
\usepackage{algorithm}%
\usepackage{algorithmicx}%
\usepackage{algpseudocode}%
\usepackage{listings}%
\usepackage{subcaption}
\usepackage{tabularx}
\definecolor{changesred}{RGB}{128, 0, 32}
\definecolor{changesblue}{rgb} {0.0, 0.5, 1.0}
\newcommand\ming[1]{{\color{black}#1}}

\newcommand\revone[1]{{\color{black}#1}}
\newcommand\revtwo[1]{{\color{black}#1}}

\theoremstyle{thmstyleone}%
\theoremstyle{thmstyletwo}%

\theoremstyle{thmstylethree}%

\begin{document}

\title[Wind Direction Effects on Urban Flow and Turbulent Statistics]{Impact of Wind Direction on Flow and Turbulent Statistics Over a Realistic Urban Area: A Large-Eddy Simulation Study}


\author[1]{\fnm{J.M.} \sur{Duró}}

\author[1]{\fnm{E.} \sur{Mestres}}

\author[2]{\fnm{M.} \sur{Teng}}

\author[2]{\fnm{O.} \sur{Lehmkuhl}}

\author*[1]{\fnm{I.} \sur{Rodríguez}}\email{ivette.rodriguez@upc.edu}

\affil*[1]{\orgname{TUAREG, Turbulence and aerodynamics research group, Universitat Politècnica de Catalunya, Colom 11, 08222 Terrassa}, \country{Spain}}

\affil[2]{\orgname{Large-scale Computational Fluid Dynamics, Barcelona Supercomputing Center, Plaça Eusebi Güell, 1-3 08034 Barcelona }, \country{Spain}}


\abstract{\revone{Effects of wind direction in realistic urban canopies remain difficult to characterize systematically because local flow patterns, building-height variability, and turbulent statistics within a realistic urban canopy may respond differently to changes in the approaching wind. To investigate these multi-scale directional impacts,} we conducted high-resolution large-eddy simulations of the atmospheric boundary layer over the Zona Universitària Pedralbes district in Barcelona. \revtwo{ The computational meshes  employ fourth-order spectral elements, yielding pedestrian-level resolutions below 1 m and resulting in approximately $5.0\times 10^8$ degrees of freedom}. Sixteen wind directions uniformly distributed over 360$^\circ$ are simulated to capture the full directional response of the urban fabric. At pedestrian level, the flow exhibits \ming{strong} directional sensitivity, \ming{with} moderate rotations of the incoming wind \ming{reorganizing} the dominant flow pathways from continuous accelerated corridors, when winds align with major street axes, to fragmented cellular patterns under oblique inflows. These directionally activated channeling corridors govern local ventilation and sheltering. Yet, despite this pronounced local variability, double averaged vertical profiles of mean velocity and turbulence intensity collapse remarkably across all wind directions. Two distinct inflection points are consistently identified in the mean velocity profiles: one located slightly below the average building height, \ming{$H_\mathrm{avg}$}, confirming the persistence of a shear-driven mixing-layer regime even in highly heterogeneous urban morphology, and a second, deeper inflection point near pedestrian level, \ming{approximately 0.08–0.10 $H_\mathrm{avg}$ above the ground,}
that marks the transition between recirculating near-ground flow and more connected \ming{transport within the urban canopy}. 
\revone{Subdomain analysis further shows that this lower inflection point is morphology- and direction-dependent, reflecting changes in the transition between sheltered near-ground recirculation and transport within the urban canopy as preferential flow routes are activated or suppressed by the incoming wind direction.}
It is shown that urban aerodynamic response is simultaneously robust at the neighbourhood scale, behaving quasi-isotropically in the double-averaged sense, and strongly directional at the local scale, where the activation of preferential ventilation pathways dictates the spatial distribution of momentum and turbulence.}

\keywords{large-eddy simulation, urban canopy flow, wind direction, pedestrian-level wind, double-averaging, turbulence}

\maketitle

\section{Introduction}\label{sec1}
The  growth of the global  population in cities has made the study of urban microclimates a critical priority for public health and sustainable design.  At the heart of this field is the interaction between the atmospheric boundary layer and the urban canopy layer (UCL), where the transport of pollutants, heat, and moisture is governed by complex fluid dynamics. 
Unlike smooth surfaces, urban environments are characterised by complex interactions between buildings and the atmospheric flow, leading to highly unsteady and three-dimensional \ming{(3D)} wind patterns, including separation at building edges, wake formation, and intense shear layers. Thus, understanding the effect of wind direction on urban airflow is crucial for a wide range of applications, including air quality assessment, pedestrian comfort,  urban resilience to extreme weather events, among others.

Traditionally, urban flow research has relied on the Monin–Obukhov similarity theory (MOST) to describe the wind profile near the surface. However, the extreme morphological heterogeneity of real cities often violates the assumption of horizontal homogeneity required by MOST, limiting its applicability in realistic urban environments \citep{Franke2007,Barlow2009}. To overcome these limitations and to account for the complex interactions between atmospheric flow and urban geometry, numerical simulations have progressively become essential tools for the study of urban aerodynamics. 

Owing to the intrinsic complexity of the flow and the high computational cost associated with resolving a wide range of spatial and temporal scales, most urban flow studies conducted to date have relied on Reynolds-Averaged Navier–Stokes (RANS) approaches (e.g. \cite{Toparlar2015,Brozovsky2021,Antoniou2019}). However, while RANS models are frequently used for their efficiency, they often fail to capture the unsteady nature of urban turbulence \citep{xie2006,GarciaSanchez2018}. In this sense, large-eddy simulation (LES) has emerged as a preferred high-fidelity tool, as it directly resolves the large, energy-containing turbulent structures that dominate momentum and scalar transport (e.g., \cite{GarciaSanchez2018, Yan2022,Teng2025}).

The current understanding of turbulent flow in urban canopies has been largely built upon idealised configurations such as regular arrays of cubes, isolated buildings, and two-dimensional street canyons. These studies identified distinct flow regimes, i.e., isolated roughness, wake interference, and skimming flow \citep{oke1988}, according to the building-height-to-street-width aspect ratio. Early experimental and numerical studies established the fundamental role of obstacle layout and packing density in controlling mean flow, turbulence structure, and momentum transport within and above the canopy, while highlighting the importance of dispersive stresses and the limitations of steady RANS modelling in highly inhomogeneous flows \citep{coceal2006,xie2006}. Key geometric parameters, such as frontal  ($\lambda_f$) and plan ($\lambda_p$) area ratios have been correlated to drag, ventilation efficiency, and pedestrian-level flow organisation \citep{AbdRazak2013}. These parameters have subsequently been applied across a broad range of idealised configurations to investigate the links between urban packing and the dynamical organisation of the intra-canopy flow \citep{Lin2021,Blunn2022}.

 More recent work has addressed  complex architectural features such as roof shapes \citep{LlagunoMunitxa2017} or  façade elements like balconies  \citep{Zheng2022} in shaping the  turbulent exchange,  highlighting how local architectural modifications can substantially alter near-field flow structures, motivating a broader investigation of geometric complexity and introducing more realistic urban scenarios \citep{akinlabi2022,Wang2025}.
A defining characteristic of realistic urban areas is height variability, which significantly alters turbulence dynamics compared to uniform-height arrays. Research indicates that the presence of high-rise buildings disproportionately contributes to surface drag and triggers strong updrafts on their windward faces, creating elongated wakes that can extend kilometres downstream \citep{Tian2024, akinlabi2022}. In these heterogeneous environments, the mean velocity profile often lacks the distinct inflection point typical of uniform canopies \citep{Tian2024}. Turbulent kinetic energy (TKE) production is thus highly localised around building edges and separation zones \citep{Shui2024, Wang2023}.

Despite the depth of existing literature, a significant portion of urban flow research assumes an ambient wind direction perpendicular to building faces. In reality, wind is changing directions and gives  rise to a more complex phenomena within the canopy.
 \cite{Kim2004}  presented simulations examining how varying wind directions influence airflow over arrays of cubic obstacles, representative of urban canopies. The findings highlighted that even slight deviations in wind direction can significantly alter flow patterns, which has important implications for urban planning and pollutant dispersion modelling.
 Other studies, such as those of \cite{Claus2012} and \cite{Monnier2018} demonstrated that oblique inflow dismantles the two-dimensional recirculation cells typical of perpendicular flow, replacing them with complex \ming{3D} vortices and intensified lateral channelling that enhances mixing between adjacent canyons. This has been corroborated by high-resolution simulations such as those by \cite{Castro2017} and \cite{Shui2024}, which show that oblique angles generate asymmetric recirculation and unique pollutant pathways.
 
 However, morphological complexity, further complicates the flow dynamics.  \cite{Giometto2016} demonstrated that while wind aligned with main streets promotes deep penetration and coherent jets, oblique directions trigger lateral channelling and recirculation that degrade ventilation. This critical dependence was further refined by \cite{Ricci2018}, whose wind-tunnel and RANS study of a historical district showed that oblique inflows generate complex crossflows that disrupt coherent ventilation. The practical consequence is a dynamic, wind-directed ventilation network, a concept mapped in downtown Beijing by \cite{Wang2020}, who showed that the connectivity of pedestrian-level ventilation corridors is entirely dependent on inflow angle. The practical importance of capturing this directional dependence is confirmed by the study of \cite{Hagbo2024a}, which indicate that reliable urban wind assessments require simulations across a wide spectrum of wind directions. 
 \subsection{\revone{Scope}}
\revone{Despite substantial progress in urban flow simulations, important gaps remain in the understanding of how full-scale urban geometry interacts with the {ABL} and shapes flow dynamics within the urban canopy. In particular, the combined effects of building height variability, street layout, spatial heterogeneity \ming{and wind direction} on the flow are still not fully characterised in full-scale urban environments. Although previous studies have shown that wind direction can modify flow organisation in idealised arrays and selected urban configurations, its role in linking local urban-canopy flow structures with spatially averaged turbulence statistics remains insufficiently understood in realistic heterogeneous neighbourhoods.

In this context, we perform building-resolving \ming{LES} of an urban 
neighbourhood in Barcelona, considering multiple inflow conditions, to shed 
light on these processes and to improve our understanding of the organisation 
of the mean flow and turbulence statistics within the urban canopy. 
The 
novelty of the present study lies in the systematic assessment of sixteen wind 
directions over the full $(360^\circ)$ range within the same realistic urban 
morphology, allowing us to distinguish between direction-dependent local flow 
organisation and the behaviour of double-averaged statistics. Particular attention is given to the flow routes at pedestrian level, the different impact of wind direction at local and spatially averaged scales, and the emergence of a lower near-ground inflection point associated with the transition from recirculating flow to connected transport within the urban canopy.}

\section{Definition of the cases and computational domain}

\begin{figure*}
  \centering
    \includegraphics[width=0.85\textwidth]
    {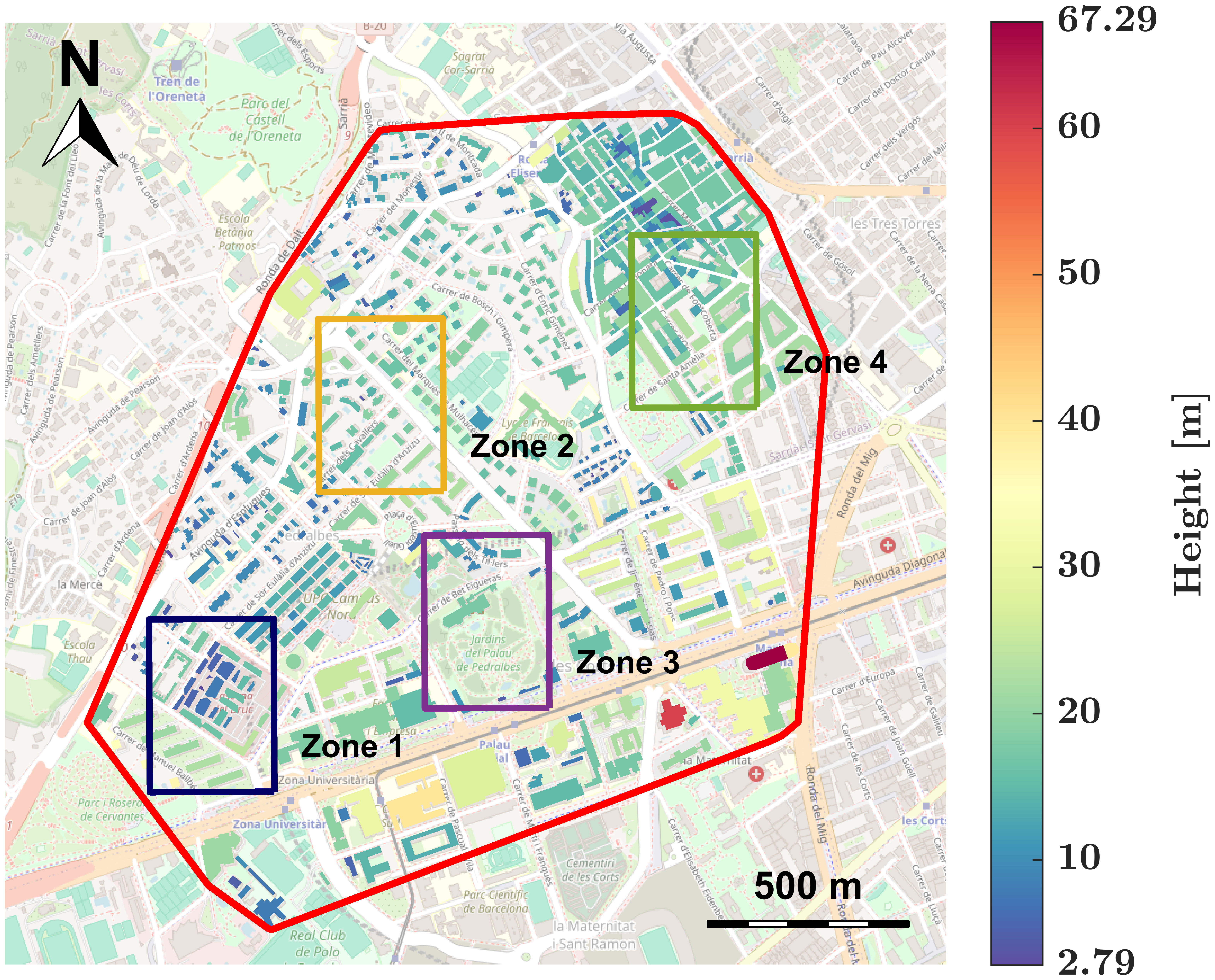}
  \caption{ \revtwo{Map of Barcelona showing the study area. Map data © OpenStreetMap contributors (\href{https://www.openstreetmap.org}{openstreetmap.org}) }.The area under study is enclosed in red, while the zones with different morphological parameters are also shown in the figure. Building height is also given.} 
  \label{fig:city}
\end{figure*}

\begin{figure*}
  \centering
    \includegraphics[width=0.95\textwidth]{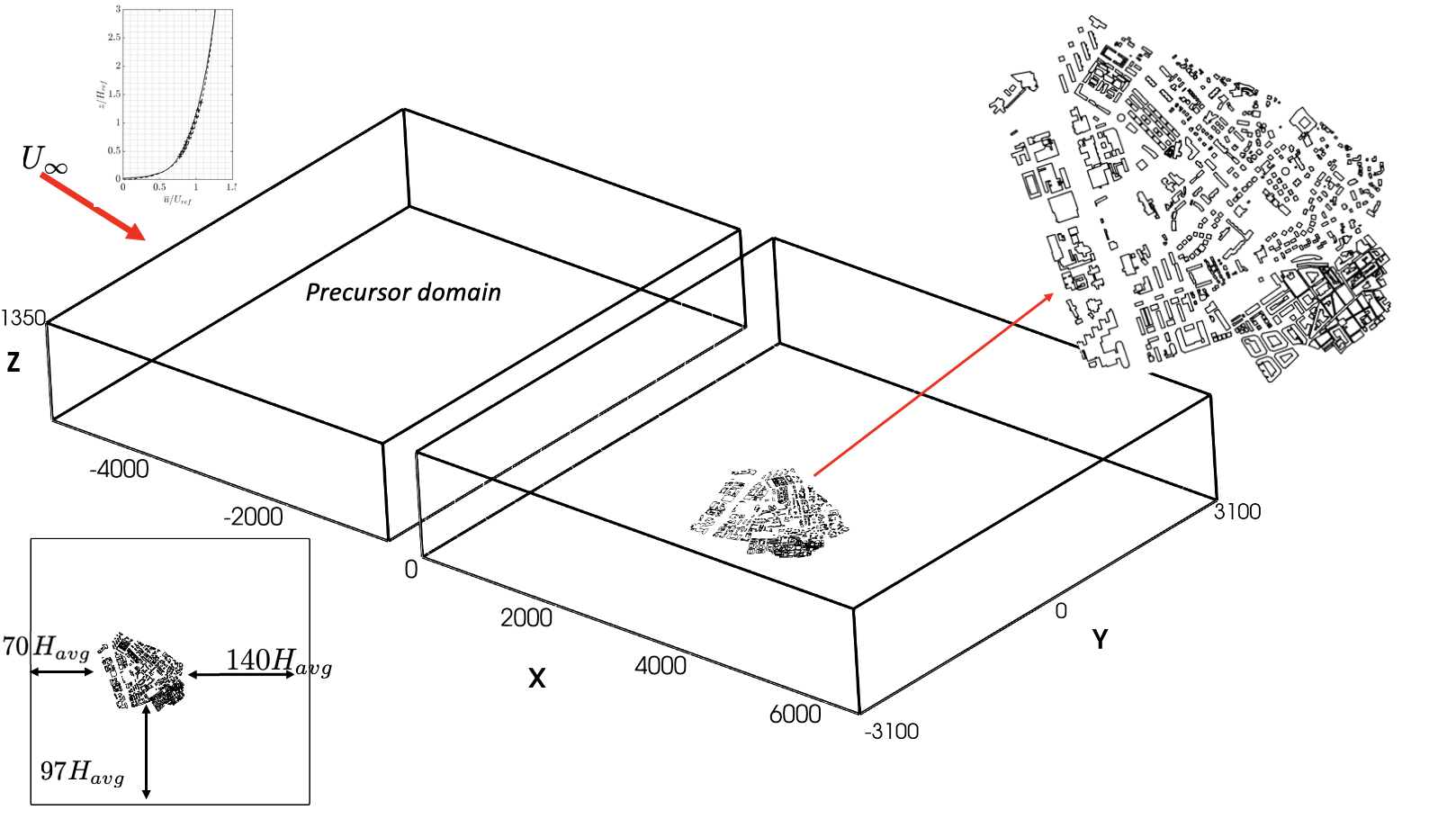}
  \caption{Computational domain \revtwo{with sketches of urban geometries. $U_{\infty}$ and $H_{avg}$ denote freestream velocity and average building height, respectively.}}
  \label{fig:domain}
\end{figure*}

To evaluate the impact of the wind direction on the urban flow statistics, the present study focuses on a neighbourhood located in the western part of Barcelona, covering an area of approximately $1.7 \times 1.9 \mathrm{km}^2$. The selected area corresponds to the \textit{Zona Universitària--Pedralbes} district, which is characterised by pronounced morphological complexity, with a mixture of academic buildings, residential blocks, open spaces and tree-lined avenues (see Fig. \ref{fig:city}). The maximum building height within the domain is $H_{\max} = 67.29 \mathrm{m}$, whereas the average building height is $H_{\mathrm{avg}} = 19.27 \mathrm{m}$, with a corresponding standard deviation of $H_{\mathrm{std}} = 8.19 \mathrm{m}$.
The topographic setting of Barcelona exerts a strong influence on its local wind regime. The city is situated between the Mediterranean Sea to the southeast and the Collserola mountain range to the northwest, forming a coastal-to-mountain transition that channels the dominant winds along the northeast–southwest direction.  This position, together with the irregular and non-orthogonal street network, results in a wide range of flow configurations depending on the approaching wind direction.
 
  To systematically assess the sensitivity of the urban flow to the incoming wind direction, a total of sixteen inflow angles are therefore considered, uniformly distributed over the full $360^\circ$ range with increments of $22.5^\circ$. These directions, denoted hereafter by $\Phi$, enable a comprehensive evaluation of directional effects on both the mean flow organisation and the turbulent statistics within the urban canopy layer. This combination of topographic exposure, geometric variability, and directional sampling makes the district a representative case for analysing the influence of wind direction on  canopy flow structures in a  Mediterranean neighbourhood.

The urban morphology is further characterised using standard urban density metrics, specifically, the plan area density and frontal area density. The former is defined as $\lambda_p = A_{\mathrm{buildings}} / A_{\mathrm{domain}}$,
where $A_{\mathrm{buildings}}$ is the total horizontal projected area of all buildings and $A_{\mathrm{domain}}$ is the plan area of the district. For the present configuration, $\lambda_p = 0.24$, indicating a relatively compact urban fabric. The latter,  $\lambda_f$,  quantifies the effective blockage of the flow in the direction normal to the incoming wind, and is defined as
$\lambda_f = A_{\mathrm{frontal}} / A_{\mathrm{domain}}$,
 $A_{\mathrm{frontal}}$ being the total frontal area of the buildings exposed to the flow. Due to the strong anisotropy of the urban layout, $\lambda_f$ varies with the wind direction and ranges between 0.165 and 0.193 across the 16 simulated inflow angles. This directional dependence is expected to influence momentum extraction, turbulence production and the redistribution of flow structures within the urban canopy layer.

Figure \ref{fig:domain} provides a street-level perspective of the study area together with the computational domain employed in the simulations. Although the example shown corresponds to the inflow angle $\revtwo{\Phi = 180^\circ}$,  which represents a wind blowing from the South, the domain setup is representative of all simulated wind directions. To account for changes in wind direction, the city geometry is rotated progressively, while maintaining the inlet boundary fixed from left to right. In this convention, for instance, an inflow angle of \revtwo{$\Phi = 225^\circ$} corresponds to a wind blowing from the southwest. This ensures consistency in the implementation of boundary conditions across all cases and avoids numerical artefacts associated with rotating the flow field.

Following our previous work on full-scale realistic urban configurations \citep{Teng2025}, the distances between the urban area and the domain boundaries have been set to allow the correct development of the approaching atmospheric boundary layer and to minimise unwanted blockage or reflection effects. Overall, the computational domain spans over $337H_{\mathrm{avg}} \times 321H_{\mathrm{avg}} \times 70H_{\mathrm{avg}}$ in the streamwise, lateral, and vertical directions \ming{(denoted as $x$, $y$ and $z$ hereafter)}, respectively. The city is placed at a distance of 70$H_{\mathrm{avg}}$ from the inlet, whereas outlet is at $140H_{\mathrm{avg}}$ downstream the city. The lateral boundaries are located at $97H_{\mathrm{avg}}$ from the city outskirts.
In fact, all these  distances  are set to values that are significantly larger than those recommended in the best practice guidelines for urban CFD simulations ~\citep{Franke2007}. 
\revtwo{ For wide, extended urban models, \cite{Blocken2015} notes that the standard COST732 blockage criteria can be insufficient to prevent artificial lateral flow acceleration even when nominally satisfied, and recommends larger lateral clearances. The lateral boundaries placed at $91\,H_\mathrm{avg}$ from the city outskirts are well in excess of the COST732 minimum of $5\,H_\mathrm{max} \approx 17.5\,H_\mathrm{avg}$, ensuring that the free-slip lateral condition does not artificially constrain the flow within the urban canopy. Similarly, the upstream distance of $65\,H_\mathrm{avg}$ ($>5\,H_\mathrm{max}$) provides sufficient fetch for the turbulent ABL inflow to develop and for the flow deceleration that occurs as the approaching boundary layer encounters the urban roughness to take place without distorting the inflow conditions at the domain inlet. The downstream distance of $124\,H_\mathrm{avg}$ ($>15\,H_\mathrm{max}$) ensures full wake development downstream of the city.

 Furthermore, the COST732 guidelines explicitly state that for LES the domain must be large enough to contain the largest energetically relevant flow structures. This makes the vertical extent particularly critical, as it must be selected sufficiently high to allow for the full development of the inertial sublayer.}
The upper limit of the roughness sublayer is typically located at approximately $2{-}5\,H_{\mathrm{avg}}$ \citep{Raupach1991a}. Above this region, the flow progressively transitions into the inertial sublayer, where the mean velocity profile follows a logarithmic law. According \cite{Jimenez2004b}, to ensure a fully developed rough turbulent flow that is reasonably free from direct roughness effects, the condition $\delta / H_{\mathrm{avg}} > 50$ \ming{($\delta$ denotes domain height)} should be satisfied. In the present configuration, the top boundary is located at approximately $70\,H_{\mathrm{avg}}$, which is well above this threshold, thus providing sufficient vertical extent for the correct establishment of the inertial sublayer and minimising potential contamination of the flow statistics by top boundary effects.
 This vertical extent clearly exceeds the domain heights adopted in similar urban flow studies, such as \cite{Giometto2016}, who employed a domain height of about $10.6\,H_{\mathrm{avg}}$, \cite{Cheng2023a} ($11{-}30\,H_{\mathrm{avg}}$), and  \cite{Wang2025}($6.7-14\,H_{\mathrm{avg}}$).

\section{\ming{Methodology}}
\label{math}
\revone{The mathematical and numerical description of the LES formulation employed here, including the governing equations, the subgrid-scale closure and the numerical scheme, as well as its validation against wind-tunnel measurements for urban flow configurations, are presented in detail in our previous work \cite{Teng2025}.} For the sake of completeness, a brief overview is provided here. 

The neutral atmospheric boundary layer over the urban environment is modelled within a \ming{LES}  framework, in which the filtered governing incompressible Navier--Stokes equations are solved. Here, the subgrid-scale stresses are modelled using the \cite{vreman2004} model. 

The filtered equations are solved using the SOD2D solver (Spectral high-Order coDe for solving partial Differential equations), a low-dissipation spectral-element method designed for both direct and large-eddy simulations of turbulent flows \citep{GASPARINO2024109067}. The numerical framework relies on a continuous Galerkin formulation with projection stabilisation, which efficiently suppresses spurious oscillations in convection-dominated regimes while preserving the low-dissipation properties required for accurate turbulence representation. Spatial discretisation is performed on Gauss--Lobatto--Legendre nodes, providing high accuracy through their favourable interpolation and quadrature characteristics. To mitigate aliasing errors arising from the nonlinear convective terms, a skew-symmetric splitting strategy is employed  \citep{kennedy2008reduced}, effectively reducing numerical artefacts without the need for costly over-integration.

Time integration is carried out using a third-order backward differentiation formula combined with extrapolation (BDF--EXT3) within an operator-splitting framework, allowing the velocity and pressure fields to be solved in a decoupled manner while maintaining good stability and accuracy  \citep{Karniadakis1991}. The resulting pressure Poisson equation is solved using a conjugate gradient method, ensuring efficient convergence for large-scale problems.

\subsection{\ming{Numerical setup}}
\subsubsection{\ming{Mesh resolution}}
\begin{figure*}
  \centering
  \begin{subfigure}[b]{0.55\textwidth}
    \includegraphics[width=\textwidth]{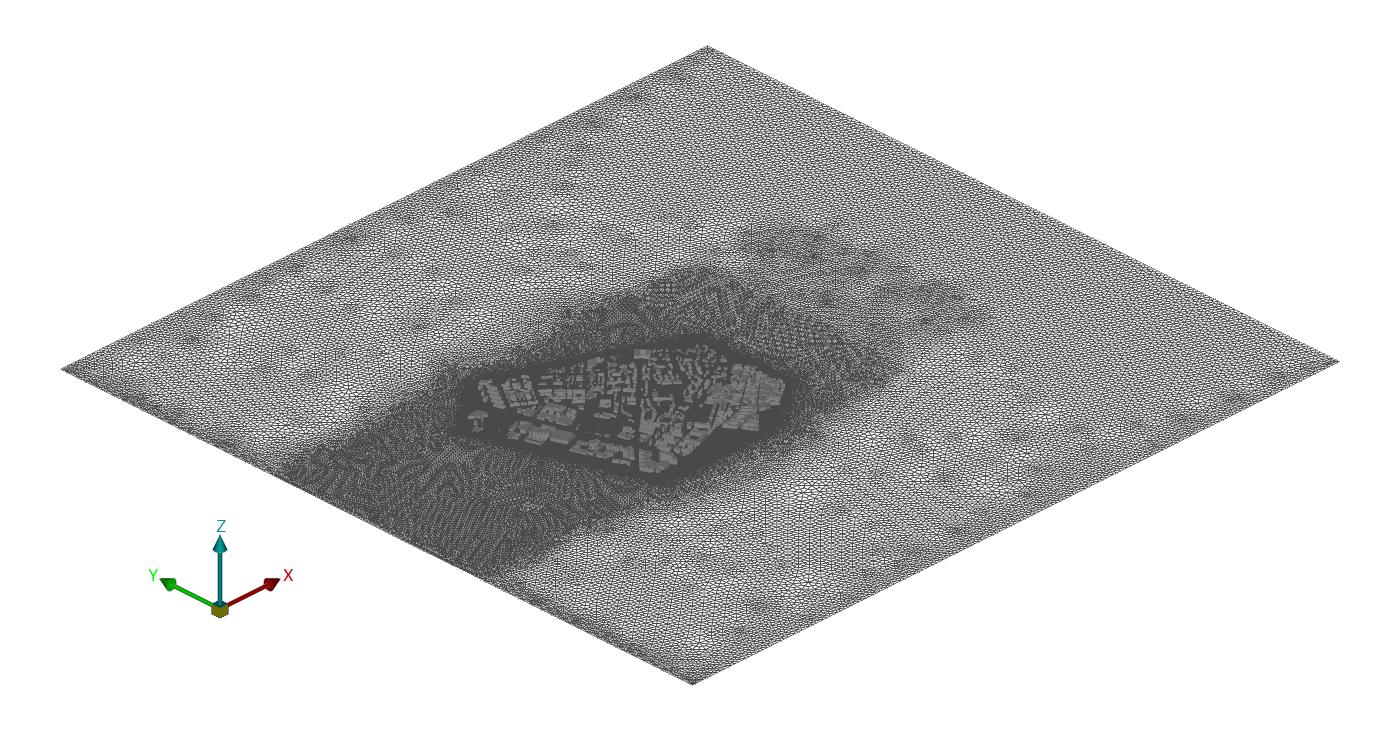}
        \caption{}
    \label{fig:mesh_total}
  \end{subfigure}
  \hfill
  \begin{subfigure}[b]{0.4\textwidth}
    \includegraphics[width=\textwidth]{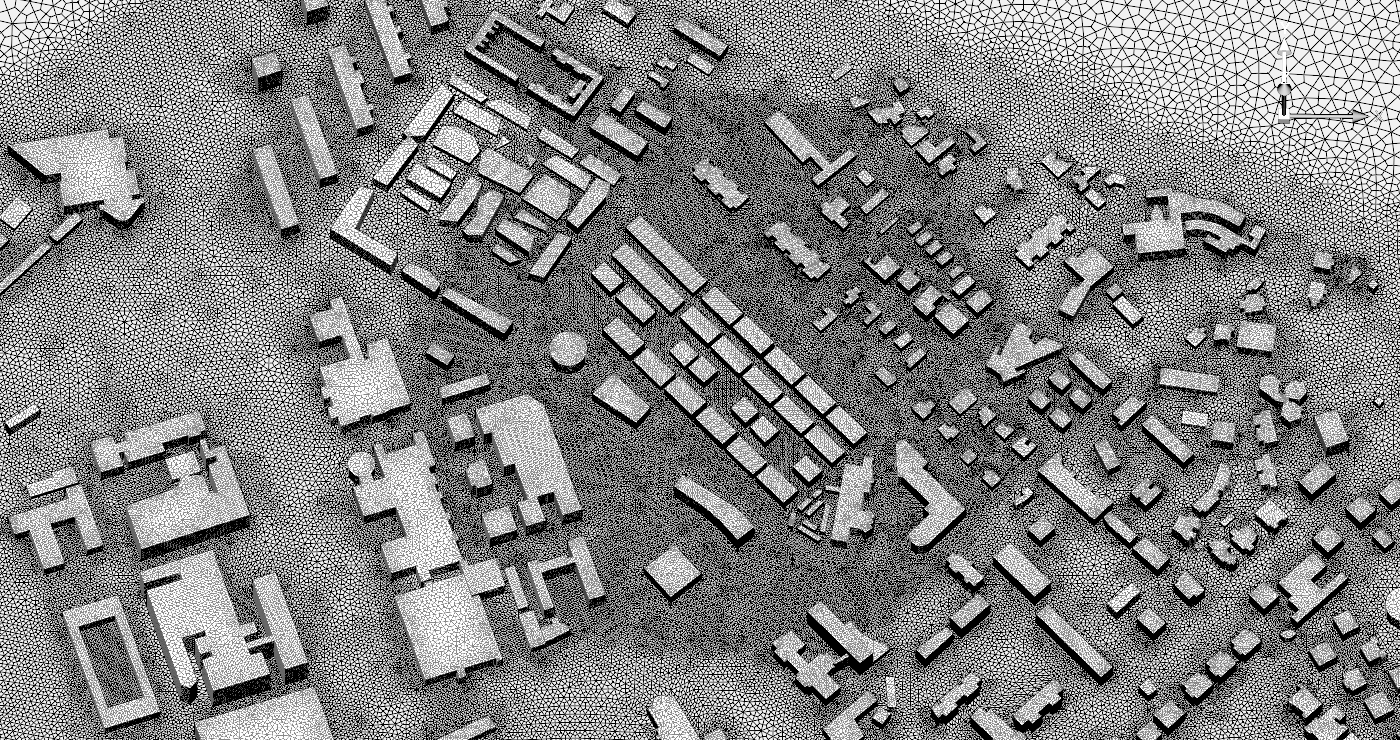}
    \caption{}
    \label{fig:mesh_detail}
  \end{subfigure}
  \caption{(a) Overview of the \revtwo{surface mesh (the surface discretization of the geometry)} and the nested refinement zones. (b) Detail of the mesh resolution around the building surfaces. For clarity, only the \revtwo{surface} mesh elements are shown.  
  }
  \label{fig:mesh}
\end{figure*} 

To accurately resolve the flow around the complex urban geometry, the computational grid employs unstructured hexahedral elements supporting fourth-order polynomial basis functions within each element. This approach allows the high accuracy of spectral element methods to be retained even in regions of strong geometric irregularity. The mesh is organised through a hierarchical system of nested refinement zones, ensuring that spatial resolution is concentrated where flow gradients are strongest, namely in the vicinity of buildings, within street canyons, and at pedestrian height, while progressively coarser elements are used farther from the ground and toward the domain boundaries.
The refinement strategy and overall mesh layout are illustrated in Fig.~\ref{fig:mesh}, which presents an overview of the computational grid and the nested refinement zones (Fig.~\ref{fig:mesh_total}), together with a detailed view of the mesh resolution around the building surfaces (Fig.~\ref{fig:mesh_detail}). For clarity, only the \revtwo{surface} mesh elements are shown in the close-up view. Each \ming{3D} element contains 125 solution points, corresponding to the fourth-order polynomial discretisation.
At the finest level of refinement, the mesh spacing reaches approximately 1 m at pedestrian height. This local resolution allows the main flow features associated with separation, reattachment, and channeling between buildings to be explicitly captured. The mesh is progressively coarsened in the vertical direction, with refinement extending up to several building heights above the mean roof level to ensure an adequate description of the flow region of interest and the transition toward the overlying inertial sublayer. The combination of high-order spatial discretisation and locally refined unstructured topology enables an accurate representation of both the complex building morphology and the multi-scale flow features.
The total number of points used in each simulation, depending on the specific orientation of the city model and the extent of the refined regions, resulted in meshes containing between $4.8\times10^8$ and {$5.2\times10^8$} grid points. This set-up is consistent with that employed in previous work \citep{Teng2025}, where the methodology was extensively validated against wind tunnel measurements and shown to provide an accurate representation of the flow within the urban canopy layer. As detailed in Annex~\ref{app:mesh}, a mesh resolution assessment was conducted by comparing two grids with identical element counts but using second- and fourth-order spatial discretisation. Although both meshes yielded comparable results, the fourth-order configuration was selected for its enhanced accuracy and its improved capability to capture the complex flow dynamics.

\subsubsection{\ming{Boundary conditions}}
As commented before, the numerical strategy adopted follows that tested and validated for urban flow simulations in  \cite{Teng2025}, but to provide a self-contained description, the key aspects of this setup are \ming{described} below.
A critical aspect of the simulation is the generation of a realistic, turbulent inflow that characterises a neutral atmospheric boundary layer (ABL). To achieve this, an online precursor method is employed. This approach utilises two concurrently running computational domains: a precursor domain dedicated to the autonomous development of a fully-developed ABL, and a main domain containing the urban geometry of interest. The precursor domain was configured to replicate a flow environment over very rough terrain, characterised by a surface roughness length of $z_0 = 1.53$ m. The flow was driven by a constant pressure gradient \ming{in the streamwise $(x)$ direction}, which was calibrated to maintain a target friction velocity of $u_* = 0.596$ m/s. This setup resulted in a reference velocity of $U_{\text{ref}} = 6.1$ m/s at a height of $H_{\text{ref}} = 100$ m, establishing a representative ABL profile for the inflow\revone{, which corresponds to a Reynolds number based on the reference velocity and average building height of $Re_{H}=U_{ref}\:H_{avg}/\nu\approx6.5\times10^6$.  The use of a precursor simulation (dual-domain) } strategy eliminates the need for storing and recycling precomputed inflow data, thereby providing a dynamically consistent and temporally evolving turbulent inflow boundary condition for the primary simulation. In the precursor \ming{domain, periodic} boundary conditions are applied in both the \ming{$x-$ and $y-$}directions. The coupling between the domains is handled dynamically at each time step, where the instantaneous \ming{3D} velocity field at the outflow plane of the precursor domain is directly imposed as a Dirichlet boundary condition at the inlet of the main domain. This ensures that the main simulation receives a physically realistic, time-resolved turbulent inflow. 

For the lateral boundaries in the main domain, as well as the top boundary of both domains, free-slip conditions are prescribed. This implies a zero-stress condition, where the normal velocity component is zero and the tangential components have a zero normal gradient. A fixed static pressure condition is applied at the outlet of the main domain to allow for the undisturbed exit of flow structures. 

Resolving the complex flow dynamics within the urban canopy requires special attention to near-wall modeling. The flow around buildings is characterised by strong pressure gradients, flow separation, and vigorous three-dimensionality, which violate the equilibrium assumptions underlying standard wall-function approaches. Consequently, a no-slip boundary condition is applied directly on all building surfaces. This ensures that the separation points, recirculation zones, and wake dynamics are captured directly by the solver without relying on potentially inaccurate empirical models. \revtwo{At the ground surface, however,  an equilibrium  wall model based on the logarithmic law for rough walls and neutral stability \citep{Owen2020WMLES} is applied to mitigate the prohibitive computational cost of resolving the viscous sublayer across the entire domain, thus striking a balance between physical fidelity and computational expense}.

\subsubsection{\ming{Statistical convergence}}
To ensure the flow reached a statistically steady state, all simulations were initialised on a coarser, second-order mesh and advanced in time for over  10000 \ming{sec.} of physical time. This initial spin-up phase, equivalent to approximately 56 eddy turnover times (ETT, defined as $ ETT = H_{\text{avg}} / u_* $), allowed for the full development of turbulent structures. Following this, the flow field was interpolated onto the final, higher-fidelity fourth-order computational mesh. The simulations were then advanced for an additional 30 ETT to allow the flow to adjust to the refined grid before time-averaging began. Statistics were subsequently collected over a period exceeding 160--170 ETT, depending on the case, to ensure the convergence of first- and second-order moments.

The total sampling duration employed in this study is thus substantial and aligns with, or exceeds, the standards of high-fidelity urban LES. It is larger than the durations reported in other studies, such as the approximately 40 ETT in \cite{Gronemeier2021} or the sampling periods of approximately 70 ETT used by \cite{Wang2025}, while it is comparable to those used by \cite{Auvinen2020} (190 ETT) or \cite{Giometto2016} (150 ETT). While canonical studies on idealised arrays can employ extended averaging periods (e.g., 400 ETT in \cite{coceal2006}, 650 ETT in \cite{Castro2017}), such durations are often computationally prohibitive for large-scale, realistic urban simulations. Thus, the sampling period implemented in this work represents a carefully balanced approach, providing well-converged statistics while maintaining computational feasibility given the substantial resources required for high-fidelity simulation of a realistic urban domain.

\section{Results and discussion}

\subsection{Aerodynamic parameters}

Above the roughness sublayer, the mean velocity profile  can  be approximated as, 

\begin{equation}
\langle \bar{u} \rangle(z) = \frac{u_*}{\kappa} \ln \left( \frac{z - d}{z_0} \right),
\end{equation}

where $\langle \bar{u} \rangle(z)$ \ming{($\langle \cdot \rangle$ and $\overline{\cdot}$ denoting the spatial and temporal averaging operator, respectively)} is the mean streamwise wind speed at height $z$, $d$ is the displacement height,   $z_0$ the aerodynamic roughness length, $u_*$  the friction velocity  and $\kappa$ is the von Kármán constant.  The  estimation of these parameters is a non-trivial task, primarily due to the complex and vertically variable nature of momentum fluxes within the urban roughness sublayer (see for instance \cite{Kanda2013,Sutzl2021}).

\begin{table*}[t]
\centering
\caption{Parameters for the fitted logarithmic-law normalised by the average building height, $H_{\mathrm{avg}}$, for each wind direction, $\Phi$. \revone{$u_*$, $d$ and $z_0$ denote friction velocity,  displacement height, and aerodynamic roughness length, respectively. $Re_*$ is the roughness Reynolds number ($Re_* = u_* \cdot z_0 / \nu$) with $\nu = 1.81 \times 10^{-5} m^2/s$.}}
{\small
\renewcommand{\arraystretch}{1.25}
\begin{tabular}{rcccc}
\hline
\revone{$\Phi$ (°)} & \textbf{$u_*$ (m/s)} & \textbf{$d/H_{\mathrm{avg}}$} & \textbf{$z_0/H_{\mathrm{avg}}$} & \revone{\textbf{$Re_* \times 10^3$}} \\
\hline
0     & 0.6232 & 0.6200 & 0.1029 & 68.28 \\
22.5  & 0.5892 & 0.7420 & 0.0920 & 57.73 \\
45    & 0.5728 & 0.8270 & 0.0689 & 42.02 \\
67.5  & 0.5686 & 0.5079 & 0.0785 & 47.53 \\
90    & 0.5954 & 0.8554 & 0.0774 & 49.08 \\
112.5 & 0.6163 & 0.5172 & 0.1225 & 80.40 \\
135   & 0.6233 & 0.2746 & 0.1462 & 97.04 \\
180   & 0.6145 & 0.7756 & 0.1115 & 72.95 \\
202.5 & 0.5877 & 0.6754 & 0.1043 & 65.28 \\
225   & 0.5160 & 1.0012 & 0.0504 & 27.69 \\
247.5 & 0.5562 & 0.9059 & 0.0579 & 34.29 \\
270   & 0.5861 & 0.8788 & 0.0759 & 47.37 \\
292.5 & 0.6362 & 0.4411 & 0.1351 & 91.53 \\
315   & 0.5762 & 0.7618 & 0.0972 & 59.64 \\
337.5 & 0.6410 & 0.6226 & 0.1079 & 73.65 \\ \hline
\end{tabular}
}
\label{tab:loglaw_params}
\end{table*}

\begin{figure}[t]
    \centering
    \includegraphics[width=0.65\textwidth]{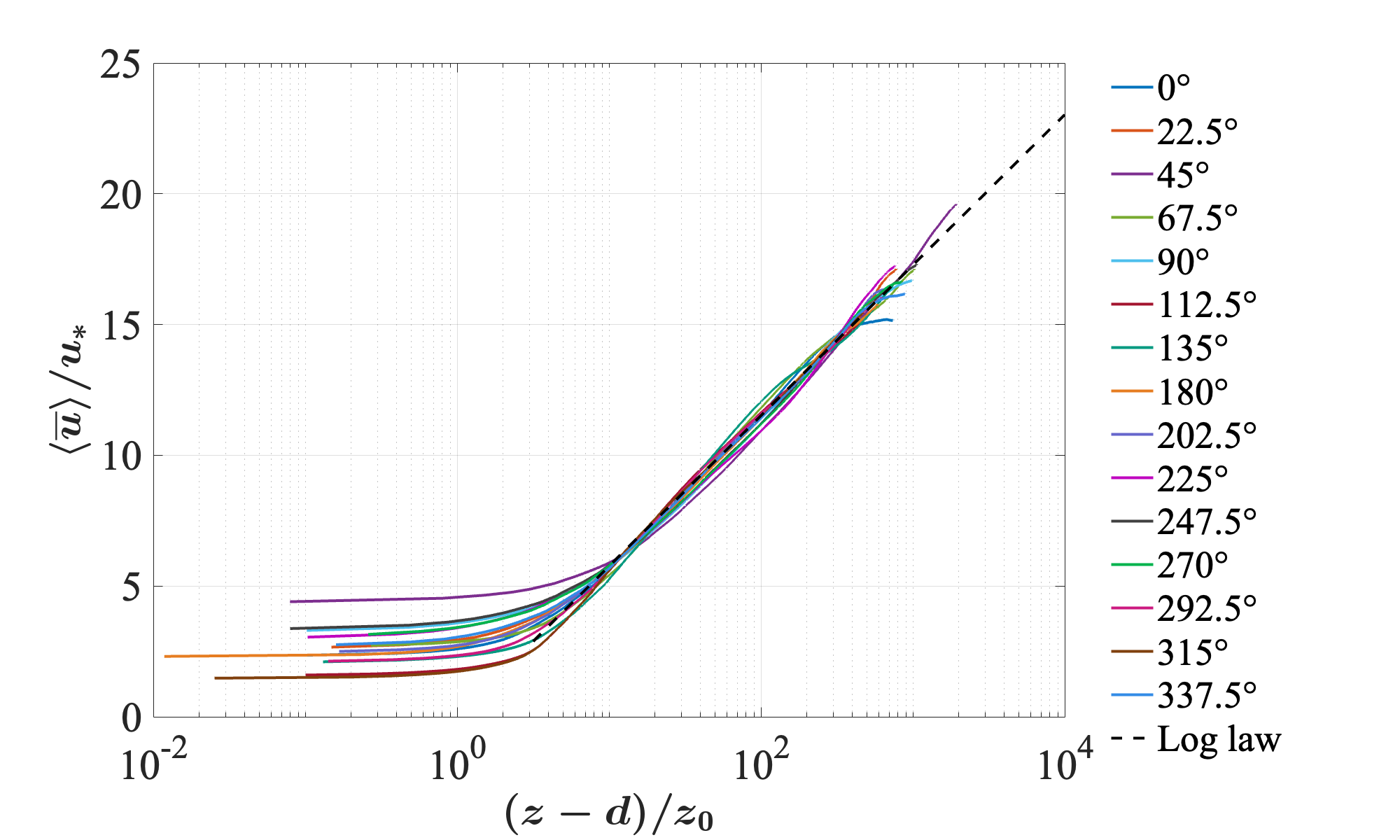}
    \caption{Normalised mean velocity profiles in inner scale, \ming{$\langle \overline{u} \rangle/u_*$}, for all simulated wind directions. \ming{$d$ and $z_0$ denote displacement height and aerodynamic roughness length}. The black dashed line represents the theoretical logarithmic law of the wall.}
    \label{fig:loglaw_profiles}
\end{figure}

In this study, $u_*$  is determined directly from the simulation results as the spatially- and time-averaged total drag force for the entire ground and building surfaces, i.e. considering both form and viscous drag. With $u_*$ determined directly from the simulations, the aerodynamic parameters $z_0$ and $d$  are obtained by fitting the logarithmic profile to the mean wind data within a specific vertical region of the inertial sublayer. The selection of this fitting region is critical, as it significantly influences the resulting parameter values. \revone{This sensitivity has been discussed in \cite{Kanda2013}, \cite{Sutzl2021} and \cite{Duan2021}, showing that a higher fitting range generally results in greater displacement heights and smaller roughness lengths, while a lower range produces the opposite effect. In simulations with uniform building heights, the differences between fitting ranges are minor, and the log-law provides a good approximation even within the roughness sublayer. However, for more realistic cases with heterogeneous building heights, the estimated logarithmic profiles vary strongly depending on the chosen range, particularly for layouts with high frontal area density and large building-height standard deviation. Actually, \cite{Duan2021} report that $\sigma_H$ has considerably less influence on $z_0$ and $d$ than $H_\mathrm{avg}$ and $H_\mathrm{max}$, highlighting the importance of a fitting criterion based on to these two parameters. In this work, we have adopted the vertical range proposed by Kanda et al.\ (2013), $z \in [H_\mathrm{max} + 0.2H_\mathrm{avg},\, H_\mathrm{max} + H_\mathrm{avg}]$. For the present geometry, this places the lower bound of the fitting region above the tallest building in the domain ($H_\mathrm{max} = 67.29$\,m), ensuring that the fit is performed outside the roughness sublayer even for the tallest structures, while remaining within the inertial sublayer where the constant-flux assumption and the log-law are expected to be valid.
}

The resulting parameters for all simulations are compiled in Table \ref{tab:loglaw_params}. \revone{In the table, the roughness Reynolds number $Re_* = u_* z_0/\nu$ is also reported. As can be seen,  values are ranging from $2.8 \times 10^4$ to $9.7 \times 10^4$, confirming that all simulations are in the fully aerodynamically rough regime.} The resulting \ming{profiles of $\langle \overline{u} \rangle/u_*$} for all simulated wind directions are plotted in Fig.~\ref{fig:loglaw_profiles} \ming{against $(z-d)/z_0$}. The profiles exhibit a consistent collapse within the logarithmic region, confirming the validity of the aerodynamic parameters over the different wind directions.

\subsection{Wind flow dynamics}

\begin{figure}[]
  \centering
  \begin{subfigure}[b]{0.45\textwidth}
    \includegraphics[width=\textwidth]{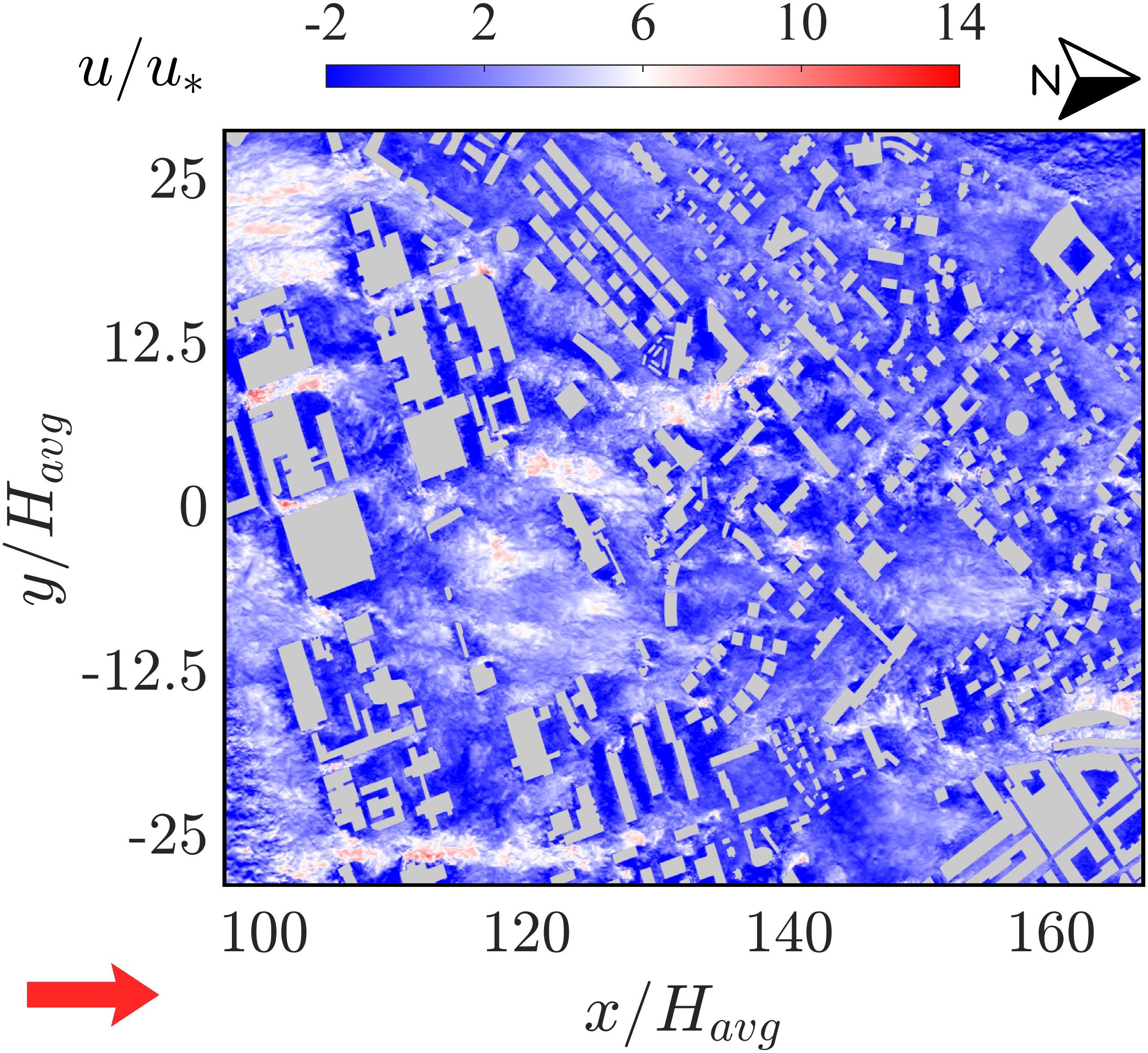}
    \caption{}
    \label{fig:subfig_a}
  \end{subfigure}
  \hfill
     \begin{subfigure}[b]{0.45\textwidth}
    \includegraphics[width=\textwidth]{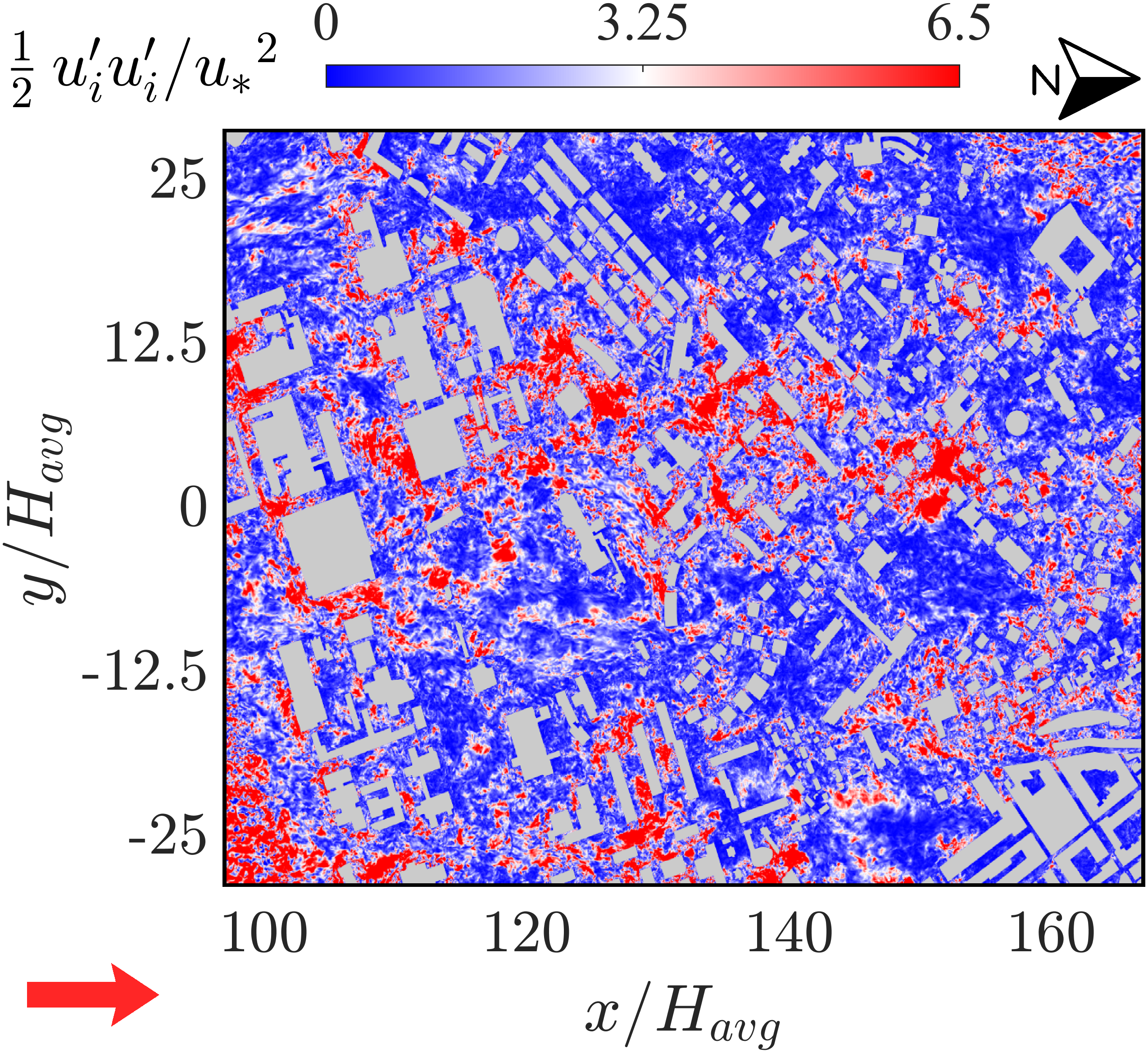}
    \caption{}
    \label{fig:subfig_b}
  \end{subfigure}
  \hfill
  \begin{subfigure}[b]{0.45\textwidth}
    \includegraphics[width=\textwidth]{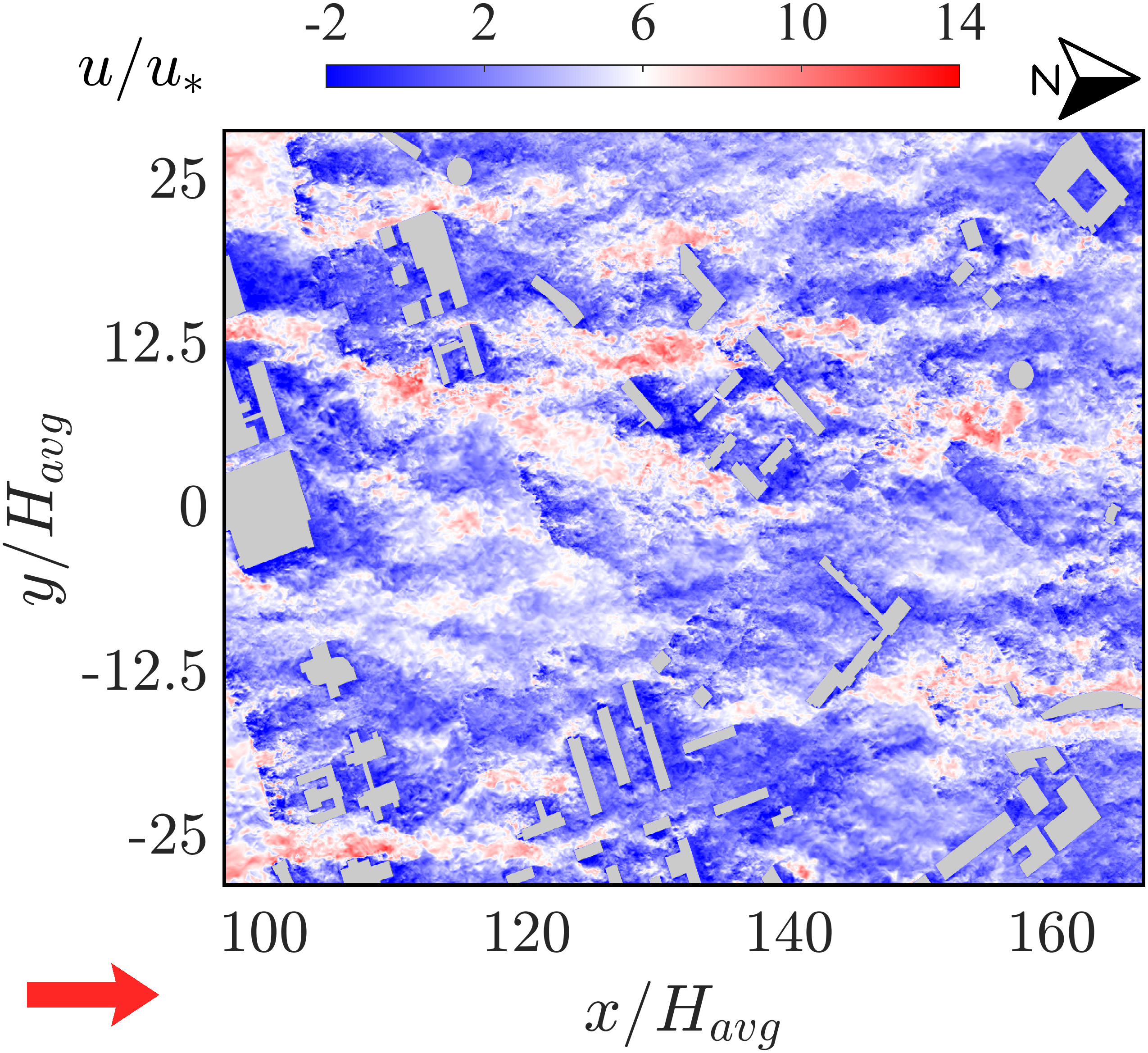}
    \caption{}
    \label{fig:subfig_c}
  \end{subfigure}
    \hfill
  \begin{subfigure}[b]{0.45\textwidth}
    \includegraphics[width=\textwidth]{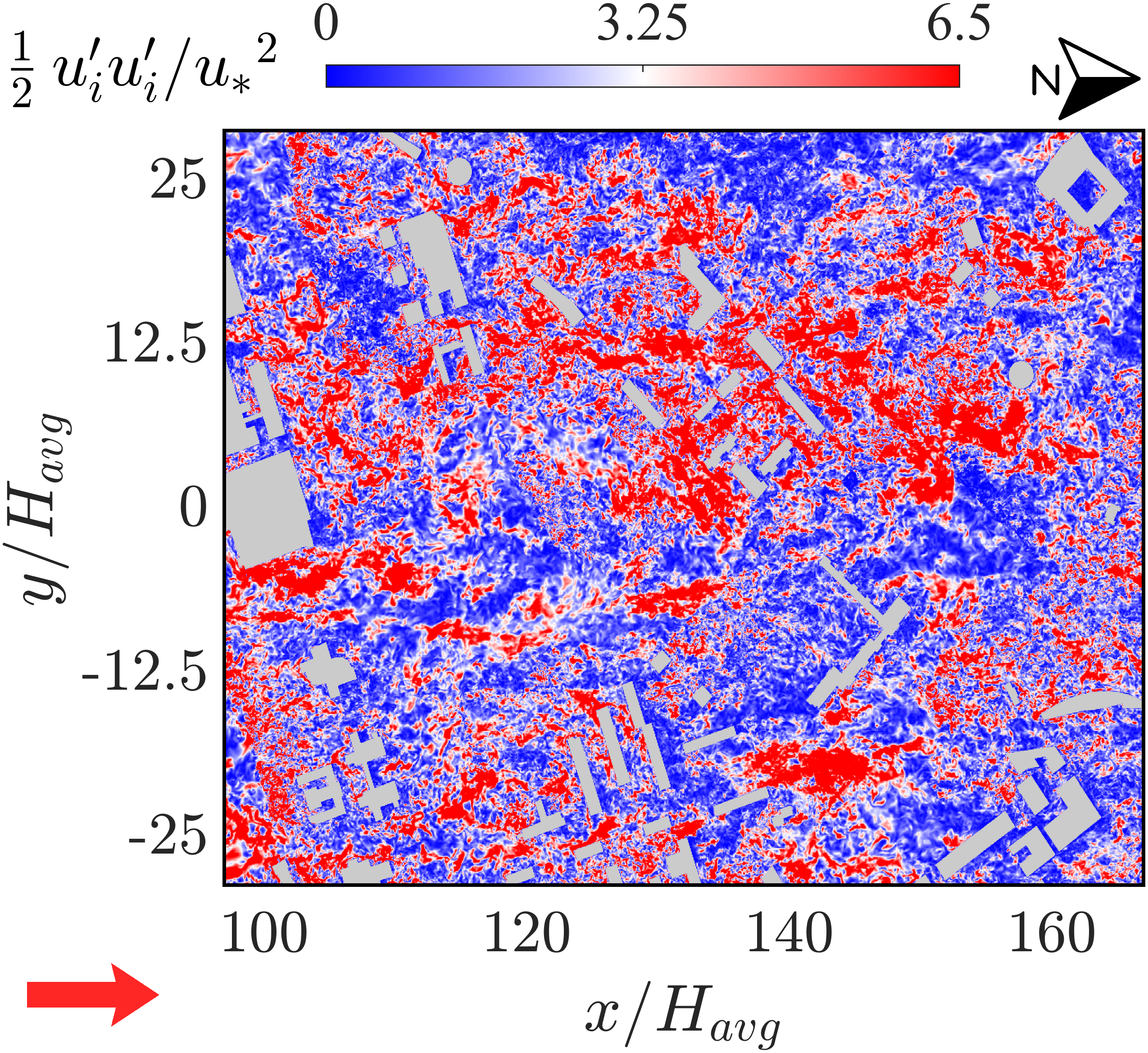}
    \caption{}
    \label{fig:subfig_d}
      \end{subfigure}
  \begin{subfigure}[b]{0.45\textwidth}
    \includegraphics[width=\textwidth]{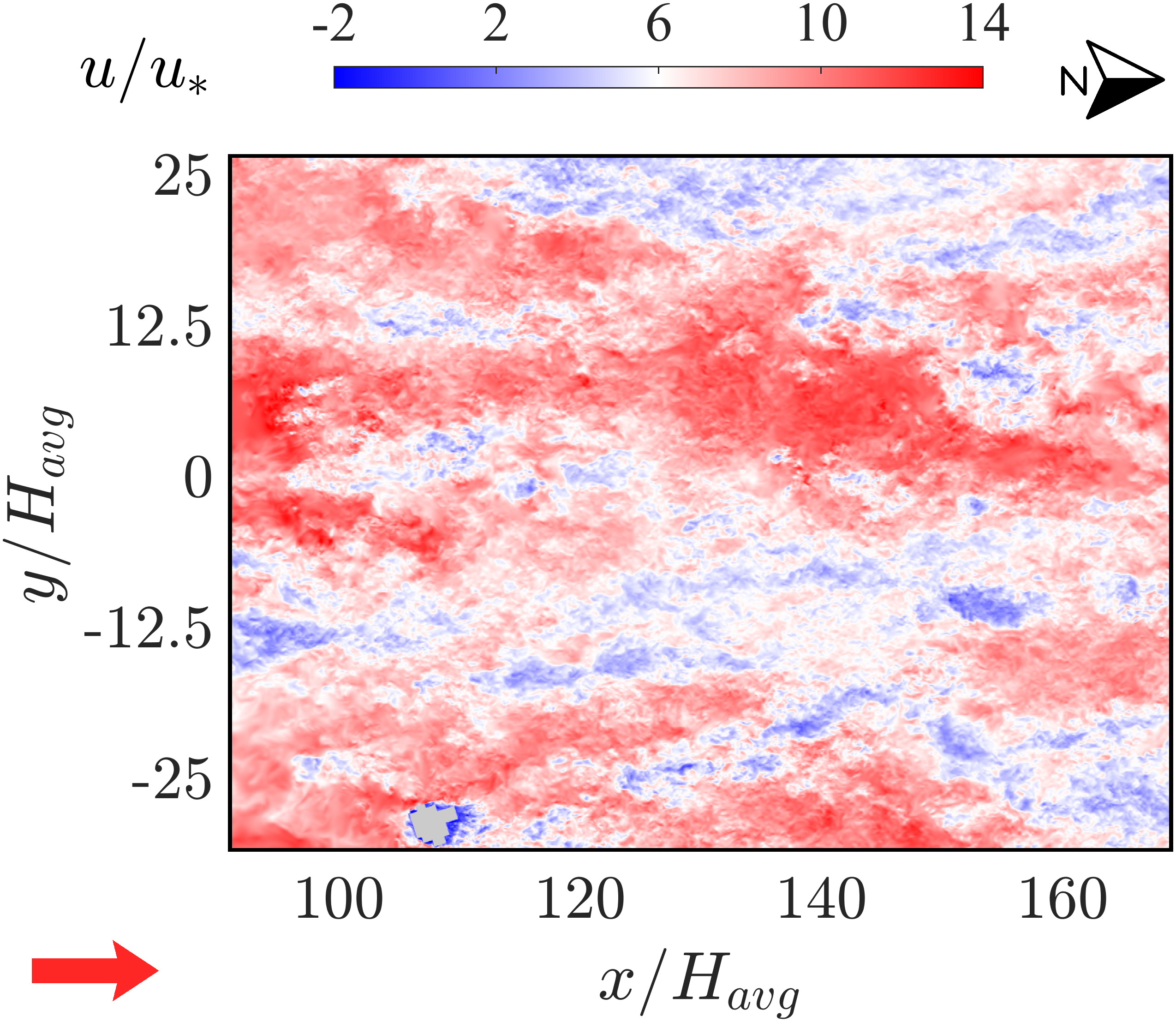}
    \caption{}
    \label{fig:subfig_e}
  \end{subfigure}
      \hfill
     \begin{subfigure}[b]{0.45\textwidth}
    \includegraphics[width=\textwidth]{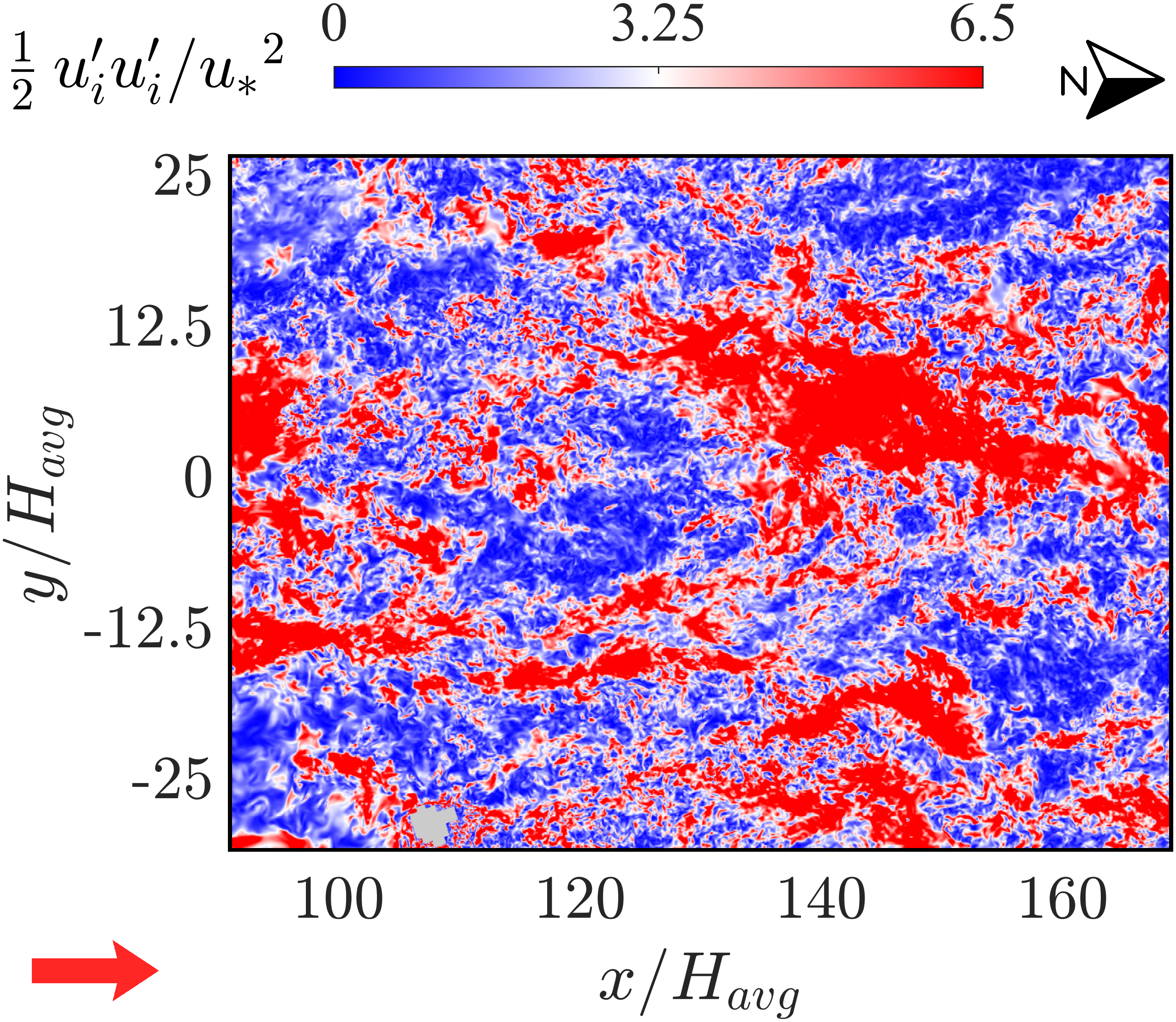}
    \caption{}
    \label{fig:subfig_f}
  \end{subfigure}
\caption{Instantaneous streamwise velocity\ming{, ${u/u_*}$, }(left panels)  and TKE\ming{, $\frac{1}{2} u'_iu'_i/u^2_*$, }(right panels) at wind direction \revtwo{$\Phi=180^\circ$ (South)}. (a, b) at pedestrian level $z/H_\mathrm{avg}=0.09$; (c, d) at $z/H_\mathrm{avg}=1$;  (e, f) at $z/H_\mathrm{avg}=3$. \ming{$H_\mathrm{avg}$ denotes average building height.}}
  \label{fig:instant_hor}
\end{figure}

\begin{figure}[]
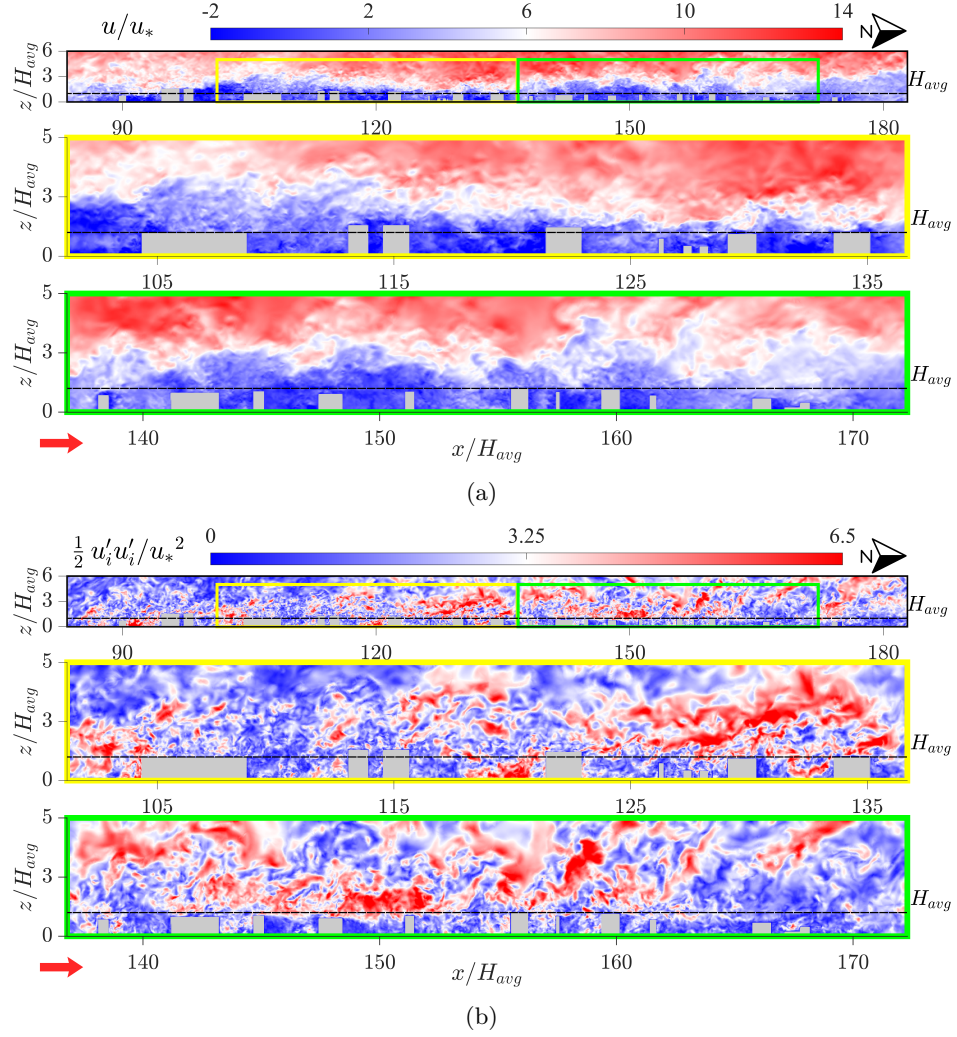

  \centering
       \begin{subfigure}[b]{0.95\textwidth}
    \includegraphics[width=\textwidth]{figures/velocity_vertical}
    \caption{}
   \end{subfigure}
          \begin{subfigure}[b]{0.95\textwidth}
    \includegraphics[width=\textwidth]{figures/tke_vertical}
    \caption{}
   \end{subfigure}
    \caption{Vertical slice of $(a)$ the instantaneous streamwise velocity\ming{, ${u/u_*}$,} and $(b)$ TKE\ming{, $\frac{1}{2} u'_iu'_i/u^2_*$,} taken at $y/H_{avg}=12$, at wind direction \revtwo{$\Phi=180^\circ$ (South)}. The horizontal dashed line represents the average building height, $H_{avg}$.}
  \label{fig:instant_ver}
\end{figure}

Figure \ref{fig:instant_hor} illustrates the impact of the urban canopy morphology on the flow structure at different heights (pedestrian level $z/H_\mathrm{avg}=0.09$, average building  height $z/H_\mathrm{avg}=1$ and at $z/H_\mathrm{avg}=3$, capturing instantaneous  streamwise velocity and \ming{TKE}. Additionally, Fig. \ref{fig:instant_ver} depicts these quantities in a vertical plane taken at  $y/H_{avg}=12$. Across the domain, the complex building layout generates a highly heterogeneous flow field, characterised by distinct flow regimes including recirculation zones, channeling effects, and intermittent high-velocity gusts.

At pedestrian level, the instantaneous flow  exhibits different structural patterns,  isolated vortices and unstable recirculation regions in the wake of the  buildings, while longitudinal streaks develop, channeling through aligned street canyons. Similar features were reported by  \cite{akinlabi2022} in their study on a real urban flow, where complex layouts and high-rise buildings were found to significantly enhance the spatial heterogeneity of the flow.
The instantaneous fields highlight the coexistence of stagnation regions in deeply sheltered courtyards and localised jets at building gaps and intersections. Moreover, the impingement of incoming wind on the windward façades of buildings diverts the fluid downwards, which amplifies local mean wind speeds at the pedestrian level. 
A similar feature was reported by \cite{OH2024105682}, noting that this amplification can increase wind speeds by a factor of 1.8 to 2 via the Venturi effect. 

\ming{As height increases}, i.e., at $z/H_\mathrm{avg}=1$ and $z/H_\mathrm{avg} = 3$ (see Fig. \ref{fig:instant_hor}c--f) the flow shifts from canopy-dominated dynamics to a shear-driven regime. Unlike the localised jets at pedestrian level, the field here becomes spatially coherent, forming elongated high- and low-speed streaks aligned with the mean wind. These high-speed patches downstream of building clusters are the footprint of momentum carried upward from rooftop and corner shear layers, governing the vertical exchange with the inertial sublayer (Fig. \ref{fig:instant_ver}).  This transition shows that individual buildings no longer dictate the flow pattern; instead, the dynamics are dominated by large-scale coherent structures that can extend for hundreds of meters, as observed by \cite{Giometto2016, akinlabi2022}.

\begin{figure}[]
  \centering
  \begin{subfigure}[b]{0.45\textwidth}
    \includegraphics[width=\textwidth]{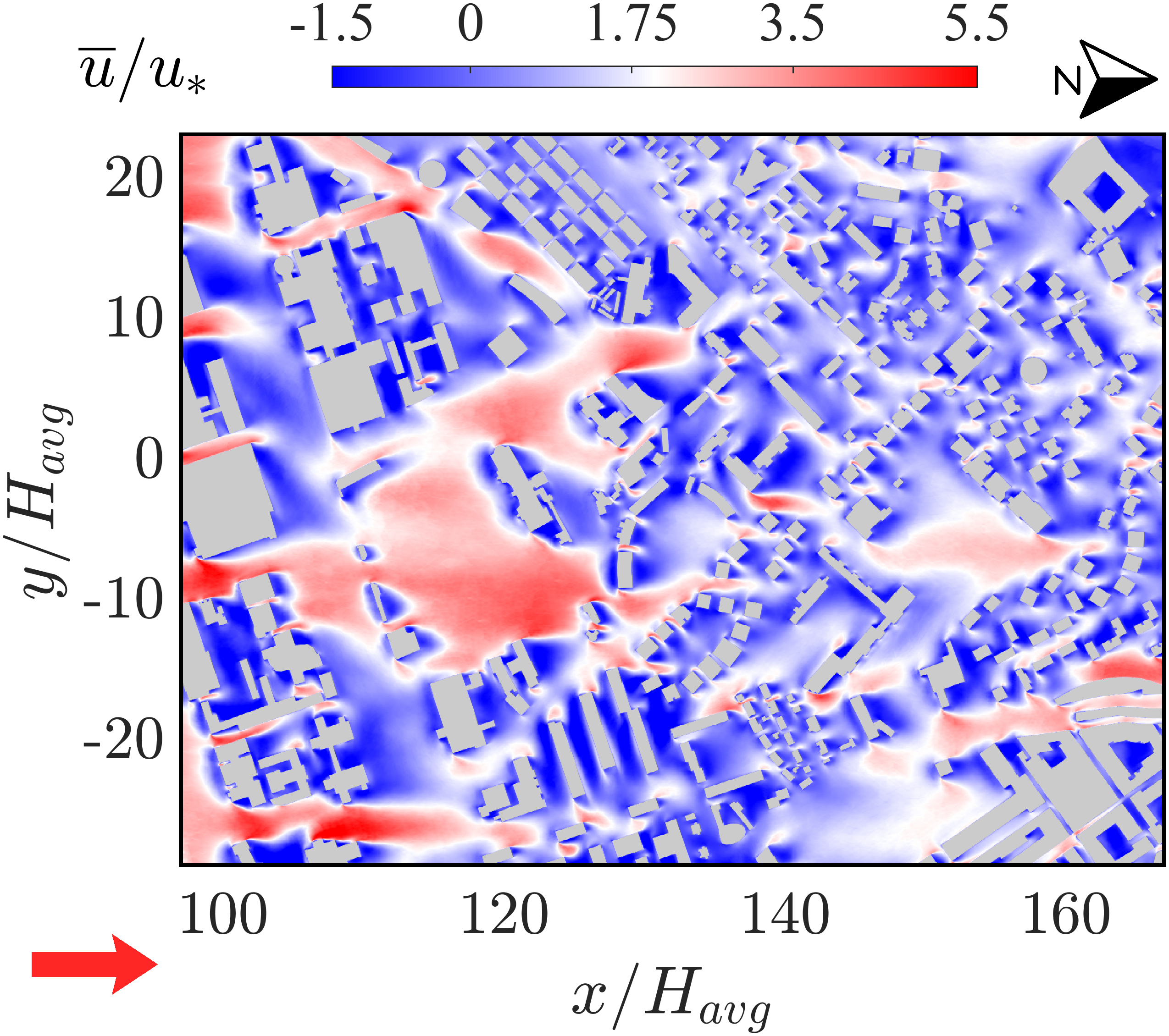}
    \caption{}
    \label{fig:subfig_a}
  \end{subfigure}
  \hfill
  \begin{subfigure}[b]{0.45\textwidth}
    \includegraphics[width=\textwidth]{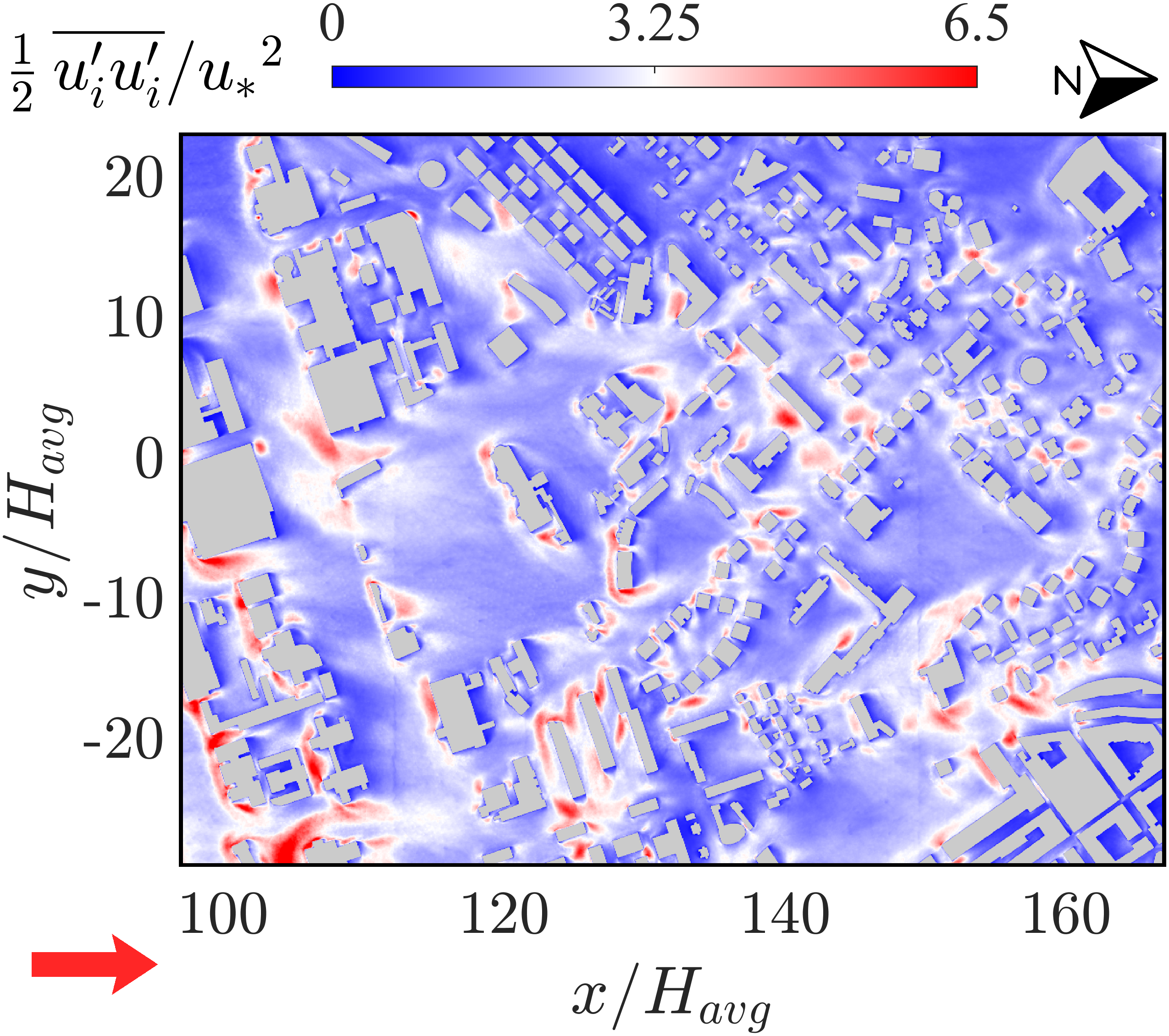}
    \caption{}
    \label{fig:subfig_b}
  \end{subfigure}
\caption{\ming{Time-averaged} (a) streamwise velocity, $\overline{u}/u_*$, and (b) TKE\ming{, $\frac{1}{2} \overline{u'_iu'_i}/u^2_*$},  at wind direction \revtwo{$\Phi=180^\circ$ (South)} at pedestrian level $z/H_{avg}=0.09$.}
  \label{fig:ave_hor}
\end{figure}

\begin{figure}[]
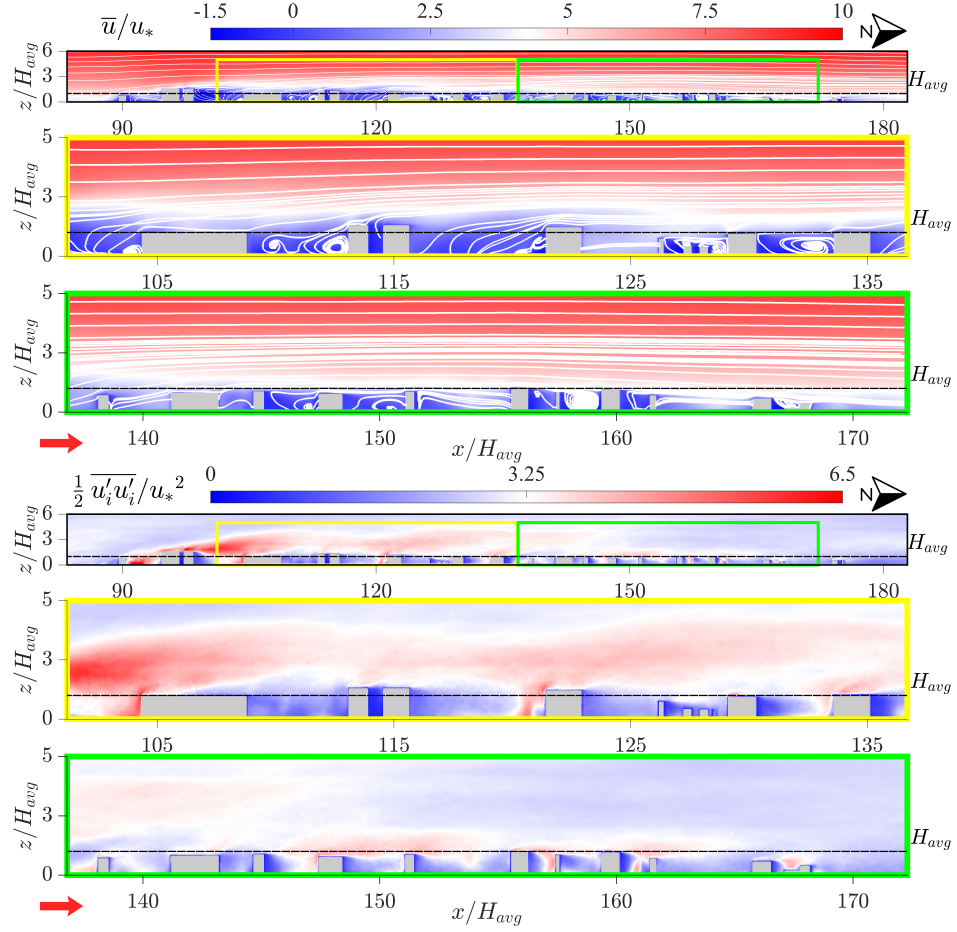

  \centering
   \begin{subfigure}[b]{0.95\textwidth}
    \includegraphics[width=\textwidth]{figures/avg_velocity_ver}
      \end{subfigure}
         \begin{subfigure}[b]{0.95\textwidth}
    \includegraphics[width=\textwidth]{figures/avg_tke_ver}
      \end{subfigure}
 \caption{Vertical slice of the average streamwise velocity, $\overline{u}/u_*$, and TKE, $\frac{1}{2} \overline{u'_iu'_i}/u^2_*$, taken at $y/H_\mathrm{avg}=12$, at wind direction \revtwo{$\Phi=180^\circ$ (South)}. The horizontal dashed line represents the average building height, $H_\mathrm{avg}$.}
  \label{fig:ave_ver}
\end{figure}

All the features described above, are  also evident on the mean profiles (see \ming{Figs.} \ref{fig:ave_hor} and \ref{fig:ave_ver}), where the mean streamwise velocity shows the organisational patterns imposed by the urban geometry. 
The overall neighbourhood flow is a multi-regime system. Isolated roughness flow occurs around taller structures, while wake interference and skimming flow prevail across most of the built area. Building height heterogeneity plays a key role, as taller structures deflect the flow and create extensive areas of weak or recirculating wind within deep canyons. This aligns with \cite{Cheng2023a} and \cite{Giometto2016}, who noted that while recirculations occupy most of the canopy, a wide variance in roof heights, similar to our study 
($H_{std}/H_{avg}=0.42$),  prevents flow homogeneity by creating a patchwork of wake and non-wake regions that increases wind penetration from above.
As the channeling discussed in the instantaneous flow at the pedestrian level, the mean field exhibits strong acceleration along primary street axes aligned with the inflow. These south-north oriented axes form well-defined corridors of high momentum that separate larger sheltered zones. These corridors are critical for urban ventilation, as highlighted by \cite{Wang2020} in their study of downtown Beijing.

The \ming{TKE} field is intrinsically linked to these mean structures. Regions with high mean TKE are concentrated in shear layers at the interfaces between high- and low-momentum zones. Specifically, high TKE is found along the lateral edges of channeling corridors, in the near-wake regions behind sharp building corners, and at intersection hotspots where multiple flows interact. 
These are zones of continuous turbulent production due to strong mean shear. The instantaneous TKE snapshots reveal this energy is delivered intermittently, with intense bursts likely associated with vortex shedding from upstream buildings or the break-down of larger instabilities.

The vertical structure of the flow, see Fig.  \ref{fig:ave_ver},  reveals the interaction between the overriding flow and the canopy layer, leading to distinct vertical regimes. Within the lower part of the canopy, the flow is highly disrupted, and here, wake interference dominates; the flow behind each building is strongly influenced by its immediate upstream neighbour, leading to complex, {3D} recirculations and stagnant zones with very low mean velocity but significant turbulent fluctuations generated locally. Near the roof level, a skimming flow regime becomes established over large building blocks limiting the vertical exchange. However,  building height variation allows the flow to dip into the canopy, creating local impingement zones, downward drafts, and contributing to the ventilation of interior streets.  \revone{The mean streamlines in the vertical plane highlight that the geometry of recirculation within the canopy is controlled by the interaction between the overlying flow and the local building arrangement. The organisation of these recirculation patterns is sensitive to wind direction, as evidenced by the substantial reorganisation of flow pathways and wake structures documented in Section~\ref{pedestrian_wind} for different inflow orientations. This has direct implications for scalar transport: regions dominated by weak mean exchange and persistent recirculation may increase scalar residence time and favour pollutant trapping, whereas regions with stronger vertical exchange can promote scalar removal from the canopy.}%

\subsection{Wind effects at pedestrian level}\label{pedestrian_wind}

\begin{figure}[h!]
  \centering
     \begin{subfigure}[b]{0.45\textwidth}
    \includegraphics[width=\textwidth]{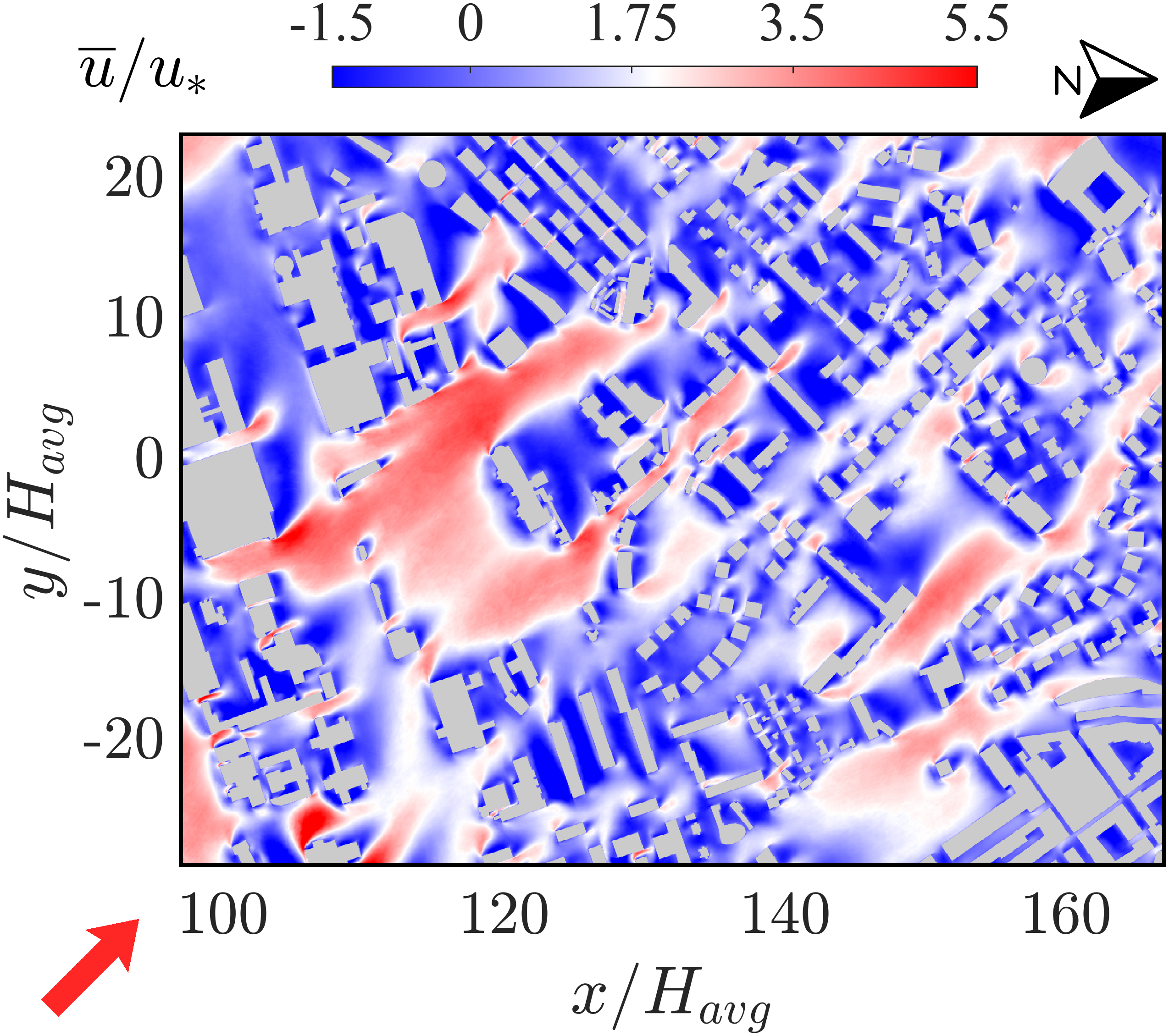}
    \caption{}
    \label{fig:subfig_h1}
  \end{subfigure}
      \hfill
    \begin{subfigure}[b]{0.45\textwidth}
    \includegraphics[width=\textwidth]{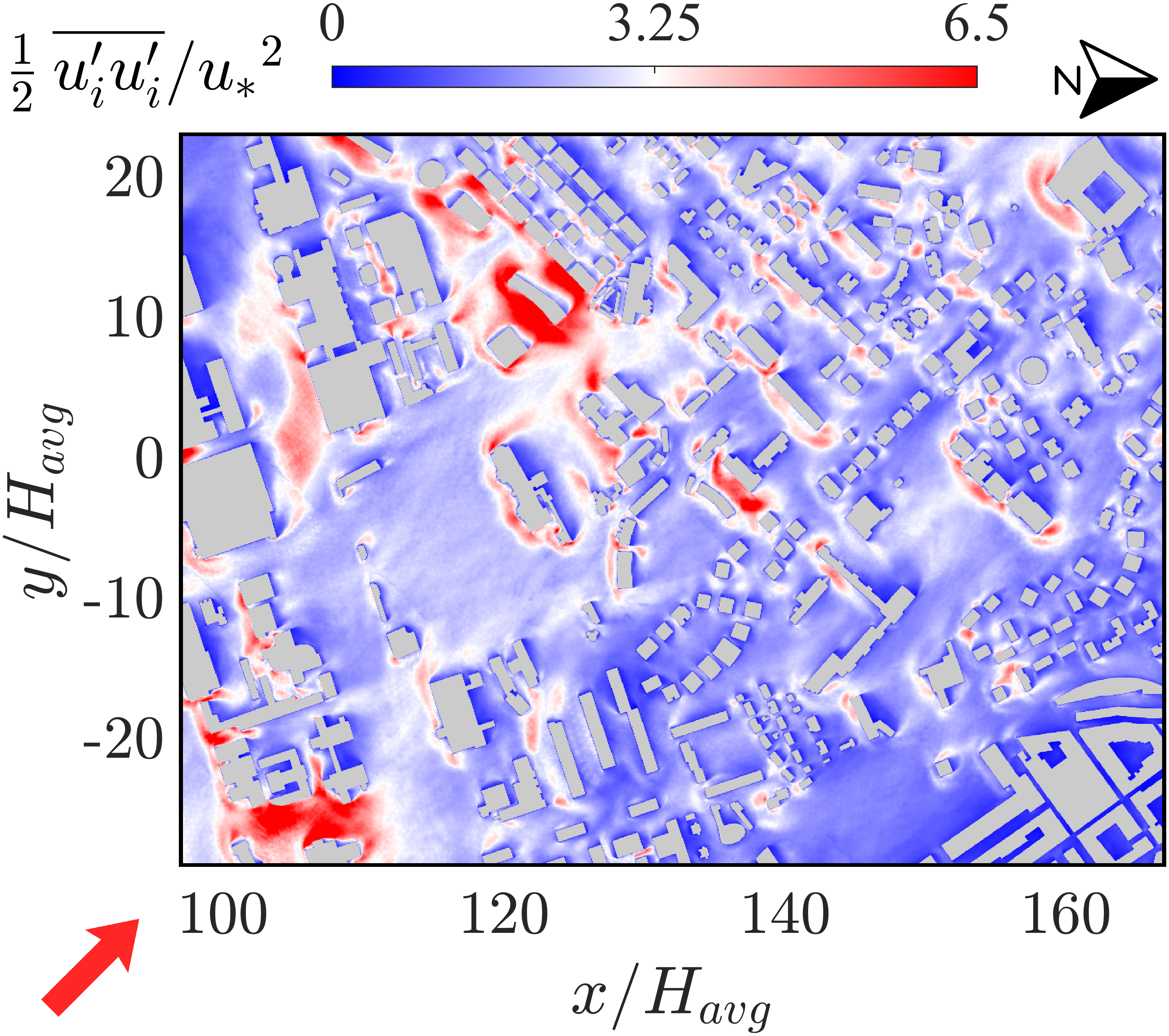}
    \caption{}
    \label{fig:subfig_g1}
  \end{subfigure}
  \begin{subfigure}[b]{0.45\textwidth}
    \includegraphics[width=\textwidth]{figures/bcn_avvel_adim_0deg}
    \caption{}
    \label{fig:subfig_a1}
  \end{subfigure}
  \hfill
  \begin{subfigure}[b]{0.45\textwidth}
    \includegraphics[width=\textwidth]{figures/avg_tke_pedestrian}
    \caption{}
    \label{fig:subfig_b1}
  \end{subfigure}
    \begin{subfigure}[b]{0.45\textwidth}
    \includegraphics[width=\textwidth]{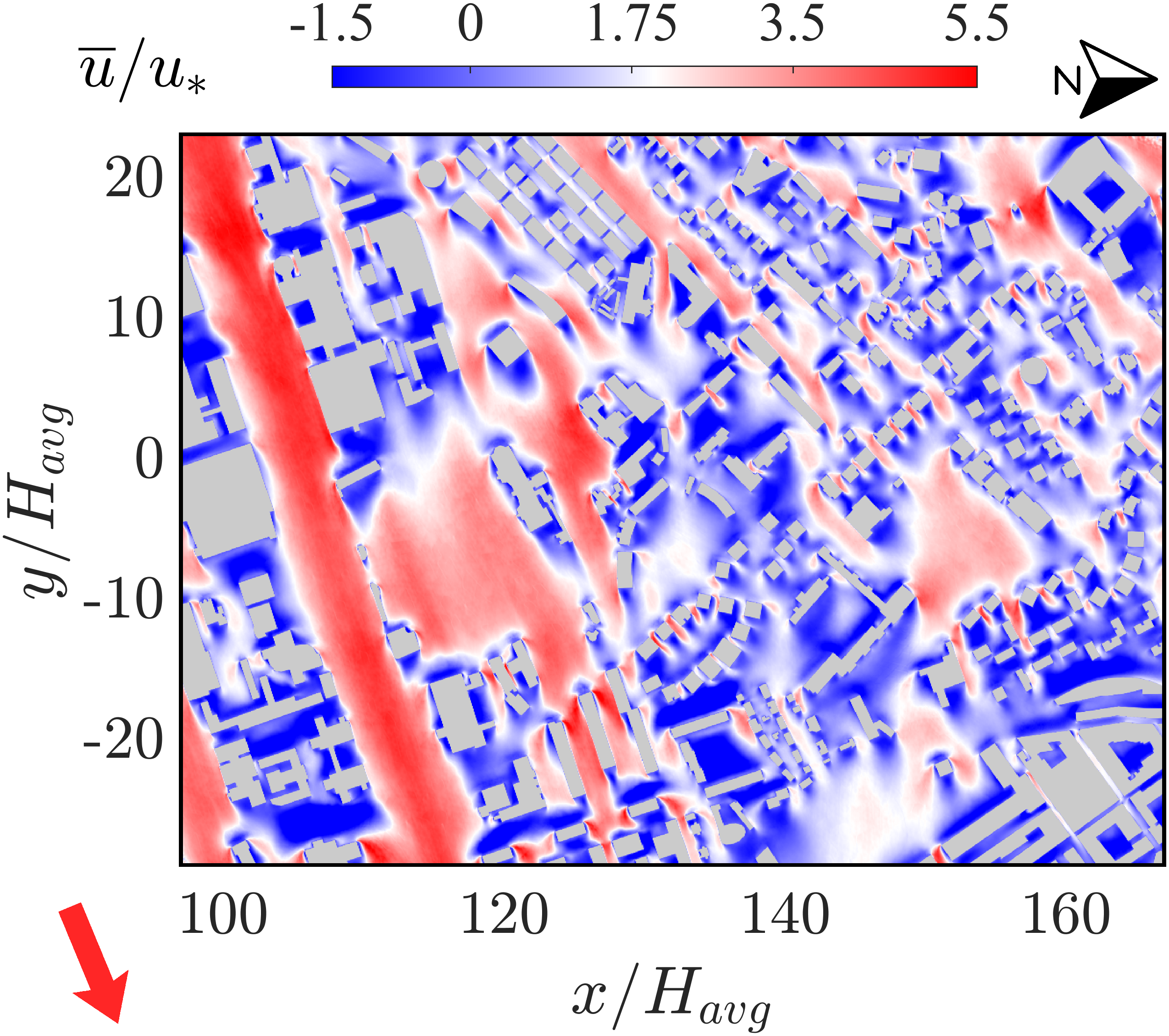}
    \caption{}
    \label{fig:subfig_c1}
  \end{subfigure}
  \hfill
  \begin{subfigure}[b]{0.45\textwidth}
    \includegraphics[width=\textwidth]{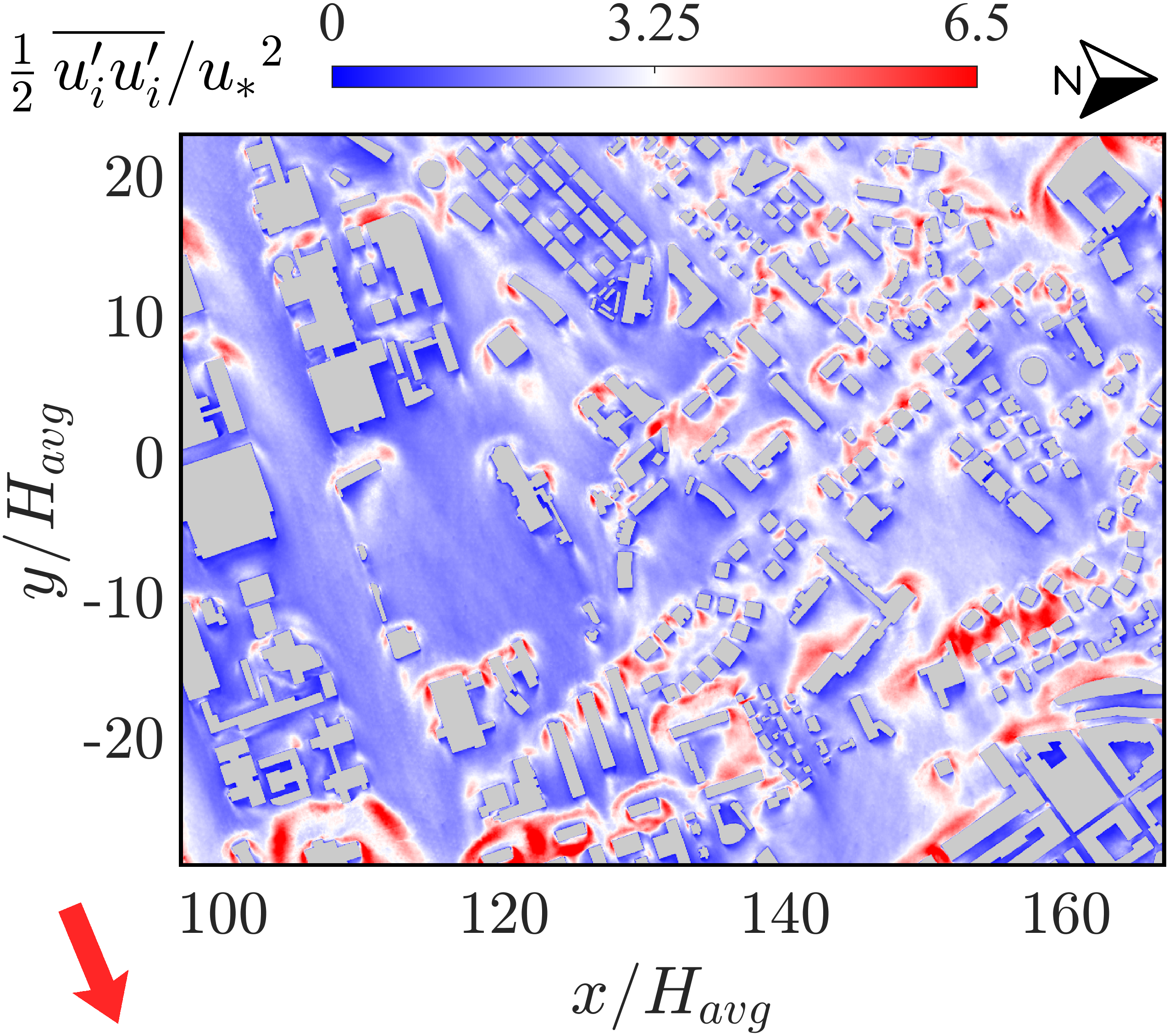}
    \caption{}
    \label{fig:subfig_d1}
  \end{subfigure}
  \end{figure}
    \begin{figure}
   \ContinuedFloat
   \begin{subfigure}[b]{0.45\textwidth}
    \includegraphics[width=\textwidth]{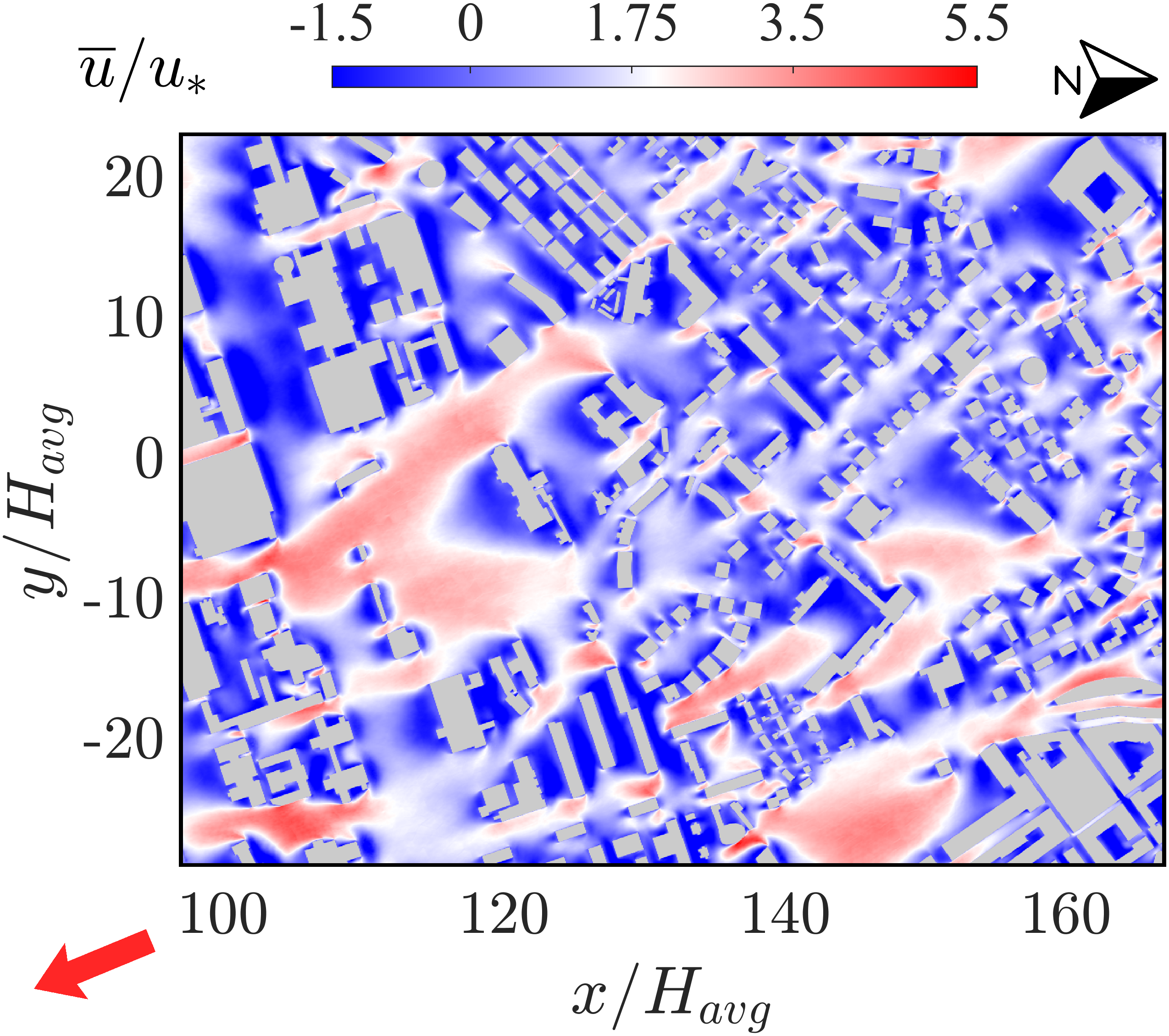}
    \caption{}
    \label{fig:subfig_e1}
  \end{subfigure}
      \hfill
    \begin{subfigure}[b]{0.45\textwidth}
    \includegraphics[width=\textwidth]{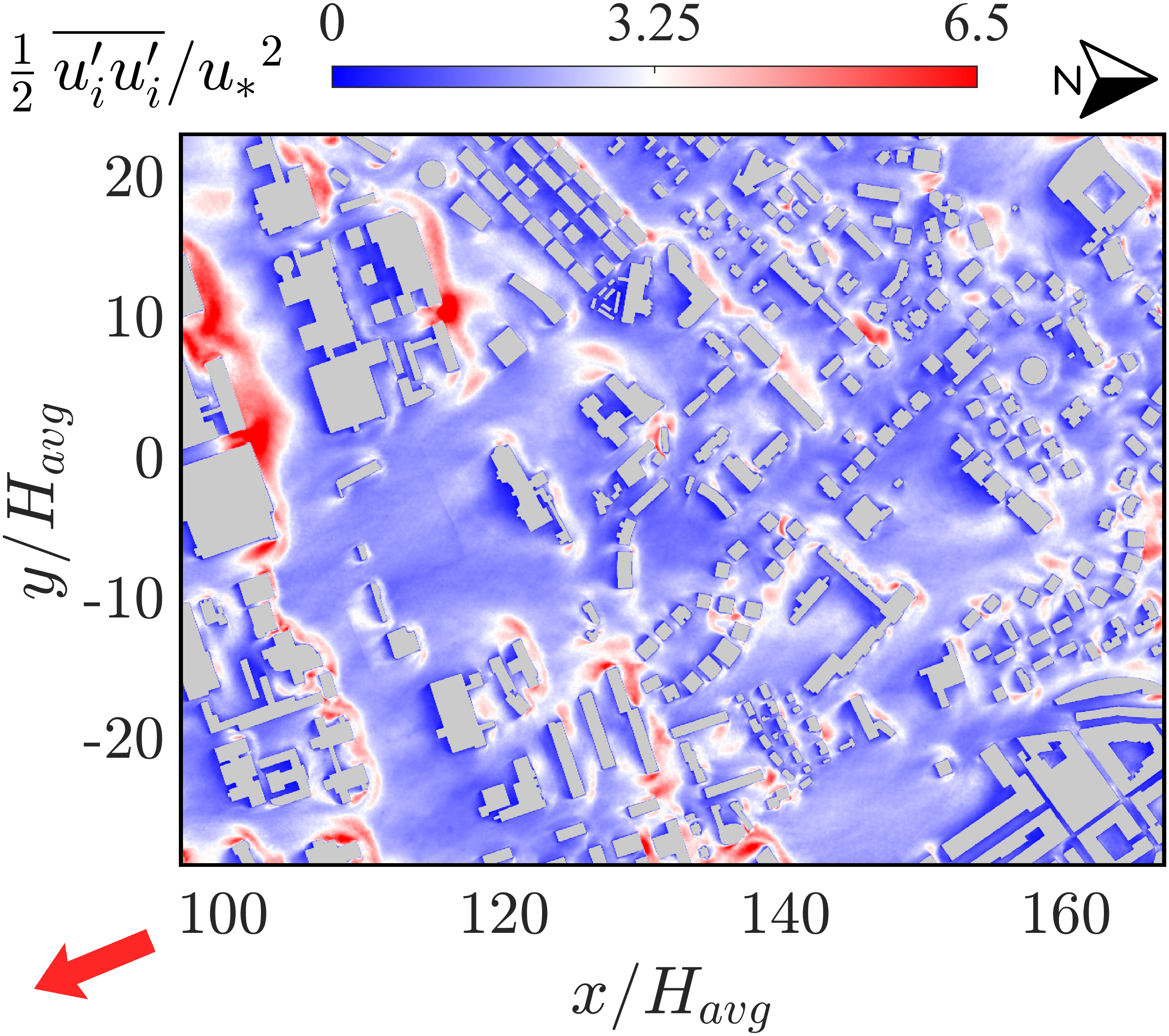}
    \caption{}
    \label{fig:subfig_f1}
  \end{subfigure}
  \caption{Influence of the wind direction on the \ming{time-averaged} streamwise velocity, $\overline{u}/u_*$, (left panels) and TKE, $\frac{1}{2} \overline{u'_iu'_i}/u^2_*$, (right panels) (a, b) \revtwo{$\Phi=135^\circ$}, (c, d)  \revtwo{$\Phi=180^\circ$}, (e,f) \revtwo{$\Phi=247.5^\circ$}, (g, h) \revtwo{$\Phi=337.5^\circ$}. Wind direction is marked with a red arrow in each panel.}
  \label{fig:ave_wind}
\end{figure}

Figure \ref{fig:ave_wind} shows the effect of wind direction on the  mean streamwise velocity  and turbulent kinetic energy  at the  pedestrian level  for $\revtwo{\Phi=135^\circ,\:180^\circ,\:247.5^\circ,\:337.5^\circ}$.  
The figure demonstrates a high sensitivity to the incoming wind direction, which triggers substantial reorganisations of dominant flow pathways, wake structures, and turbulence distribution within the canopy. This behaviour reflects the intrinsic anisotropy of the urban roughness and the strong directional dependence of momentum transport in such densely built environments. Similar to the observations for $\revtwo{\Phi=180^\circ}$, across all inflow orientations, the urban canopy promotes a highly heterogeneous mean flow controlled by the morphology of the city, but  the spatial organisation and continuity of features vary markedly with $\Phi$. 

Depending on the wind direction, corridors give way to zones with cellular structures or the activation of alternative pathways, reflecting the directional sensitivity already documented in canonical and realistic canopy studies. All these observations are also consistent with the mechanisms reported in idealised street canyons \cite{Kim2004} and regular block arrays \cite{Lin2014}, where the specific alignment of the geometry produces either long acceleration bands or large sheltered areas with quasi-steady vortex structures in the lee of compact clusters.

Indeed, wind direction plays an important role in the change of flow patterns. 
For instance, rotating the inflow toward oblique orientations, e.g., $\revtwo{\Phi=247.5^\circ}$ see Fig. \ref{fig:subfig_c1}, fragments the dominant corridors into multiple branches, marking a shift toward a cellular flow organisation. This resembles the multi-cell patterns and oblique exchange mechanisms reported by \cite{Santiago2013}, where cross-street pressure gradients drive lateral momentum transfer and the breakup of canonical wakes. In this realistic geometry, the process activates alternative ventilation pathways and prevents the formation of single dominant recirculation cells, resulting in a more distributed momentum field than in simplified arrays. Further rotation to $\revtwo{\Phi=337.5^\circ}$ shifts acceleration bands laterally, reconnecting distant open spaces via new preferential channels.
The resulting flow is highly anisotropic;  for $\revtwo{\Phi=135^\circ}$, momentum concentrates within a few dominant corridors while adjacent areas remain deeply sheltered. This localisation, as identified in canonical blocks  by \cite{Lin2014}, characterises inefficient ventilation regimes where the urban fabric restricts momentum dispersion to a limited number of pathways.

The \ming{TKE} fields further reveal that wind direction modulates not only the mean ventilation efficiency but also the spatial structure of turbulence production and dissipation. Confirming the instantaneous flow observations, elevated turbulence levels systematically develop near windward building edges, sharp corners, and intersection regions, where strong velocity gradients and flow separation prevail. As the inflow orientation changes, these energetic zones migrate across the canopy, reflecting shifts in the location of dominant shear layers and impingement regions. 

A recurrent feature across all cases is the emergence of preferential ventilation corridors whose orientation and continuity depend on the alignment between wind direction and the underlying street network. These corridors act as conduits for momentum transport and constitute the primary pathways for ventilation at pedestrian level. 
At the same time, large sheltered regions persist in the wake of dense building clusters and within enclosed courtyards, where weak mean flow and reduced turbulence can favour pollutant accumulation or thermal discomfort.

\begin{table*}[h!]
\caption{Morphological parameters for each zone and wind direction. \revone{$H_{\mathrm{avg}}$ and $H_{\mathrm{std}}$ denote the average building height and standard deviation of building height. $\lambda_f$ and $\lambda_p$ are the frontal and plan area. Values in parentheses indicate the effective frontal density area $\lambda_{f,\mathrm{eff}} \equiv \lambda_f H_{\mathrm{avg}}/H_{\mathrm{std}}$.} The location of each zone is shown in \ming{Fig.~\ref{fig:city}}.}
\centering
{\small
\renewcommand{\arraystretch}{1.25}
\begin{tabular}{cccccccc}
\hline
\textbf{Zone} 
& $H_{\mathrm{avg}}\ming{(m)}$ & $H_{\mathrm{std}}\ming{(m)}$ 
& $\lambda_p$ 
& \multicolumn{4}{c}{$\lambda_f$ $(\lambda_{f,\mathrm{eff}})$} \\
\cline{5-8}
& & & & $135^\circ$ & $180^\circ$ & $247.5^\circ$ & {$337.5^\circ$} \\
\hline
1 & 15.21 & 6.54 & 0.25 & 0.22 (0.51) & 0.22 (0.51) & 0.18 (0.42) & 0.23 (0.53) \\
2 & 17.34 & 3.11 & 0.19 & 0.18 (1.00) & 0.21 (1.17) & 0.18 (1.00) & 0.20 (1.12) \\
3 & 14.70 & 4.04 & 0.10 & 0.07 (0.25) & 0.08 (0.29) & 0.07 (0.25) & 0.08 (0.29) \\
4 & 18.69 & 3.56 & 0.47 & 0.29 (1.52) & 0.34 (1.79) & 0.32 (1.68) & 0.31 (1.63) \\
\hline
\end{tabular}
}
\label{tab:morpho_params}
\end{table*}

\revone{

To further quantify the effects of wind direction and the impact of
morphological characteristics on the wind at pedestrian level, four
representative subdomains with distinct morphological characteristics are
selected. The location of each zone is shown in Fig.~\ref{fig:city}, while the
corresponding morphological parameters are summarised in Table~\ref{tab:morpho_params}, revealing
substantial contrasts in packing density, ranging from a relatively open
configuration in Zone~3 ($\lambda_p = 0.10$) to a very dense urban fabric
in Zone~4 ($\lambda_p = 0.47$), with consistent variations in frontal area
density.

Two metrics are evaluated at $z = 1.75$~m for each subdomain and the four
inflow angles of Fig.~\ref{fig:ave_wind}. The first is the 90th percentile of the normalised
velocity magnitude, $U_{P90}/U_\infty$, where $U_{P90}$ is the value of the
time-averaged velocity magnitude below which 90\% of the pedestrian-level
grid points of the subdomain fall and $U_\infty$ is the freestream velocity
at the boundary layer edge. The second is the velocity ratio
$v_R = \overline{U}/U_\infty$, where $\overline{U}$ is the spatially
averaged, time-averaged velocity magnitude. The
results are summarised in Table~\ref{tab:pedestrian_metrics}.}

\begin{table*}[h]
\centering
\caption{Pedestrian-level metrics  ($z = 1.75$~m) for the four zones in Table~\ref{tab:morpho_params} and the four wind directions in  Fig.~\ref{fig:ave_wind}. The value of the metrics over the whole domain is also given (All). $U_{P90}/U_\infty$ denotes the 90th-percentile normalised velocity and
$v_R = \overline{U}/U_\infty$ the velocity ratio.}
\label{tab:pedestrian_metrics}
{\small
\renewcommand{\arraystretch}{1.25}
\begin{tabular}{llcccc}
\toprule
 & & \multicolumn{4}{c}{$\Phi$} \\
\cmidrule(lr){3-6}
Zone & Metric & $135^\circ$ & $180^\circ$ & $247.5^\circ$ & $337.5^\circ$ \\
\midrule
\multirow{2}{*}{All}
  & $U_{P90}/U_\infty$      & 0.25 & 0.24 & 0.26 & 0.24 \\
  & $v_R$ & 0.142 & 0.144 & 0.147 & 0.142 \\[4pt]
\multirow{2}{*}{Zone~1}
  & $U_{P90}/U_\infty$      & 0.22 & 0.24 & 0.22 & 0.25 \\
  & $v_R$ & 0.127 & 0.128 & 0.134 & 0.126 \\[4pt]
\multirow{2}{*}{Zone~2}
  & $U_{P90}/U_\infty$      & 0.19 & 0.18 & 0.21 & 0.17 \\
  & $v_R$ & 0.113 & 0.104 & 0.124 & 0.100 \\[4pt]
\multirow{2}{*}{Zone~3}
  & $U_{P90}/U_\infty$      & 0.30 & 0.29 & 0.27 & 0.24 \\
  & $v_R$ & 0.198 & 0.202 & 0.192 & 0.161 \\[4pt]
\multirow{2}{*}{Zone~4}
  & $U_{P90}/U_\infty$      & 0.18 & 0.21 & 0.17 & 0.21 \\
  & $v_R$ & 0.106 & 0.125 & 0.097 & 0.121 \\
\bottomrule
\end{tabular}
}
\end{table*}
\revone{
A clear and consistent hierarchy emerges with respect to plan area density.
Zone~3, the most open area, records the highest pedestrian-level
velocities across all wind directions, with $U_{P90}/U_\infty$ ranging from 0.24
to 0.30 and $v_R$ between 0.16 and 0.20. Zone~4, the densest
one, records among the lowest values, with $U_{P90}/U_\infty$ between 0.17
and 0.21 and $v_R$ between 0.097 and 0.125, although Zone~2
($\lambda_p = 0.19$) reaches comparably low levels for certain inflow
orientations. Within Zone~4, the highest $U_{P90}/U_\infty$ (0.21) occurs at
$\Phi = 337.5^\circ$, which nearly aligns with the dominant NNW--SSE street
orientation of the study area, while the lowest $v_R$ (0.097)
occurs at $\Phi = 247.5^\circ$, close to the perpendicular ENE--WSW
direction where building fronts present maximum blockage. The
$\Phi = 135^\circ$ direction, on the opposite end of the same NNW--SSE
axis, yields a similarly low $U_{P90}/U_\infty$ to that observed at $\Phi = 247.5^\circ$. This is probably due to the upstream sheltering of buildings when the flow approaches from the southeast.
Note also that the roughly 1.5- to 2-fold difference in $v_R$ between Zone~3 and Zone~4 
is consistent across all four wind directions, which confirms that local packing
density has a determinant effect on  pedestrian-level wind
speeds.  However, this local variability is not reflected in the neighbourhood-averaged
$v_R$, which remains near 0.14 for all inflow angles,
consistent with the quasi-isotropic behaviour of the double-averaged
statistics as will be discussed in Section \ref{sec:rough}.

It should also be pointed out  that the plan area density
$\lambda_p$ ranking alone does not fully explain the zone-to-zone
differences. Zone~1 ($\lambda_p = 0.25$) achieves higher $v_R$
than Zone~2 ($\lambda_p = 0.19$) despite being denser in plan view, and
Zone~2 occasionally performs worse than Zone~4 for certain wind directions,
suggesting that building height variability and street-corridor alignment
also modulate pedestrian-level wind exposure beyond what plan density alone
captures.

Such inconsistencies between the observed metric hierarchy and the $\lambda_p$ ordering are in agreement with the findings of \cite{Duan2023}, who showed that the frontal area density $\lambda_f$ provides a more robust scaling than $\lambda_p$ for wind exposure in real urban districts. For instance, a closer inspection of the $\lambda_f$ values at $\Phi = 135^\circ$ reported in Table~\ref{tab:morpho_params} indeed reveals a hierarchy more consistent with the velocity observations, with Zone~3 recording the lowest $\lambda_f$ ($= 0.07$) and the highest pedestrian-level velocities and Zone~4 on the opposite side ($\lambda_f = 0.29$). However, this still does not resolve the Zone~1 / Zone~2 anomaly.  On the contrary, Zone~1 presents a slightly higher $\lambda_f$ ($= 0.22$) than Zone~2 ($= 0.18$), which would predict the opposite velocity ordering to that observed. Following  \cite{Duan2023}, we have used here  an effective frontal density $\lambda_{f,\mathrm{eff}} \equiv \lambda_f H_\mathrm{avg}/H_\mathrm{std}$ to account for building-height variability. For $\Phi = 135^\circ$, one obtains $\lambda_{f,\mathrm{eff}} = 0.25$, $0.51$, $1.00$, and $1.52$ for Zones~3, 1, 2, and~4, respectively, an ordering fully consistent with the observed velocity hierarchy at that direction. The much larger building-height variability of Zone~1 ($H_\mathrm{std} = 6.54$~m versus $3.11$~m in Zone~2) yields a substantially lower $\lambda_{f,\mathrm{eff}}$ despite the higher $\lambda_f$, reflecting the stronger vertical momentum exchange associated with a more heterogeneous roofline. A similar correspondence is found at $\Phi = 247.5^\circ$, while at $\Phi = 180^\circ$ and $337.5^\circ$ the agreement is less clear, with Zone~4 recording a slightly higher $v_R$ than Zone~2 despite a larger $\lambda_{f,\mathrm{eff}}$. This behaviour possibly can be attributed to the street-corridor alignment effects which become relevant for certain inflow orientations. 
}
\subsection{Flow statistics in the roughness sublayer}
\label{sec:rough}

In urban canopy flows, building-induced spatial heterogeneity necessitates specialised averaging to analyse turbulence statistics. Following the double-averaging method \citep{Raupach1982}, the instantaneous variable  $\phi(\mathbf{x}, t)$  is first decomposed using the traditional Reynolds decomposition, as $
\phi(\mathbf{x}, t) = \overline{\phi}(\mathbf{x}) + \phi'(\mathbf{x}, t),
$
 where \ming{$\cdot'$ represents the fluctuating component}.  Applying a spatial average to the time-averaged field then yields, 
 $ 
\phi(\mathbf{x}, t) = \langle \overline{\phi} \rangle + \phi''(\mathbf{x}) + \phi'(\mathbf{x}, t).
$ 
Here, $\langle \overline{\phi} \rangle$ is the {superficial} double average, 
$\phi'' = \overline{\phi} - \langle \overline{\phi} \rangle$ is the spatial deviation, and  $\phi'$  remains the temporal fluctuation.

 \definecolor{wd01}{rgb}{0.193,0.302,0.550}
\definecolor{wd02}{rgb}{0.286,0.800,0.280}
\definecolor{wd03}{rgb}{0.800,0.506,0.280}
\definecolor{wd04}{rgb}{0.360,0.800,0.789}
\definecolor{wd05}{rgb}{0.800,0.280,0.515}
\definecolor{wd06}{rgb}{0.599,0.280,0.800}
\definecolor{wd07}{rgb}{0.439,0.550,0.275}
\definecolor{wd08}{rgb}{0.550,0.358,0.367}
\definecolor{wd09}{rgb}{0.799,0.800,0.280}
\definecolor{wd10}{rgb}{0.720,0.520,0.800}
\definecolor{wd11}{rgb}{0.440,0.629,0.800}
\definecolor{wd12}{rgb}{0.280,0.800,0.552}
\definecolor{wd13}{rgb}{0.800,0.280,0.280}
\definecolor{wd14}{rgb}{0.800,0.717,0.520}
\definecolor{wd15}{rgb}{0.280,0.361,0.800}

\begin{figure}[h!]
  \centering
  \begin{subfigure}[b]{0.49\textwidth}
    \includegraphics[width=\textwidth]{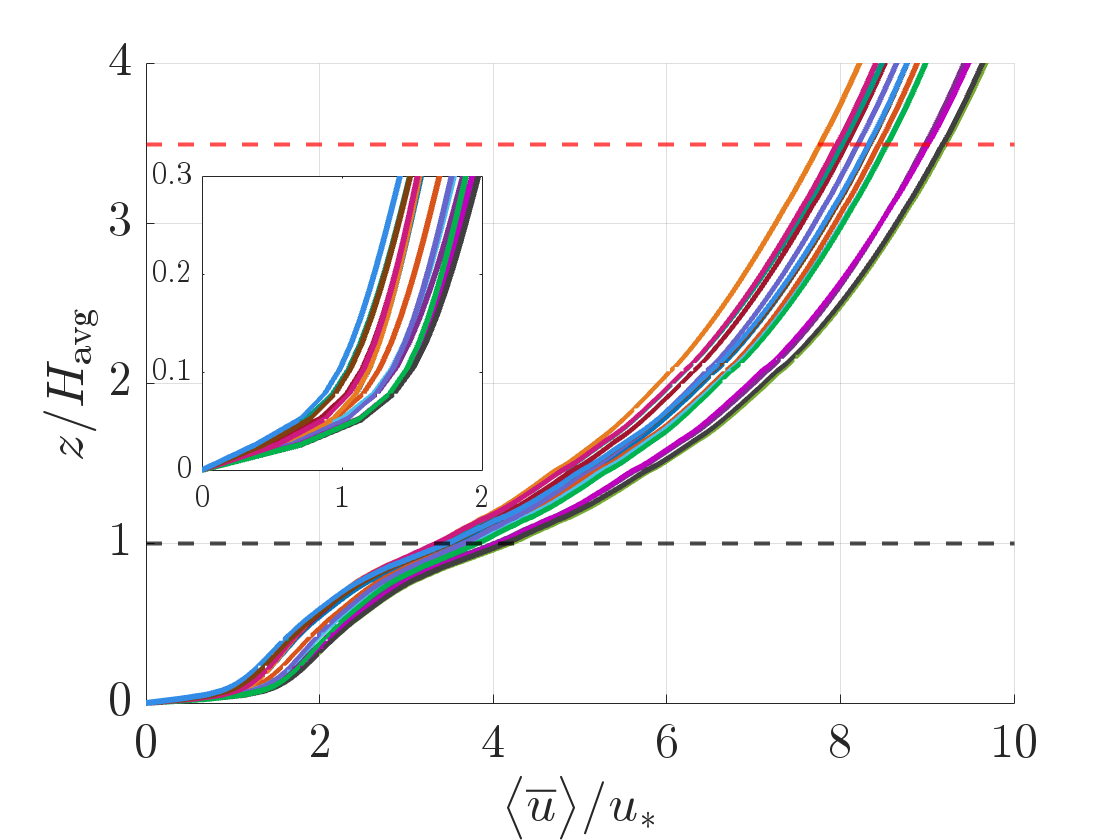}
    \caption{}
    \label{fig:subfig_a1}
  \end{subfigure}
  \begin{subfigure}[b]{0.49\textwidth}
    \includegraphics[width=\textwidth]{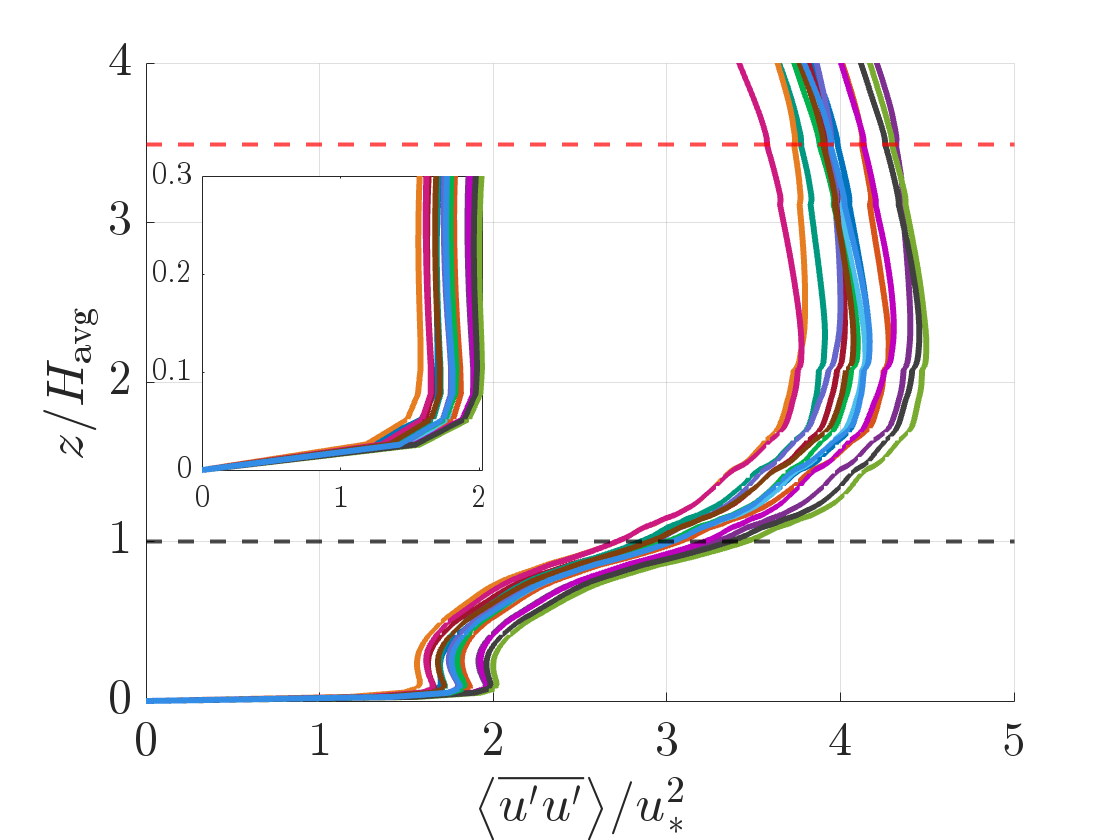}
    \caption{}
    \label{fig:subfig_b1}
  \end{subfigure}
\caption{Double-average vertical profiles of  \ming{$(a)$ streamwise velocity, $\langle \overline{u} \rangle/u_*$, and $(b)$ the variance of the streamwise velocity, $\langle \overline{u'u'}\rangle/u_*^2$,} for the different wind directions: 
\textcolor{wd01}{\rule{0.5cm}{1pt}} \textcolor{black}{0°}, 
\textcolor{wd02}{\rule{0.5cm}{1pt}} \textcolor{black}{22.5°}, 
\textcolor{wd03}{\rule{0.5cm}{1pt}} \textcolor{black}{45°}, 
\textcolor{wd04}{\rule{0.5cm}{1pt}} \textcolor{black}{67.5°}, 
\textcolor{wd05}{\rule{0.5cm}{1pt}} \textcolor{black}{90°}, 
\textcolor{wd06}{\rule{0.5cm}{1pt}} \textcolor{black}{112.5°}, 
\textcolor{wd07}{\rule{0.5cm}{1pt}} \textcolor{black}{135°}, 
\textcolor{wd08}{\rule{0.5cm}{1pt}} \textcolor{black}{180°}, 
\textcolor{wd09}{\rule{0.5cm}{1pt}} \textcolor{black}{202.5°}, 
\textcolor{wd10}{\rule{0.5cm}{1pt}} \textcolor{black}{225°}, 
\textcolor{wd11}{\rule{0.5cm}{1pt}} \textcolor{black}{247.5°}, 
\textcolor{wd12}{\rule{0.5cm}{1pt}} \textcolor{black}{270°}, 
\textcolor{wd13}{\rule{0.5cm}{1pt}} \textcolor{black}{292.5°}, 
\textcolor{wd14}{\rule{0.5cm}{1pt}} \textcolor{black}{315°}, 
\textcolor{wd15}{\rule{0.5cm}{1pt}} \textcolor{black}{337.5°}.
The black dashed line denotes the average building height, \ming{$H_\mathrm{avg}$}, and the red dashed line the maximum building height,  \ming{$H_\mathrm{max}$}.}
  \label{fig:DA}
\end{figure}

Figure \ref{fig:DA} presents the vertical profiles of $\langle \overline u \rangle/u_*$ and its fluctuations $\langle \overline{u'u'}\rangle/u_*^2$ for all wind directions. 
Despite the significant directional sensitivity observed in instantaneous and time-averaged fields, where localised acceleration and sheltering are important (Fig. \ref{fig:ave_wind}), these features are attenuated by spatial averaging. The resulting double-averaged profiles indicate that 
the influence of wind direction at the ensemble scale  is relatively modest, with all orientations yielding similar velocity shapes, particularly within the canopy  $z/H_{\text{avg}} < 1$. This consistency suggests that the effective aerodynamic roughness is not strongly directional in the studied area. A possible explanation is that this results from the disordered distribution of building heights and orientations, which effectively dampens directional sensitivity at the neighbourhood scale when double-averaging is applied. This suggests that simplified, direction-agnostic parameterisations may be sufficient for urban flow modelling in neighbourhoods that do not exhibit extreme morphological anisotropy.

A salient feature in Fig. \ref{fig:DA} is the presence of two distinct inflection points in the  $\langle \overline u \rangle/u_*$ profiles, which reveals a complex vertical organisation. The primary inflection point is located near
 $H_{\mathrm{avg}}$  while a secondary point is observed deep within the canopy at pedestrian level 
 ($z \approx 0.08\text{--}0.10\,H_{\mathrm{avg}}$). The coexistence of these two inflection points indicates that momentum exchange in realistic urban environments such as the one examined here is governed by multiple, vertically separated dynamical mechanisms.

The upper inflection point near $H_{\mathrm{avg}}$, is typically interpreted as the effective canopy height, following the framework of \cite{Oke2017}. This feature is associated with a mixing-layer-like regime where strong velocity gradients drive shear-driven turbulence and vertical momentum exchange \citep{Raupach1996}. However, the emergence and location of this point depend heavily on morphology. While idealised, uniform configurations produce sharp inflection points due to localised shear \citep{xie2006}, realistic geometries exhibit more variability. \cite{Giometto2016} reported inflection points located near $H_{\mathrm{avg}}$, while \cite{Cheng2023a} observed inflection points spanning the range between $H_{\mathrm{avg}}$ and $H_{\mathrm{max}}$.  In contrast,  \cite{akinlabi2022} reported the absence of a clear inflection point  in cases with extreme height variability and intense wake interactions.

The present configuration, with $H_\mathrm{avg}/H_{max}=0.42$, allows for significant downward momentum transport and wake-induced mixing. Evidence of this down-washing effect is visible in the vertical velocity streamlines (Fig. \ref{fig:ave_ver}), where taller structures deflect high-momentum fluid into the canyons, effectively diluting the roof-level shear layer. Furthermore, the TKE distribution confirms that turbulence production is not restricted to a uniform interface at $H_\mathrm{avg}$
 but is fragmented across a range of building heights.  This is in contrast to the conclusions of  \cite{Tian2024} that suggested that inflection points may be absent in realistic urban flows. The present results demonstrate that realistic canopies can indeed exhibit a discernible upper inflection point, provided that roof-level shear generation remains sufficiently organised to overcome wake-induced mixing, its existence and position being an outcome of the specific balance between building height heterogeneity and the resulting vertical momentum transport.

Beyond this canopy-scale feature, a second inflection point is observed at much lower heights, around $z \approx 0.08\text{--}0.10\,H_{\mathrm{avg}}$. 
This low-level inflection point reflects a pronounced change in the vertical shear distribution and marks a transition between two dynamically distinct intra-canopy flow regimes. Below this elevation, the mean flow is dominated by near-ground processes such as persistent recirculation within street canyons and sheltered zones where low velocities limit horizontal momentum transport. Consequently, the double-averaged mean velocity exhibits only a weak increase with height in this near-wall region. Above this lower inflection point, the mean velocity increases more steadily as the influence of near-ground sheltering weakens and more connected intra-canopy flow pathways become dominant. This point represents an effective dynamical height associated with internal canopy momentum exchange, and its emergence is a direct consequence of spatial averaging blending heterogeneous flow structures with different dynamical characteristics at the same elevation.

Unlike the inflection points identified in the mean velocity profiles, the peak in turbulent intensity occurs in between $H_{\text{avg}}$ and $H_{\text{max}}$ in all wind directions. This systematic shift in the location of peak turbulence intensity highlights the different physical mechanisms governing mean flow and turbulence generation. While the inflection point marks the top of the shear-driven mixing layer within the canopy, the $\langle \overline {u'u'}\rangle/u^2_*$ peak reflects regions of intense turbulent production and wake interaction.  The fact that the maximum  consistently occurs above the canopy layer suggests that turbulent stresses are not solely generated within the canopy, but are strongly influenced by advection and upward diffusion of wake structures originating on the roof of taller buildings. These elements introduce large-scale flow disturbances that penetrate above canopy layer. 
Moreover, the magnitude of the peak is significant in all cases, indicating robust and organised turbulent activity above the urban roughness elements. Compared to previous studies, these peaks occur at lower heights (below $H_{max}$), which can be attributed to the relatively homogeneous building height in the studied area and the limited presence of very tall structures. As a result, turbulence fluctuations appear to be governed by a combination of the production at $H_{avg}$ and the interaction with the few taller buildings.

\begin{figure}[]
  \centering
    \includegraphics[width=\textwidth]{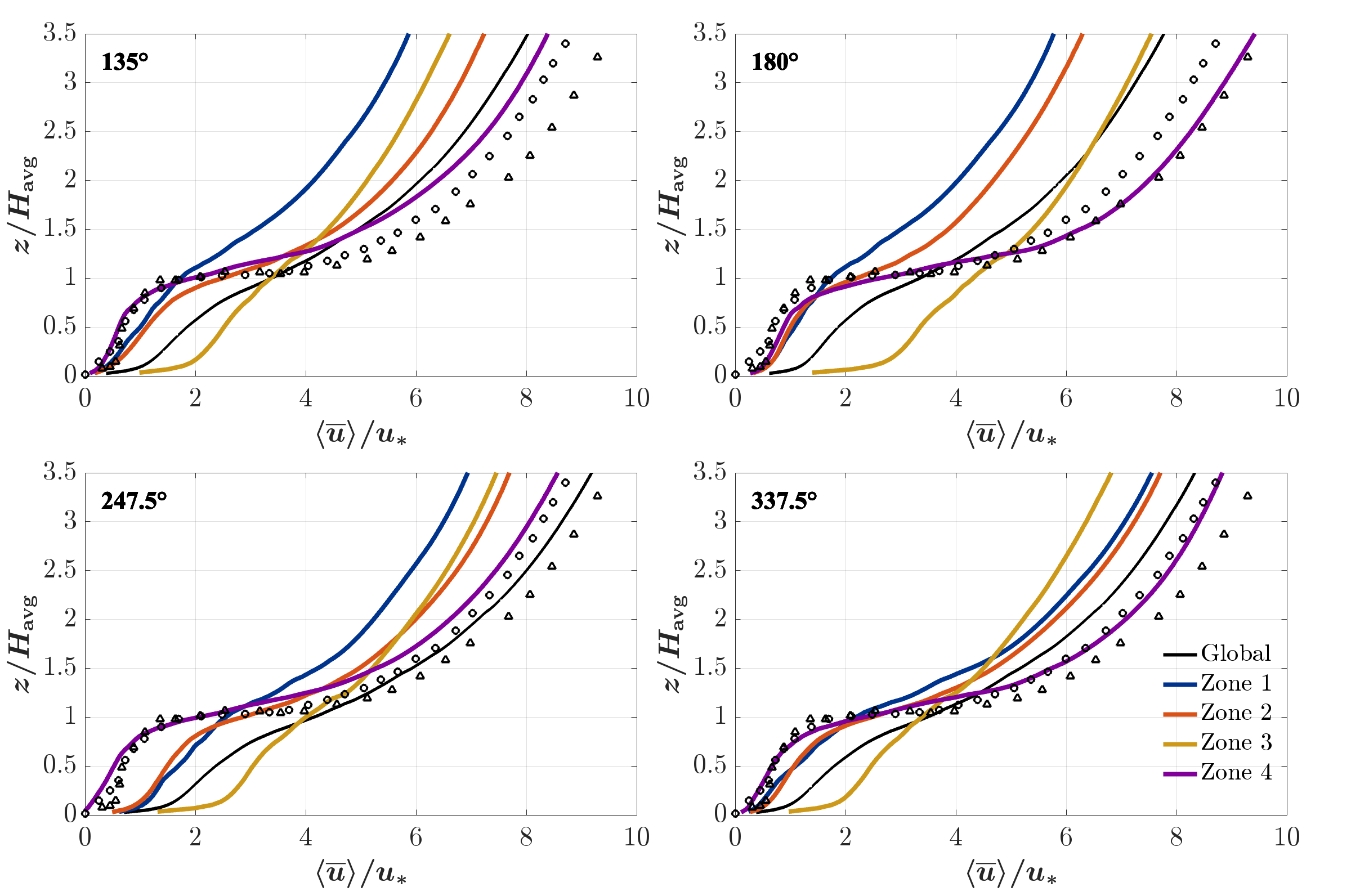}
 \caption{Variation of the double-average streamwise velocity profiles, $\langle \overline{u} \rangle/u_*$, across the different zones for the different wind directions. The black solid line corresponds to the neighbourhood profile, open circles \cite{coceal2006}, open triangles \cite{Castro2017} }\label{profile_zones}
\end{figure}

Although the neighbourhood-scale profiles show only a weak sensitivity to wind direction, this apparent robustness masks significant local variability. To explore this effect, \ming{ double-average quantities are computed for the four representative zones in Table \ref{tab:morpho_params}. The results are plotted in Fig.~\ref{profile_zones}.}  

Figure \ref{profile_zones} reveals that the neighbourhood-scale average masks significant spatial heterogeneity, which becomes apparent only when analysing the domain as distinct morphological sub-blocks. For instance, the canopy-scale transition near $H_{avg}$ is different  across zones and  larger deviations   occur within the canopy.  An example of this is zone 3, the most open configuration ($\lambda_p=0.10$); its profile exhibits only a weak or barely identifiable inflection around building height, suggesting that the transition between intra-canopy and above-canopy flow is more gradual and less dominated by a sharply localised roof-level shear layer. 

Conversely, denser zones exhibit a sharper separation between a sheltered near-ground region and a rapidly accelerating mid-canopy layer, which is consistent with stronger sheltering and wake interactions. Wind direction modulates these trends in a zone-dependent manner. Zone 4, the densest region ($\lambda_p=0.47$), shows limited sensitivity deep inside the canopy, indicating flow constrained by a wake-dominated regime. However, clear directional differences emerge just above the canopy, where the velocity deficit increases significantly when the wind aligns with the street network. This behaviour suggests changes in transport organisation at the canopy top rather than simple variations in drag magnitude. When inflow aligns with dominant corridors, momentum is preferentially channeled through connected pathways while adjacent areas remain sheltered. In a double-averaged sense, this corridor-dominated regime reduces lateral exchange and weakens vertical entrainment, leading to a more pronounced mean velocity deficit above the canopy despite local ventilation along aligned streets. The comparison with \cite{Castro2017} (open triangles) and with the \cite{coceal2006} data (open circles) on idealised cube-array reinforces this interpretation. 
 In particular, for the \revtwo{180$^\circ$}  inflow case the Zone 4 profile aligns closely with \cite{Castro2017} results at 45$^\circ$ incidence, due to local geometric alignment that causes the zone to behave like an ordered array under oblique inflow. This supports the observation that morphological regularity produces more canonical canopy-layer behaviour, whereas less ordered zones depart from idealised references through stronger local sheltering and a more diffuse intra-canopy transition.

\begin{figure}[]
  \centering
  \begin{subfigure}[b]{0.48\textwidth}
    \includegraphics[width=\textwidth]{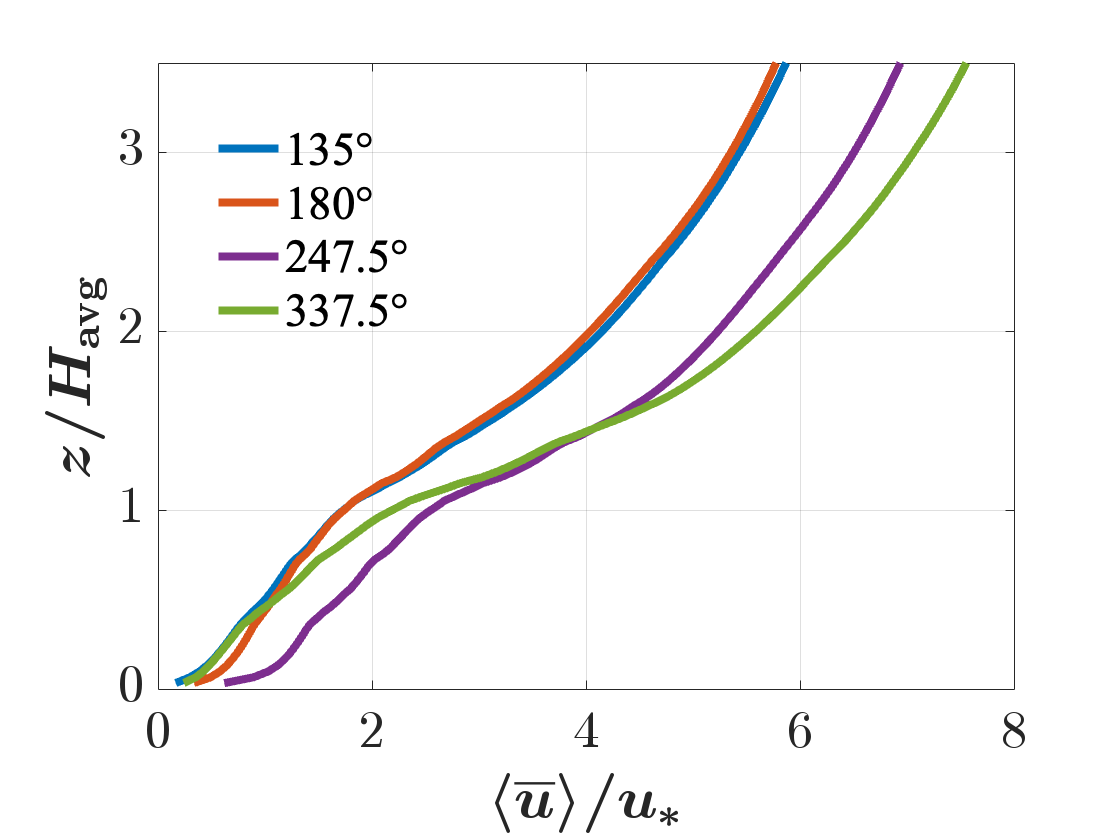}
    \caption{}
    \label{fig:subfig11_a}
  \end{subfigure}
  \hfill
  \begin{subfigure}[b]{0.48\textwidth}
    \includegraphics[width=\textwidth]{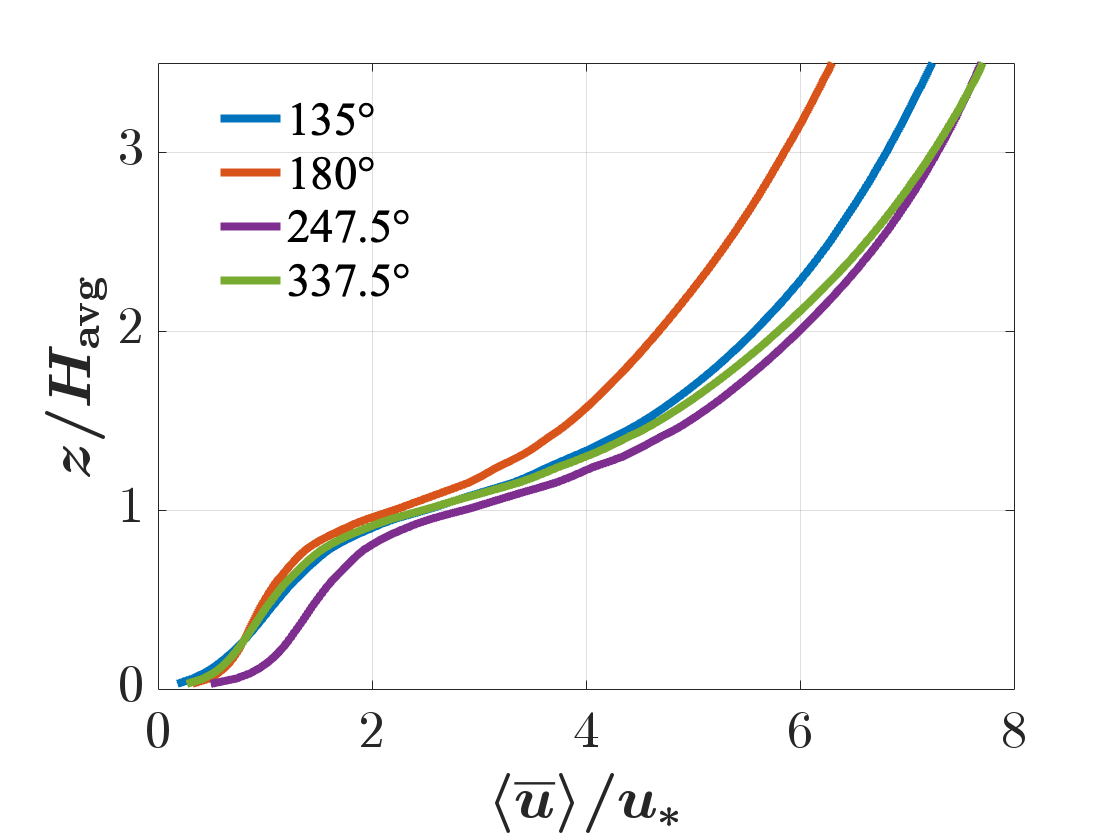}
    \caption{}
    \label{fig:subfig11_b}
  \end{subfigure}
\caption{Double-average streamwise velocity profiles, $\langle \overline{u} \rangle/u_*$, for different wind directions (a) Zone 1 ; (b) Zone 2. }
  \label{fig:zone1_winddir}
\end{figure}

 This mechanism is further highlighted by Zone 1 \ming{(Fig. \ref{fig:subfig11_a})}, where changes in inflow alignment lead to a  reorganisation of intra-canopy momentum penetration, as preferential ventilation pathways are activated or suppressed. Actually, in this zone, the shape of the profile at pedestrian and mid-canopy heights varies noticeably with wind direction (Fig. \ref{fig:zone1_winddir}), despite nearly identical $\lambda_p$ and only moderate variations of $\lambda_f$ across directions. 
 This indicates that for a fixed morphological density, the dominant driver of momentum exchange is not solely roughness magnitude but the existence of corridors that become active when the inflow aligns with the local urban texture. Consequently, the lower inflection point can be interpreted as a morphology-dependent and direction-dependent marker of the transition between sheltered recirculating flow and corridor-driven transport, whereas the upper inflection point remains primarily associated with canopy-scale shear and the effective canopy height.

\section{Conclusions}

High-resolution large-eddy simulations \ming{(LES)} were conducted over the Zona Universitària-Pedralbes district in Barcelona to investigate the impact of wind directions on urban flow and turbulence statistics. The \ming{domain} employed unstructured hexahedral meshes with fourth-order polynomial discretisation, achieving pedestrian-level resolutions on the order of 1 m and total grid sizes exceeding \ming{$5.0 \times 10^8$} degrees of freedom. \revone{The present work provides a systematic assessment of wind-direction effects over a realistic full-scale urban morphology using sixteen inflow directions spanning the full $360^\circ$ range. This enables the influence of wind direction to be examined consistently across both pedestrian-level flow organisation and double-averaged vertical statistics.}

The \ming{results} show that pedestrian-level flow exhibits a marked and systematic sensitivity to the incoming wind direction. Rotations in the approaching wind produce substantial reorganisations of the dominant flow pathways, wake structures, and turbulence distribution within the first few meters above ground. When the inflow aligns with the primary street axes, long and continuous acceleration bands develop, channeling momentum efficiently through the urban fabric and creating well-defined corridors of high velocity. Conversely, oblique inflows dismantle these coherent structures, replacing them with fragmented, cellular patterns where multiple branches emerge and recirculation zones become more distributed. Central to this behaviour is the activation of channeling corridors, i.e., preferential pathways that become operative only when the wind direction is favourably oriented with respect to the local street network.

Contrary to the strong directionality observed in the planar flow fields, the double-averaged vertical profiles exhibit a low-directional sensitivity across all simulated wind directions, remaining mostly invariant to the orientation of the incoming flow. This apparent robustness suggests that, despite the pronounced local anisotropy induced by individual buildings and streets, the effective aerodynamic response of the district as a whole behaves in a quasi-isotropic manner. The double-averaged profiles consistently reveal two distinct inflection points: one located slightly below the average building height, marking the top of a shear-driven mixing layer analogous to flows over vegetation canopies, and a second, deeper inflection point near pedestrian level that delineates the transition between near-ground recirculating flow and more connected intra-canopy transport. Turbulence intensity maxima, however, are systematically found between the average and maximum building heights, implicating the tallest structures as dominant sources of turbulent kinetic energy production through wake shedding and roof-edge shear.

Yet, this global robustness masks important local variability, as clear directional signatures are observed when the flow is analysed in different sub-zones with distinct morphological parameters. In denser, more regularly planned subdomains, the velocity profiles align closely with reference data from idealised cube arrays. In contrast, more open or irregular zones display weaker inflection signatures and reduced directional sensitivity. Critically, the lower inflection point, when present, is found to be both morphology-dependent and direction-dependent, acting as a dynamical marker for the transition between sheltered, recirculating flow and corridor-driven advection. This directional activation of preferential pathways, already evident at the pedestrian level, thus leaves a measurable imprint on the vertical structure of the mean flow when examined at the appropriate spatial scale.

Overall, the results suggest that at the city or neighbourhood scale, the use of aerodynamic parameterisations independently of the wind direction is feasible. \revone{The study also demonstrates that wind-direction effects in a realistic urban canopy are scale-dependent: they strongly reorganise pedestrian-level flow and local vertical structure, while producing only weak changes in double-averaged mean and turbulent statistics. This distinction is important for linking high-resolution urban LES to reduced-order urban parameterisations and practical wind-environment assessments.} However, for local studies involving pedestrian comfort, ventilation, or contaminant dispersion, the interaction between local morphology and wind direction remains critically relevant.

\backmatter

\bmhead{Acknowledgements}

This work has been partially financially supported by 'Agència de Gestió d'Ajuts Universitaris i de Recerca' under the call CLIMA 2023 (ref. 2023 CLIMA 00097). The authors acknowledge the support
of the Departament de Recerca i Universitats de la Generalitat de Catalunya through the research
group Large-scale Computational Fluid Dynamics (ref.: 2021 SGR 00902) and the Turbulence and
Aerodynamics Research Group (ref.: 2021 SGR 01051). 
The authors also acknowledge Red Española de Supercomputacion the resources provided in Marenostrum V Supercomputer (IM-2024-2-0006, IM-2024-3-0006).

\section*{Declarations}

\indent \textbf{Conflict of interest:} The authors have no conflicts to disclose.

\bigskip
\textbf{Author Contributions}
Josep M. Dur\' o: Data collection and curation (equal);  Formal analysis (equal); Visualization (equal); Writing – original draft (equal). Writing – review\& editing (equal). Ernest Mestres: Data collection and  curation (equal);  Formal analysis (equal);  Visualization (equal);   Writing – review \& editing (equal). Ming Teng: Formal analysis (equal); Visualization (equal); Writing – review \& editing (equal). Oriol Lehmkuhl:  Software (equal); Funding acquisition (equal); Investigation (equal); Methodology (equal); Project administration (equal); Resources (equal);  Writing – review \& editing (equal). 
  Ivette Rodriguez: Conceptualization (equal);  Formal analysis (equal); Methodology (equal);  Investigation (equal); Visualization (equal);  Funding acquisition (equal);  Project administration (equal); Resources (equal); Supervision (equal);Writing - original draft (equal); Writing – review\& editing (equal)
  
\bigskip
\textbf{Funding}
  Partial financial support was received from 'Agència de Gestió d'Ajuts Universitaris i de Recerca' under the call CLIMA 2023 (ref. 2023 CLIMA 00097). 
 Research groups Large-scale Computational Fluid Dynamics and Turbulence and Aerodynamics Research Group received financial support by the SGR grants of the Departament de Recerca i Universitats of the Generalitat de Catalunya (2021 SGR 00902 and 2021 SGR 01051, respectively).

  \bigskip
\textbf{Data availability} The data that support the findings of this study are available from
the corresponding author upon reasonable request.

\bigskip

\begin{appendices}



\section{Grid sensitivity analysis}
\label{app:mesh}
\begin{figure}[]
  \centering
    \includegraphics[width=0.7\textwidth]{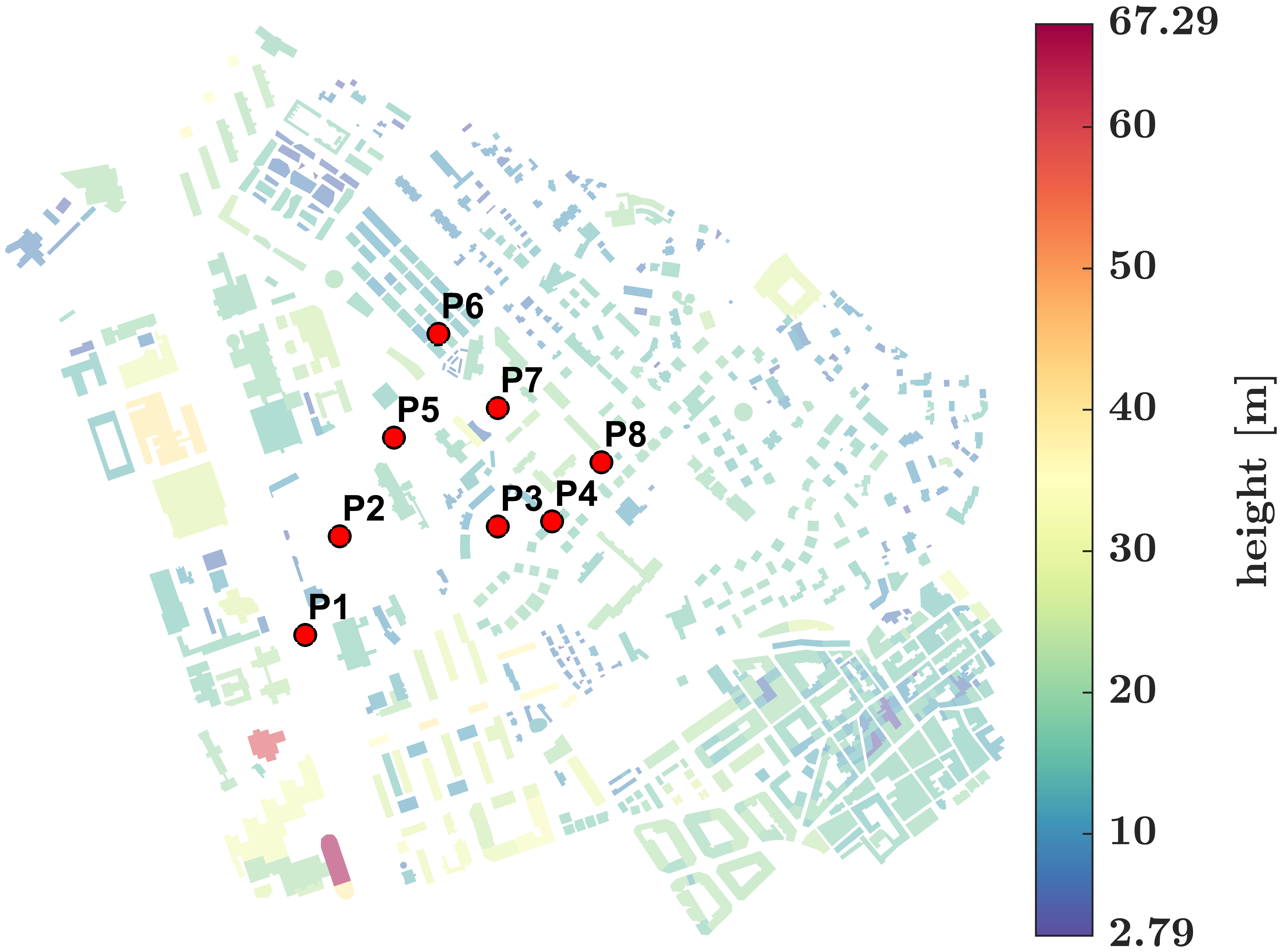}
  \caption{Location of the vertical probes used in the mesh assessment. }
  \label{sondas}
\end{figure}

\begin{figure}[]
  \centering
    \includegraphics[width=0.7\textwidth]{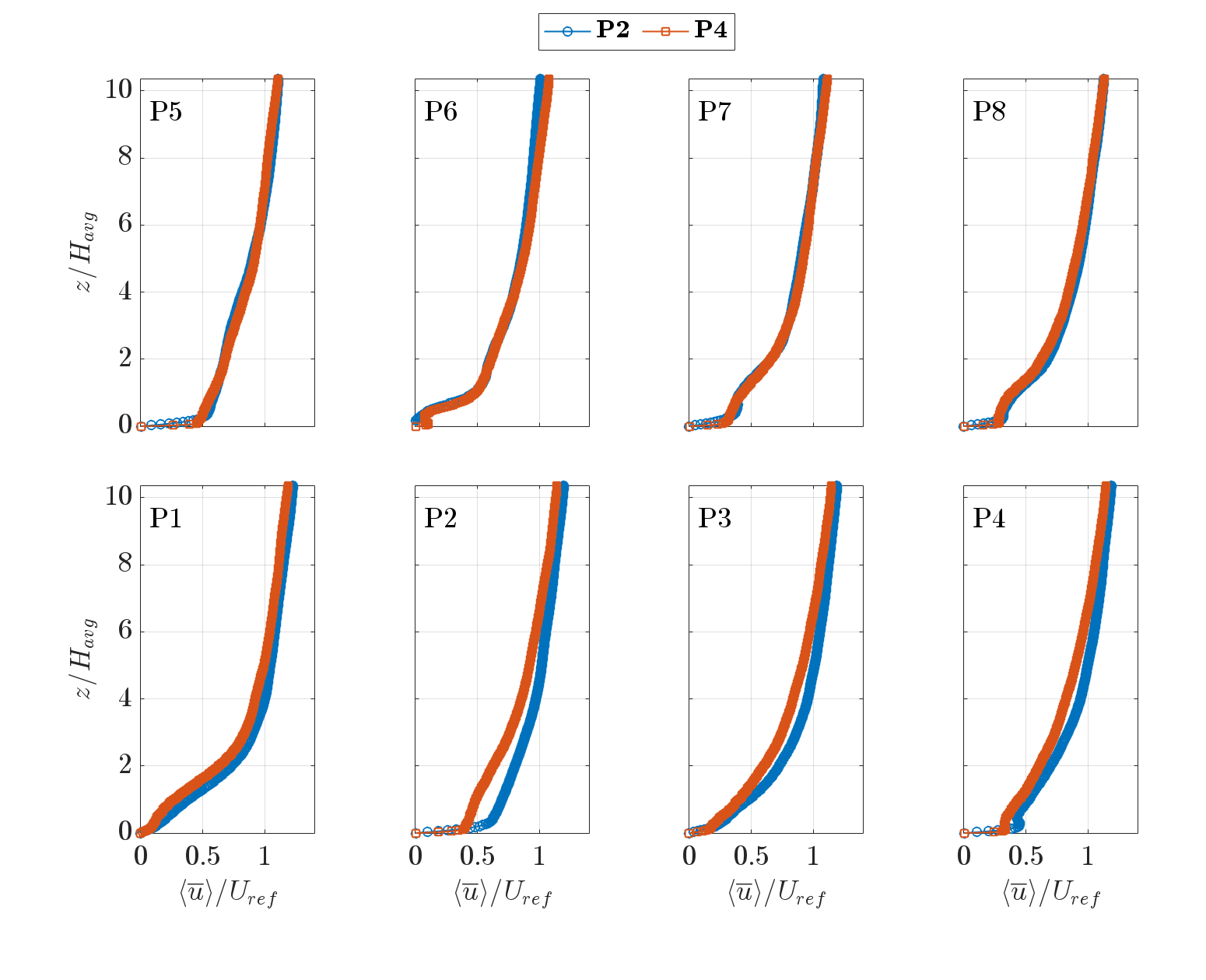}
 \caption{Mesh assessment. \ming{Double-average streamwise velocity profiles, $\langle \overline{u} \rangle/U_{ref}$,} at eight stations for wind direction $\revtwo{\Phi=180^\circ}$. Comparison between the \ming{second order} mesh and the \ming{fourth order} mesh.}
  \label{velocity_assessment}
\end{figure}

  \begin{figure}[]
  \centering 
    \includegraphics[width=0.7\textwidth]{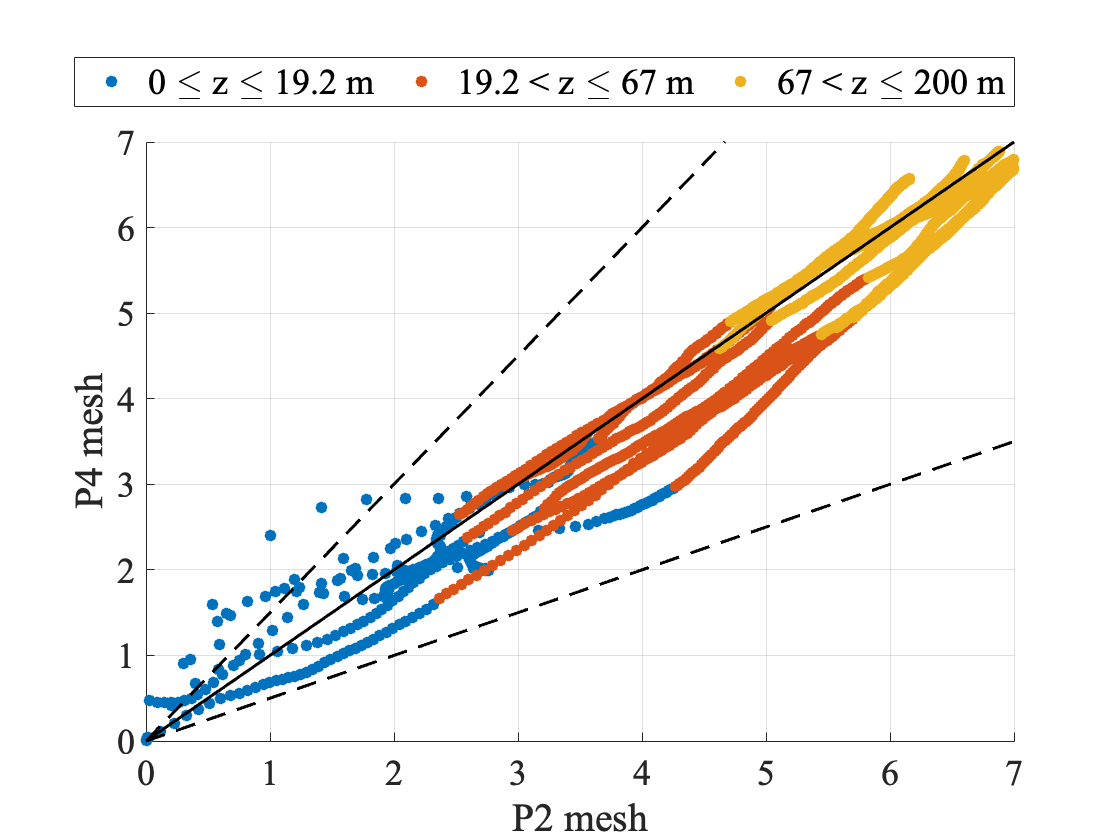}
  \caption{Scatter plot for both mesh resolutions comparing the \ming{streamwise velocity} at eight stations. Points are coloured according their height: $0\leq z\leq H_{avg}$ (urban canopy layer), $H_{avg}\leq z\leq H_{max}$ (roughness sublayer), $H_{max}\leq z \leq 5H_{max}$ (outer layer). The solid line denotes the one to one relation, while the dashed lines indicate deviations of $\pm50\%$. \ming{$H_\mathrm{avg}$ and $H_\mathrm{max}$ denote the average and maximum building height, respectively.}}
  \label{velocity_scatter}
\end{figure}

To assess the impact of mesh resolution on the predicted urban wind field a grid sensitivity analysis was carried out for the reference wind direction of \revtwo{180°}. The two simulations considered use the same unstructured mesh of spectral elements and identical boundary conditions at inflow, outlet, top and building walls. The only difference is the polynomial order of the solution within each element. In the coarse configuration the flow variables are approximated with \ming{second-order} polynomials ($p=2$), which leads to approximately \ming{$5.45 \times 10^7$} grid points. In the fine configuration the \ming{fourth-}order mesh  ($p=4$) is used, which increases the number of grid points to about \ming{$5.0 \times 10^8$} while keeping the number of elements unchanged. 

Comparisons have been performed at eight different stations distributed across the district, \ming{covering a range of} building heights  and local morphologies (see Fig. \ref{sondas}). Figure \ref{velocity_assessment} presents the normalised vertical profiles these  locations. In all cases the overall shape of the profile and the transition from the roughness sublayer near roof level to the outer flow are very similar between the two resolutions. The $p=4$ solution tends to be slightly smoother in the near canopy region, but the differences 
 are small compared with the total velocity variation across the profile. Above the average building height a rather good agreement is observed. Differences are consistent with the fact that the high-order mesh resolves more accurately the steep velocity gradients and recirculation features induced by the buildings. 
 
The agreement between the two meshes is further quantified in Fig. \ref{velocity_scatter}, which shows a scatter plot of the  pairs of mean velocity sampled at all vertical levels and all probes. The points are coloured according to height, distinguishing three ranges: 
$0\leq z\leq H_\mathrm{avg}$, $H_\mathrm{avg}\leq z\leq H_\mathrm{max}$, and $H_\mathrm{max}\leq z \leq 5H_\mathrm{max}$ \ming{($H_\mathrm{avg}$ and $H_\mathrm{max}$ denoting the average and maximum building height, respectively)}. The solid line corresponds to the one to one relation, while the dashed lines indicate the deviations of $\pm 50\%$ with respect to the reference. Most points cluster tightly around the one to one line, especially for heights above the roughness sublayer. In the intermediate band the scatter increases but remains largely within the $\pm 50\%$ bounds. The largest dispersion is observed within the urban canopy layer where strong local velocity gradients and flow separation around buildings enhance the sensitivity to grid resolution. Even in this region, however, the majority of points remain within the $\pm50\%$ interval and no systematic bias of the coarse mesh is observed.

\end{appendices}

\providecommand{\noopsort}[1]{}\providecommand{\singleletter}[1]{#1}%


\begin{thebibliography}{48}
\ifx \bisbn   \undefined \def \bisbn  #1{ISBN #1}\fi
\ifx \binits  \undefined \def \binits#1{#1}\fi
\ifx \bauthor  \undefined \def \bauthor#1{#1}\fi
\ifx \batitle  \undefined \def \batitle#1{#1}\fi
\ifx \bjtitle  \undefined \def \bjtitle#1{#1}\fi
\ifx \bvolume  \undefined \def \bvolume#1{\textbf{#1}}\fi
\ifx \byear  \undefined \def \byear#1{#1}\fi
\ifx \bissue  \undefined \def \bissue#1{#1}\fi
\ifx \bfpage  \undefined \def \bfpage#1{#1}\fi
\ifx \blpage  \undefined \def \blpage #1{#1}\fi
\ifx \burl  \undefined \def \burl#1{\textsf{#1}}\fi
\ifx \doiurl  \undefined \def \doiurl#1{\url{https://doi.org/#1}}\fi
\ifx \betal  \undefined \def \betal{\textit{et al.}}\fi
\ifx \binstitute  \undefined \def \binstitute#1{#1}\fi
\ifx \binstitutionaled  \undefined \def \binstitutionaled#1{#1}\fi
\ifx \bctitle  \undefined \def \bctitle#1{#1}\fi
\ifx \beditor  \undefined \def \beditor#1{#1}\fi
\ifx \bpublisher  \undefined \def \bpublisher#1{#1}\fi
\ifx \bbtitle  \undefined \def \bbtitle#1{#1}\fi
\ifx \bedition  \undefined \def \bedition#1{#1}\fi
\ifx \bseriesno  \undefined \def \bseriesno#1{#1}\fi
\ifx \blocation  \undefined \def \blocation#1{#1}\fi
\ifx \bsertitle  \undefined \def \bsertitle#1{#1}\fi
\ifx \bsnm \undefined \def \bsnm#1{#1}\fi
\ifx \bsuffix \undefined \def \bsuffix#1{#1}\fi
\ifx \bparticle \undefined \def \bparticle#1{#1}\fi
\ifx \barticle \undefined \def \barticle#1{#1}\fi
\bibcommenthead
\ifx \bconfdate \undefined \def \bconfdate #1{#1}\fi
\ifx \botherref \undefined \def \botherref #1{#1}\fi
\ifx \url \undefined \def \url#1{\textsf{#1}}\fi
\ifx \bchapter \undefined \def \bchapter#1{#1}\fi
\ifx \bbook \undefined \def \bbook#1{#1}\fi
\ifx \bcomment \undefined \def \bcomment#1{#1}\fi
\ifx \oauthor \undefined \def \oauthor#1{#1}\fi
\ifx \citeauthoryear \undefined \def \citeauthoryear#1{#1}\fi
\ifx \endbibitem  \undefined \def \endbibitem {}\fi
\ifx \bconflocation  \undefined \def \bconflocation#1{#1}\fi
\ifx \arxivurl  \undefined \def \arxivurl#1{\textsf{#1}}\fi
\csname PreBibitemsHook\endcsname

\bibitem[\protect\citeauthoryear{Auvinen et~al.}{2020}]{Auvinen2020}
\begin{barticle}
\bauthor{\bsnm{Auvinen}, \binits{M.}},
\bauthor{\bsnm{Boi}, \binits{S.}},
\bauthor{\bsnm{Hellsten}, \binits{A.}},
\bauthor{\bsnm{Tanhuanp{\"{a}}{\"{a}}}, \binits{T.}},
\bauthor{\bsnm{J{\"{a}}rvi}, \binits{L.}}:
\batitle{{Study of realistic urban boundary layer turbulence with
  high-resolution large-eddy simulation}}.
\bjtitle{Atmosphere}
\bvolume{11}(\bissue{2}),
\bfpage{1}--\blpage{41}
(\byear{2020})
\doiurl{10.3390/atmos11020201}
\end{barticle}
\endbibitem

\bibitem[\protect\citeauthoryear{{Abd Razak} et~al.}{2013}]{AbdRazak2013}
\begin{barticle}
\bauthor{\bsnm{{Abd Razak}}, \binits{A.}},
\bauthor{\bsnm{Hagishima}, \binits{A.}},
\bauthor{\bsnm{Ikegaya}, \binits{N.}},
\bauthor{\bsnm{Tanimoto}, \binits{J.}}:
\batitle{{Analysis of airflow over building arrays for assessment of urban wind
  environment}}.
\bjtitle{Building and Environment}
\bvolume{59},
\bfpage{56}--\blpage{65}
(\byear{2013})
\doiurl{10.1016/j.buildenv.2012.08.007}
\end{barticle}
\endbibitem

\bibitem[\protect\citeauthoryear{Akinlabi et~al.}{2022}]{akinlabi2022}
\begin{barticle}
\bauthor{\bsnm{Akinlabi}, \binits{E.}},
\bauthor{\bsnm{Maronga}, \binits{B.}},
\bauthor{\bsnm{Giometto}, \binits{M.G.}},
\bauthor{\bsnm{Li}, \binits{D.}}:
\batitle{{D}ispersive fluxes within and over a real urban canopy: a large-eddy
  simulation study}.
\bjtitle{Boundary-Layer Meteorology}
\bvolume{185}(\bissue{1}),
\bfpage{93}--\blpage{128}
(\byear{2022})
\end{barticle}
\endbibitem

\bibitem[\protect\citeauthoryear{Antoniou et~al.}{2019}]{Antoniou2019}
\begin{barticle}
\bauthor{\bsnm{Antoniou}, \binits{N.}},
\bauthor{\bsnm{Montazeri}, \binits{H.}},
\bauthor{\bsnm{Neophytou}, \binits{M.}},
\bauthor{\bsnm{Blocken}, \binits{B.}}:
\batitle{{CFD simulation of urban microclimate: Validation using
  high-resolution field measurements}}.
\bjtitle{Science of the Total Environment}
\bvolume{695},
\bfpage{133743}
(\byear{2019})
\doiurl{10.1016/j.scitotenv.2019.133743}
\end{barticle}
\endbibitem

\bibitem[\protect\citeauthoryear{Barlow and Coceal}{2009}]{Barlow2009}
\begin{botherref}
\oauthor{\bsnm{Barlow}, \binits{J.F.}},
\oauthor{\bsnm{Coceal}, \binits{O.}}:
A review of urban roughness sublayer turbulence.
Meteorology Research and Development Technical Report
\textbf{527}
(2009)
\end{botherref}
\endbibitem

\bibitem[\protect\citeauthoryear{Blunn et~al.}{2022}]{Blunn2022}
\begin{barticle}
\bauthor{\bsnm{Blunn}, \binits{L.P.}},
\bauthor{\bsnm{Coceal}, \binits{O.}},
\bauthor{\bsnm{Nazarian}, \binits{N.}},
\bauthor{\bsnm{Barlow}, \binits{J.F.}},
\bauthor{\bsnm{Plant}, \binits{R.S.}},
\bauthor{\bsnm{Bohnenstengel}, \binits{S.I.}},
\bauthor{\bsnm{Lean}, \binits{H.W.}}:
\batitle{{Turbulence characteristics across a range of idealized urban canopy
  geometries}}.
\bjtitle{Boundary-Layer Meteorology}
\bvolume{182}(\bissue{2}),
\bfpage{275}--\blpage{307}
(\byear{2022})
\end{barticle}
\endbibitem

\bibitem[\protect\citeauthoryear{Blocken}{2015}]{Blocken2015}
\begin{barticle}
\bauthor{\bsnm{Blocken}, \binits{B.}}:
\batitle{Computational {Fluid} {Dynamics} for urban physics: {Importance},
  scales, possibilities, limitations and ten tips and tricks towards accurate
  and reliable simulations}.
\bjtitle{Building and Environment}
\bvolume{91},
\bfpage{219}--\blpage{245}
(\byear{2015})
\doiurl{10.1016/j.buildenv.2015.02.015}
\end{barticle}
\endbibitem

\bibitem[\protect\citeauthoryear{Brozovsky et~al.}{2021}]{Brozovsky2021}
\begin{botherref}
\oauthor{\bsnm{Brozovsky}, \binits{J.}},
\oauthor{\bsnm{Simonsen}, \binits{A.}},
\oauthor{\bsnm{Gaitani}, \binits{N.}}:
{Validation of a CFD model for the evaluation of urban microclimate at high
  latitudes: A case study in Trondheim, Norway}.
Building and Environment
\textbf{205}(May)
(2021)
\doiurl{10.1016/j.buildenv.2021.108175}
\end{botherref}
\endbibitem

\bibitem[\protect\citeauthoryear{Claus et~al.}{2012}]{Claus2012}
\begin{barticle}
\bauthor{\bsnm{Claus}, \binits{J.}},
\bauthor{\bsnm{Coceal}, \binits{O.}},
\bauthor{\bsnm{Thomas}, \binits{T.G.}},
\bauthor{\bsnm{Brandford}, \binits{S.}},
\bauthor{\bsnm{Belcher}, \binits{S.E.}},
\bauthor{\bsnm{Castro}, \binits{I.P.}}:
\batitle{Wind-direction effects on urban-type flows}.
\bjtitle{Boundary-Layer Meteorology}
\bvolume{142}(\bissue{2}),
\bfpage{265}--\blpage{287}
(\byear{2012})
\doiurl{10.1007/s10546-011-9667-4}
\end{barticle}
\endbibitem

\bibitem[\protect\citeauthoryear{Coceal et~al.}{2006}]{coceal2006}
\begin{barticle}
\bauthor{\bsnm{Coceal}, \binits{O.}},
\bauthor{\bsnm{Thomas}, \binits{T.G.}},
\bauthor{\bsnm{Castro}, \binits{I.P.}},
\bauthor{\bsnm{Belcher}, \binits{S.E.}}:
\batitle{{Mean flow and turbulence statistics over groups of urban-like cubical
  obstacles}}.
\bjtitle{Boundary-Layer Meteorology}
\bvolume{121}(\bissue{3}),
\bfpage{491}--\blpage{519}
(\byear{2006})
\end{barticle}
\endbibitem

\bibitem[\protect\citeauthoryear{Castro et~al.}{2017}]{Castro2017}
\begin{barticle}
\bauthor{\bsnm{Castro}, \binits{I.P.}},
\bauthor{\bsnm{Xie}, \binits{Z.T.}},
\bauthor{\bsnm{Fuka}, \binits{V.}},
\bauthor{\bsnm{Robins}, \binits{A.G.}},
\bauthor{\bsnm{Carpentieri}, \binits{M.}},
\bauthor{\bsnm{Hayden}, \binits{P.}},
\bauthor{\bsnm{Hertwig}, \binits{D.}},
\bauthor{\bsnm{Coceal}, \binits{O.}}:
\batitle{{Measurements and Computations of Flow in an Urban Street System}}.
\bjtitle{Boundary-Layer Meteorology}
\bvolume{162}(\bissue{2}),
\bfpage{207}--\blpage{230}
(\byear{2017})
\doiurl{10.1007/s10546-016-0200-7}
\end{barticle}
\endbibitem

\bibitem[\protect\citeauthoryear{Cheng and Yang}{2023a}]{Cheng2023a}
\begin{barticle}
\bauthor{\bsnm{Cheng}, \binits{W.C.}},
\bauthor{\bsnm{Yang}, \binits{Y.}}:
\batitle{{Scaling of Flows Over Realistic Urban Geometries: A Large-Eddy
  Simulation Study}}.
\bjtitle{Boundary-Layer Meteorology}
\bvolume{186}(\bissue{1}),
\bfpage{125}--\blpage{144}
(\byear{2023})
\doiurl{10.1007/s10546-022-00749-y}
\end{barticle}
\endbibitem


\bibitem[\protect\citeauthoryear{Duan and Takemi}{2021}]{Duan2021}
\begin{barticle}
\bauthor{\bsnm{Duan}, \binits{G.}},
\bauthor{\bsnm{Takemi}, \binits{T.}}:
\batitle{{Predicting urban surface roughness aerodynamic parameters using
  random forest}}.
\bjtitle{Journal of Applied Meteorology and Climatology}
\bvolume{60}(\bissue{7}),
\bfpage{999}--\blpage{1018}
(\byear{2021})
\doiurl{10.1175/JAMC-D-20-0266.1}
\end{barticle}
\endbibitem

\bibitem[\protect\citeauthoryear{Duan et~al.}{2023}]{Duan2023}
\begin{botherref}
\oauthor{\bsnm{Duan}, \binits{G.}},
\oauthor{\bsnm{Nakamae}, \binits{K.}},
\oauthor{\bsnm{Takemi}, \binits{T.}}:
{Impacts of urban morphometric indices on ventilation}.
Building and Environment
\textbf{229}(December 2022)
(2023)
\doiurl{10.1016/j.buildenv.2022.109907}
\end{botherref}
\endbibitem

\bibitem[\protect\citeauthoryear{Franke et~al.}{2007}]{Franke2007}
\begin{botherref}
\oauthor{\bsnm{Franke}, \binits{J.}},
\oauthor{\bsnm{Hellsten}, \binits{A.}},
\oauthor{\bsnm{Schlunzen}, \binits{K.H.}},
\oauthor{\bsnm{Carissimo}, \binits{B.}}:
{ The COST 732. Best Practice Guideline for CFD simulation of flows in the
  urban environment: a summary}.
Technical report
(2007).
\doiurl{10.1504/IJEP.2011.038443}
\end{botherref}
\endbibitem

\bibitem[\protect\citeauthoryear{Giometto et~al.}{2016}]{Giometto2016}
\begin{barticle}
\bauthor{\bsnm{Giometto}, \binits{M.G.}},
\bauthor{\bsnm{Christen}, \binits{A.}},
\bauthor{\bsnm{Meneveau}, \binits{C.}},
\bauthor{\bsnm{Fang}, \binits{J.}},
\bauthor{\bsnm{Krafczyk}, \binits{M.}},
\bauthor{\bsnm{Parlange}, \binits{M.B.}}:
\batitle{Spatial characteristics of roughness sublayer mean flow and turbulence
  over a realistic urban surface}.
\bjtitle{Boundary-Layer Meteorology}
\bvolume{160}(\bissue{3}),
\bfpage{425}--\blpage{452}
(\byear{2016})
\doiurl{10.1007/s10546-016-0157-6}
\end{barticle}
\endbibitem

\bibitem[\protect\citeauthoryear{Gronemeier et~al.}{2021}]{Gronemeier2021}
\begin{barticle}
\bauthor{\bsnm{Gronemeier}, \binits{T.}},
\bauthor{\bsnm{Surm}, \binits{K.}},
\bauthor{\bsnm{Harms}, \binits{F.}},
\bauthor{\bsnm{Leitl}, \binits{B.}},
\bauthor{\bsnm{Maronga}, \binits{B.}},
\bauthor{\bsnm{Raasch}, \binits{S.}}:
\batitle{{Evaluation of the dynamic core of the PALM model system 6.0 in a
  neutrally stratified urban environment: Comparison between les and
  wind-tunnel experiments}}.
\bjtitle{Geoscientific Model Development}
\bvolume{14}(\bissue{6}),
\bfpage{3317}--\blpage{3333}
(\byear{2021})
\doiurl{10.5194/gmd-14-3317-2021}
\end{barticle}
\endbibitem

\bibitem[\protect\citeauthoryear{Gasparino et~al.}{2024}]{GASPARINO2024109067}
\begin{barticle}
\bauthor{\bsnm{Gasparino}, \binits{L.}},
\bauthor{\bsnm{Spiga}, \binits{F.}},
\bauthor{\bsnm{Lehmkuhl}, \binits{O.}}:
\batitle{SOD2D: A GPU-enabled spectral finite elements method for compressible
  scale-resolving simulations}.
\bjtitle{Computer Physics Communications}
\bvolume{297},
\bfpage{109067}
(\byear{2024})
\doiurl{10.1016/j.cpc.2023.109067}
\end{barticle}
\endbibitem

\bibitem[\protect\citeauthoryear{García-Sánchez
  et~al.}{2018}]{GarciaSanchez2018}
\begin{barticle}
\bauthor{\bsnm{García-Sánchez}, \binits{C.}},
\bauthor{\bsnm{Beeck}, \binits{J.}},
\bauthor{\bsnm{Gorlé}, \binits{C.}}:
\batitle{Predictive large eddy simulations for urban flows: Challenges and
  opportunities}.
\bjtitle{Building and Environment}
\bvolume{139},
\bfpage{146}--\blpage{156}
(\byear{2018})
\doiurl{10.1016/j.buildenv.2018.05.007}
\end{barticle}
\endbibitem

\bibitem[\protect\citeauthoryear{H{\aa}gbo and Giljarhus}{2024}]{Hagbo2024a}
\begin{barticle}
\bauthor{\bsnm{H{\aa}gbo}, \binits{T.O.}},
\bauthor{\bsnm{Giljarhus}, \binits{K.E.T.}}:
\batitle{{Sensitivity of urban morphology and the number of CFD simulated wind
  directions on pedestrian wind comfort and safety assessments}}.
\bjtitle{Building and Environment}
\bvolume{253}(\bissue{February}),
\bfpage{111310}
(\byear{2024})
\doiurl{10.1016/j.buildenv.2024.111310}
\end{barticle}
\endbibitem

\bibitem[\protect\citeauthoryear{Jimenez}{2004}]{Jimenez2004b}
\begin{barticle}
\bauthor{\bsnm{Jimenez}, \binits{J.}}:
\batitle{{Turbulent Flows Over Rough Walls}}.
\bjtitle{Annual Review of Fluid Mechanics}
\bvolume{36}(\bissue{1}),
\bfpage{173}--\blpage{196}
(\byear{2004})
\doiurl{10.1146/annurev.fluid.36.050802.122103}
\end{barticle}
\endbibitem

\bibitem[\protect\citeauthoryear{Kim and Baik}{2004}]{Kim2004}
\begin{barticle}
\bauthor{\bsnm{Kim}, \binits{J.J.}},
\bauthor{\bsnm{Baik}, \binits{J.J.}}:
\batitle{{A numerical study of the effects of ambient wind direction on flow
  and dispersion in urban street canyons using the RNG k-$\epsilon$ turbulence
  model}}.
\bjtitle{Atmospheric Environment}
\bvolume{38}(\bissue{19}),
\bfpage{3039}--\blpage{3048}
(\byear{2004})
\doiurl{10.1016/j.atmosenv.2004.02.047}
\end{barticle}
\endbibitem

\bibitem[\protect\citeauthoryear{Kennedy and Gruber}{2008}]{kennedy2008reduced}
\begin{barticle}
\bauthor{\bsnm{Kennedy}, \binits{C.A.}},
\bauthor{\bsnm{Gruber}, \binits{A.}}:
\batitle{{Reduced aliasing formulations of the convective terms within the
  Navier-Stokes equations for a compressible fluid}}.
\bjtitle{Journal of Computational Physics}
\bvolume{227}(\bissue{3}),
\bfpage{1676}--\blpage{1700}
(\byear{2008})
\end{barticle}
\endbibitem

\bibitem[\protect\citeauthoryear{Kanda et~al.}{2013}]{Kanda2013}
\begin{barticle}
\bauthor{\bsnm{Kanda}, \binits{M.}},
\bauthor{\bsnm{Inagaki}, \binits{A.}},
\bauthor{\bsnm{Miyamoto}, \binits{T.}},
\bauthor{\bsnm{Gryschka}, \binits{M.}},
\bauthor{\bsnm{Raasch}, \binits{S.}}:
\batitle{{A New Aerodynamic Parametrization for Real Urban Surfaces}}.
\bjtitle{Boundary-Layer Meteorology}
\bvolume{148}(\bissue{2}),
\bfpage{357}--\blpage{377}
(\byear{2013})
\doiurl{10.1007/s10546-013-9818-x}
\end{barticle}
\endbibitem

\bibitem[\protect\citeauthoryear{Karniadakis et~al.}{1991}]{Karniadakis1991}
\begin{barticle}
\bauthor{\bsnm{Karniadakis}, \binits{G.E.}},
\bauthor{\bsnm{Israeli}, \binits{M.}},
\bauthor{\bsnm{Orszag}, \binits{S.A.}}:
\batitle{High-order splitting methods for the incompressible navier-stokes
  equations}.
\bjtitle{Journal of Computational Physics}
\bvolume{97}(\bissue{2}),
\bfpage{414}--\blpage{443}
(\byear{1991})
\doiurl{10.1016/0021-9991(91)90007-8}
\end{barticle}
\endbibitem

\bibitem[\protect\citeauthoryear{Owen et~al.}{2020}]{Owen2020WMLES}
\begin{barticle}
\bauthor{\bsnm{Owen}, \binits{H.}},
\bauthor{\bsnm{Chrysokentis}, \binits{G.}},
\bauthor{\bsnm{Avila}, \binits{M.}},
\bauthor{\bsnm{Mira}, \binits{D.}},
\bauthor{\bsnm{Houzeaux}, \binits{G.}},
\bauthor{\bsnm{Borrell}, \binits{R.}},
\bauthor{\bsnm{Cajas}, \binits{J.C.}},
\bauthor{\bsnm{Lehmkuhl}, \binits{O.}}:
\batitle{Wall-modeled large-eddy simulation in a finite element framework}.
\bjtitle{International Journal for Numerical Methods in Fluids}
\bvolume{92}(\bissue{1}),
\bfpage{20}--\blpage{37}
(\byear{2020})
\doiurl{10.1002/fld.4770}
\end{barticle}
\endbibitem


\bibitem[\protect\citeauthoryear{Lin et~al.}{2014}]{Lin2014}
\begin{barticle}
\bauthor{\bsnm{Lin}, \binits{M.}},
\bauthor{\bsnm{Hang}, \binits{J.}},
\bauthor{\bsnm{Li}, \binits{Y.}},
\bauthor{\bsnm{Luo}, \binits{Z.}},
\bauthor{\bsnm{Sandberg}, \binits{M.}}:
\batitle{Quantitative ventilation assessments of idealized urban canopy layers
  with various urban layouts and the same building packing density}.
\bjtitle{Building and Environment}
\bvolume{79},
\bfpage{152}--\blpage{167}
(\byear{2014})
\doiurl{10.1016/j.buildenv.2014.05.008}
\end{barticle}
\endbibitem

\bibitem[\protect\citeauthoryear{Lin et~al.}{2021}]{Lin2021}
\begin{botherref}
\oauthor{\bsnm{Lin}, \binits{Y.}},
\oauthor{\bsnm{Hang}, \binits{J.}},
\oauthor{\bsnm{Yang}, \binits{H.}},
\oauthor{\bsnm{Chen}, \binits{L.}},
\oauthor{\bsnm{Chen}, \binits{G.}},
\oauthor{\bsnm{Ling}, \binits{H.}},
\oauthor{\bsnm{Sandberg}, \binits{M.}},
\oauthor{\bsnm{Claesson}, \binits{L.}},
\oauthor{\bsnm{Lam}, \binits{C.K.C.}}:
{Investigation of the Reynolds number independence of cavity flow in 2D street
  canyons by wind tunnel experiments and numerical simulations}.
Building and Environment
\textbf{201}(January)
(2021)
\doiurl{10.1016/j.buildenv.2021.107965}
\end{botherref}
\endbibitem

\bibitem[\protect\citeauthoryear{Llaguno-Munitxa
  et~al.}{2017}]{LlagunoMunitxa2017}
\begin{barticle}
\bauthor{\bsnm{Llaguno-Munitxa}, \binits{M.}},
\bauthor{\bsnm{Bou-Zeid}, \binits{E.}},
\bauthor{\bsnm{Hultmark}, \binits{M.}}:
\batitle{The influence of building geometry on street canyon air flow:
  Validation of large eddy simulations against wind tunnel experiments}.
\bjtitle{Journal of Wind Engineering and Industrial Aerodynamics}
\bvolume{165},
\bfpage{115}--\blpage{130}
(\byear{2017})
\doiurl{10.1016/j.jweia.2017.03.007}
\end{barticle}
\endbibitem

\bibitem[\protect\citeauthoryear{Monnier et~al.}{2018}]{Monnier2018}
\begin{barticle}
\oauthor{\bsnm{Monnier}, \binits{B.}},
\oauthor{\bsnm{Goudarzi}, \binits{S.A.}},
\oauthor{\bsnm{Vinuesa}, \binits{R.}},
\oauthor{\bsnm{Wark}, \binits{C.}}:
\batitle{Turbulent structure of a simplified urban fluid flow studied through
  stereoscopic particle image velocimetry}.
\bjtitle{Boundary-Layer Meteorology}
\bvolume{166}(\bissue{2}),
\bfpage{239}--\blpage{268}
(\byear{2018})
\doiurl{10.1007/s10546-017-0303-9}
\end{barticle}
\endbibitem

\bibitem[\protect\citeauthoryear{Oke}{1988}]{oke1988}
\begin{barticle}
\bauthor{\bsnm{Oke}, \binits{T.R.}}:
\batitle{Street design and urban canopy layer climate}.
\bjtitle{Energy and Buildings}
\bvolume{11}(\bissue{1-3}),
\bfpage{103}--\blpage{113}
(\byear{1988})
\end{barticle}
\endbibitem

\bibitem[\protect\citeauthoryear{Oke et~al.}{2017}]{Oke2017}
\begin{bbook}
\bauthor{\bsnm{Oke}, \binits{T.R.}},
\bauthor{\bsnm{Mills}, \binits{G.}},
\bauthor{\bsnm{Christen}, \binits{A.}},
\bauthor{\bsnm{Voogt}, \binits{J.A.}}:
\bbtitle{Urban Climates}.
\bpublisher{Cambridge University Press}
(\byear{2017})
\end{bbook}
\endbibitem

\bibitem[\protect\citeauthoryear{Oh et~al.}{2024}]{OH2024105682}
\begin{barticle}
\bauthor{\bsnm{Oh}, \binits{G.}},
\bauthor{\bsnm{Yang}, \binits{M.}},
\bauthor{\bsnm{Choi}, \binits{J.}}:
\batitle{Large-eddy simulation-based wind and thermal comfort assessment in
  urban environments}.
\bjtitle{Journal of Wind Engineering and Industrial Aerodynamics}
\bvolume{246},
\bfpage{105682}
(\byear{2024})
\doiurl{10.1016/j.jweia.2024.105682}
\end{barticle}
\endbibitem

\bibitem[\protect\citeauthoryear{Raupach et~al.}{1991}]{Raupach1991a}
\begin{barticle}
\bauthor{\bsnm{Raupach}, \binits{M.R.}},
\bauthor{\bsnm{Antonia}, \binits{R.A.}},
\bauthor{\bsnm{Rajagopalan}, \binits{S.}}:
\batitle{{Rough-wall turbulent boundary layers}}.
\bjtitle{Applied Mechanics Review}
\bvolume{44}(\bissue{1}),
\bfpage{1}--\blpage{25}
(\byear{1991})
\doiurl{10.1115/1.3119492}
\end{barticle}
\endbibitem

\bibitem[\protect\citeauthoryear{Raupach et~al.}{1996}]{Raupach1996}
\begin{barticle}
\bauthor{\bsnm{Raupach}, \binits{M.R.}},
\bauthor{\bsnm{Finnigan}, \binits{J.J.}},
\bauthor{\bsnm{Brunei}, \binits{Y.}}:
\batitle{Coherent eddies and turbulence in vegetation canopies: The
  mixing-layer analogy}.
\bjtitle{Boundary-Layer Meteorology}
\bvolume{78}(\bissue{3}),
\bfpage{351}--\blpage{382}
(\byear{1996})
\doiurl{10.1007/BF00120941}
\end{barticle}
\endbibitem

\bibitem[\protect\citeauthoryear{Ricci et~al.}{2018}]{Ricci2018}
\begin{barticle}
\bauthor{\bsnm{Ricci}, \binits{A.}},
\bauthor{\bsnm{Kalkman}, \binits{I.}},
\bauthor{\bsnm{Blocken}, \binits{B.}},
\bauthor{\bsnm{Burlando}, \binits{M.}},
\bauthor{\bsnm{Freda}, \binits{A.}},
\bauthor{\bsnm{Repetto}, \binits{M.P.}}:
\batitle{Large-scale forcing effects on wind flows in the urban canopy: Impact
  of inflow conditions}.
\bjtitle{Sustainable Cities and Society}
\bvolume{42},
\bfpage{593}--\blpage{610}
(\byear{2018})
\doiurl{10.1016/j.scs.2018.08.012}
\end{barticle}
\endbibitem

\bibitem[\protect\citeauthoryear{Raupach and Shaw}{1982}]{Raupach1982}
\begin{barticle}
\bauthor{\bsnm{Raupach}, \binits{M.R.}},
\bauthor{\bsnm{Shaw}, \binits{R.H.}}:
\batitle{{Averaging procedures for flow within vegetation canopies}}.
\bjtitle{Boundary-Layer Meteorology}
\bvolume{22}(\bissue{1}),
\bfpage{79}--\blpage{90}
(\byear{1982})
\doiurl{10.1007/BF00128057}
\end{barticle}
\endbibitem

\bibitem[\protect\citeauthoryear{Santiago et~al.}{2013}]{Santiago2013}
\begin{barticle}
\bauthor{\bsnm{Santiago}, \binits{J.L.}},
\bauthor{\bsnm{Coceal}, \binits{O.}},
\bauthor{\bsnm{Martilli}, \binits{A.}}:
\batitle{How to parametrize urban-canopy drag to reproduce wind-direction
  effects within the canopy}.
\bjtitle{Boundary-Layer Meteorology}
\bvolume{149},
\bfpage{43}--\blpage{63}
(\byear{2013})
\doiurl{10.1007/s10546-013-9833-y}
\end{barticle}
\endbibitem

\bibitem[\protect\citeauthoryear{Shui et~al.}{2024}]{Shui2024}
\begin{barticle}
\bauthor{\bsnm{Shui}, \binits{T.}},
\bauthor{\bsnm{Gu}, \binits{Z.}},
\bauthor{\bsnm{Wang}, \binits{W.}}:
\batitle{Three-dimensional large eddy simulation urban neighborhood model with
  updated building drag coefficient and universal multiscale Smagorinsky model}.
\bjtitle{Physics of Fluids}
\bvolume{36}
\bfpage{075167}
(\byear{2024})
\doiurl{0.1063/5.0216385}
\end{barticle}
\endbibitem

\bibitem[\protect\citeauthoryear{S{\"{u}}tzl et~al.}{2021}]{Sutzl2021}
\begin{barticle}
\bauthor{\bsnm{S{\"{u}}tzl}, \binits{B.S.}},
\bauthor{\bsnm{Rooney}, \binits{G.G.}},
\bauthor{\bsnm{Reeuwijk}, \binits{M.}}:
\batitle{{Drag Distribution in Idealized Heterogeneous Urban Environments}}.
\bjtitle{Boundary-Layer Meteorology}
\bvolume{178}(\bissue{2}),
\bfpage{225}--\blpage{248}
(\byear{2021})
\doiurl{10.1007/s10546-020-00567-0}
\end{barticle}
\endbibitem

\bibitem[\protect\citeauthoryear{Toparlar et~al.}{2015}]{Toparlar2015}
\begin{barticle}
\bauthor{\bsnm{Toparlar}, \binits{Y.}},
\bauthor{\bsnm{Blocken}, \binits{B.}},
\bauthor{\bsnm{Vos}, \binits{P.}},
\bauthor{\bsnm{Heijst}, \binits{G.J.F.V.}},
\bauthor{\bsnm{Janssen}, \binits{W.D.}},
\bauthor{\bsnm{Hooff}, \binits{T.V.}},
\bauthor{\bsnm{Montazeri}, \binits{H.}},
\bauthor{\bsnm{Timmermans}, \binits{H.J.P.}}:
\batitle{{CFD simulation and validation of urban microclimate : A case study
  for Bergpolder Zuid , Rotterdam}}.
\bjtitle{Building and Environment}
\bvolume{83},
\bfpage{79}--\blpage{90}
(\byear{2015})
\doiurl{10.1016/j.buildenv.2014.08.004}
\end{barticle}
\endbibitem

\bibitem[\protect\citeauthoryear{Teng et~al.}{2025}]{Teng2025}
\begin{barticle}
\bauthor{\bsnm{Teng}, \binits{M.}},
\bauthor{\bsnm{Dur\'o~Diaz}, \binits{J.M.}},
\bauthor{\bsnm{Mestres}, \binits{E.}},
\bauthor{\bsnm{Muela~Castro}, \binits{J.}},
\bauthor{\bsnm{Lehmkuhl}, \binits{O.}},
\bauthor{\bsnm{Rodriguez}, \binits{I.}}:
\batitle{Atmospheric boundary layer over urban roughness: Validation of
  large-eddy simulation}.
\bjtitle{Physics of Fluids}
\bvolume{37}(\bissue{6}),
\bfpage{065129}
(\byear{2025})
\doiurl{10.1063/5.0265556}
\end{barticle}
\endbibitem

\bibitem[\protect\citeauthoryear{Tian et~al.}{2024}]{Tian2024}
\begin{barticle}
\bauthor{\bsnm{Tian}, \binits{G.}},
\bauthor{\bsnm{Ma}, \binits{Y.}},
\bauthor{\bsnm{Chen}, \binits{Y.}},
\bauthor{\bsnm{Wan}, \binits{M.}},
\bauthor{\bsnm{Chen}, \binits{S.}}:
\batitle{Impact of urban canopy characteristics on turbulence dynamics}.
\bjtitle{Building and Environment}
\bvolume{250},
\bfpage{111183}
(\byear{2024})
\doiurl{10.1016/j.buildenv.2024.111183}
\end{barticle}
\endbibitem

\bibitem[\protect\citeauthoryear{Vreman}{2004}]{vreman2004}
\begin{barticle}
\bauthor{\bsnm{Vreman}, \binits{A.W.}}:
\batitle{An eddy-viscosity subgrid-scale model for turbulent shear flow:
  {A}lgebraic theory and applications}.
\bjtitle{Physics of Fluids}
\bvolume{16}(\bissue{10}),
\bfpage{3670}--\blpage{3681}
(\byear{2004})
\end{barticle}
\endbibitem

\bibitem[\protect\citeauthoryear{Wang et~al.}{2020}]{Wang2020}
\begin{barticle}
\oauthor{\bsnm{Wang}, \binits{W.}},
\oauthor{\bsnm{Yang}, \binits{T.}},
\oauthor{\bsnm{Li}, \binits{Y.}},
\oauthor{\bsnm{Xu}, \binits{Y.}},
\oauthor{\bsnm{Chang}, \binits{M.}},
\oauthor{\bsnm{Wang}, \binits{X.}}:
\batitle{Identification of pedestrian-level ventilation corridors in downtown
  beijing using large-eddy simulations}.
\bjtitle{Building and Environment}
\bvolume{182},
\bfpage{107169}
(\byear{2020})
\doiurl{10.1016/j.buildenv.2020.107169}
\end{barticle}
\endbibitem

\bibitem[\protect\citeauthoryear{Wang et~al.}{2023}]{Wang2023}
\begin{barticle}
\bauthor{\bsnm{Wang}, \binits{Y.}},
\bauthor{\bsnm{Li}, \binits{J.}},
\bauthor{\bsnm{Liu}, \binits{W.}},
\bauthor{\bsnm{Zhang}, \binits{S.}},
\bauthor{\bsnm{Dong}, \binits{J.}},
\bauthor{\bsnm{Liu}, \binits{J.}}:
\batitle{Prediction of urban airflow fields around isolated high-rise buildings
  using data-driven non-linear correction models}.
\bjtitle{Building and Environment}
\bvolume{246},
\bfpage{110894}
(\byear{2023})
\doiurl{10.1016/j.buildenv.2023.110894}
\end{barticle}
\endbibitem

\bibitem[\protect\citeauthoryear{Wang et~al.}{2025}]{Wang2025}
\begin{barticle}
\bauthor{\bsnm{Wang}, \binits{J.}},
\bauthor{\bsnm{Llaguno-Munitxa}, \binits{M.}},
\bauthor{\bsnm{Li}, \binits{Q.}},
\bauthor{\bsnm{Giometto}, \binits{M.}},
\bauthor{\bsnm{{Bou- Zeid}}, \binits{E.}}:
\batitle{{Wind Extremes over Built Terrain: Characterization and Geometric
  Determinants}}.
\bjtitle{Boundary-Layer Meteorology}
\bvolume{191}(\bissue{2}),
\bfpage{11}
(\byear{2025})
\doiurl{10.1007/s10546-025-00899-9}
\end{barticle}
\endbibitem

\bibitem[\protect\citeauthoryear{Xie and Castro}{2006}]{xie2006}
\begin{barticle}
\bauthor{\bsnm{Xie}, \binits{Z.}},
\bauthor{\bsnm{Castro}, \binits{I.P.}}:
\batitle{{LES} and {RANS} for turbulent flow over arrays of wall-mounted
  obstacles}.
\bjtitle{Flow, Turbulence and Combustion}
\bvolume{76}(\bissue{3}),
\bfpage{291}--\blpage{312}
(\byear{2006})
\end{barticle}
\endbibitem


\bibitem[\protect\citeauthoryear{Yan et~al.}{2022}]{Yan2022}
\begin{barticle}
\bauthor{\bsnm{Yan}, \binits{Z.}},
\bauthor{\bsnm{Chen}, \binits{R.}},
\bauthor{\bsnm{Cai}, \binits{X.C.}}:
\batitle{{Large eddy simulation of the wind flow in a realistic full-scale
  urban community with a scalable parallel algorithm}}.
\bjtitle{Computer Physics Communications}
\bvolume{270},
\bfpage{108170}
(\byear{2022})
\doiurl{10.1016/j.cpc.2021.108170}
\end{barticle}
\endbibitem

\bibitem[\protect\citeauthoryear{Zheng et~al.}{2022}]{Zheng2022}
\begin{barticle}
\bauthor{\bsnm{Zheng}, \binits{X.}},
\bauthor{\bsnm{Montazeri}, \binits{H.}},
\bauthor{\bsnm{Blocken}, \binits{B.}}:
\batitle{Impact of building façade geometrical details on pollutant dispersion
  in street canyons}.
\bjtitle{Building and Environment}
\bvolume{212},
\bfpage{108746}
(\byear{2022})
\doiurl{10.1016/j.buildenv.2021.108746}
\end{barticle}
\endbibitem

\end{thebibliography}
\end{document}